\documentclass[useAMS,usegraphicx,usenatbib,a4paper]{mn2e}

\def\kms{km\,s$^{-1}$}

\def\HII{H\,{\sc ii}}
\def\CI{C\,{\sc i}}
\def\CII{C\,{\sc ii}}
\def\OI{O\,{\sc i}}
\def\OIa{[O\,{\sc i}]}

\def\NaI{Na\,{\sc i}}

\def\SII{S\,{\sc ii}}

\def\SiII{Si\,{\sc ii}}
\def\SiIII{Si\,{\sc iii}}

\def\MgII{Mg\,{\sc ii}}
\def\CaII{Ca\,{\sc ii}}
\def\FeI{Fe\,{\sc i}}
\def\FeII{Fe\,{\sc ii}}
\def\FeIII{Fe\,{\sc iii}}
\def\CoII{Co\,{\sc ii}}
\def\CoIII{Co\,{\sc iii}}
\def\NiII{Ni\,{\sc ii}}
\def\TiII{Ti\,{\sc ii}}
\def\CrII{Cr\,{\sc ii}}
\def\CaII{Ca\,{\sc ii}}
\def\CaIIa{[Ca\,{\sc ii}]}
\def\HeI{He\,{\sc i}}
\def\Nifs{$^{56}$Ni}
\def\Cofs{$^{56}$Co}
\def\Fefs{$^{56}$Fe}
\def\Msun{$M_{\odot}$}

\def\ebv{$E(B-V)$}
\def\dm15{$\Delta m_{15}(B)$}
\def\MCh{$M_\mathrm{Ch}$}
\def\lesssim{\mathrel{\hbox{\rlap{\hbox{\lower4pt\hbox{$\sim$}}}\hbox{$<$}}}}

\def\gtrsim{\mathrel{\hbox{\rlap{\hbox{\lower4pt\hbox{$\sim$}}}\hbox{$>$}}}}

\title[SN~2009dc]{High luminosity, slow ejecta and persistent carbon lines: 
\newline SN~2009dc challenges thermonuclear explosion scenarios.
}
\author[Taubenberger et al.]{S. Taubenberger$^{1}$
S.~Benetti$^{2}$, M.~Childress$^{3,4}$, R.~Pakmor$^{1}$, S.~Hachinger$^{1}$, 
\newauthor P.~A.~Mazzali$^{1,2,5}$, V.~Stanishev$^{6}$, N.~Elias-Rosa$^{7,8}$, I.~Agnoletto$^{2}$, F.~Bufano$^{2}$, 
\newauthor M.~Ergon$^{9}$, A.~Harutyunyan$^{10}$, C. Inserra$^{11}$, E.~Kankare $^{12,13}$, M.~Kromer$^{1}$, 
\newauthor H.~Navasardyan$^{2}$,  J.~Nicolas$^{14}$,  A.~Pastorello$^{15}$, E.~Prosperi$^{16}$, F.~Salgado$^{17,18}$, 
\newauthor  J.~Sollerman$^{9}$, M.~Stritzinger$^{9,17}$,  M.~Turatto$^{11}$, S.~Valenti$^{15}$ \& W.~Hillebrandt$^{1}$\\
$^{1}$Max-Planck-Institut f\"{u}r Astrophysik, Karl-Schwarzschild-Str. 1, 85741 Garching bei M\"{u}nchen, Germany\\
$^{2}$INAF Osservatorio Astronomico di Padova, Vicolo dell'Osservatorio 5, 35122 Padova, Italy\\
$^{3}$Department of Physics, University of California Berkeley, 366 LeConte Hall MC 7300, Berkeley, CA 94720-7300, USA\\
$^{4}$Physics Division, Lawrence Berkeley National Laboratory, 1 Cyclotron Road, Berkeley, CA 94720, USA\\
$^{5}$Scuola Normale Superiore, Piazza dei Cavalieri 7, 56126 Pisa, Italy\\
$^{6}$CENTRA -- Centro Multidisciplinar de Astrof\'isica, Instituto Superior T\'ecnico, Av. Rovisco Pais 1, 1049-001 Lisbon, Portugal\\
$^{7}$Spitzer Science Center, California Institute of Technology, 1200 E. California Blvd., Pasadena, CA 91125, USA\\
$^{8}$Department of Astronomy, University of California, Berkeley, CA 94720-3411, USA\\
$^{9}$Oskar Klein Centre, Department of Astronomy, AlbaNova, Stockholm University, 106 91 Stockholm, Sweden\\
$^{10}$Fundaci\'{o}n Galileo Galilei-INAF, Telescopio Nazionale Galileo, E-38700 Santa Cruz de La Palma, Tenerife, Spain\\
$^{11}$INAF Osservatorio Astrofisico di Catania, Via S.Sofia 78, 95123 Catania, Italy\\
$^{12}$Tuorla Observatory, Department of Physics and Astronomy, University of Turku, Fl-21500, Piikki\"{o}, Finland\\
$^{13}$Nordic Optical Telescope, Apartado 474, E-38700 Santa Cruz de La Palma, Tenerife, Spain\\
$^{14}$364 chemin de Notre Dame, 06220 Vallauris, France\\
$^{15}$Astrophysics Research Centre, School of Mathematics and Physics, Queen's University Belfast, Belfast BT7 1NN, UK\\
$^{16}$Osservatorio Astronomico di Castelmartini, via Bartolini 1317, 51036 Larciano, Pistoia, Italy\\
$^{17}$Las Campanas Observatory, Carnegie Observatories, Casilla 601, La Serena, Chile\\
$^{18}$Departamento de Astronom\'ia, Universidad de Chile, Casilla 36-D, Santiago, Chile}

\begin{document}

\date{Accepted 2010 November 25. Received 2010 November 24; in original form 2010 November 9}

\pagerange{\pageref{firstpage}--\pageref{lastpage}} \pubyear{2010}

\maketitle

\label{firstpage}

\begin{abstract}
  Extended optical and near-IR observations reveal that SN~2009dc
  shares a number of similarities with normal Type Ia supernovae
  (SNe~Ia), but is clearly overluminous, with a (pseudo-bolometric)
  peak luminosity of $\log(L)=43.47$ [erg s$^-1$]. Its light curves
  decline slowly over half a year after maximum light
  (\dm15$_\mathrm{true}=0.71$), and the early-time near-IR light
  curves show secondary maxima, although the minima between the first
  and the second peaks are not very pronounced. The bluer bands
  exhibit an enhanced fading after $\sim$\,200\,d, which might be
  caused by dust formation or an unexpectedly early IR
  catastrophe. The spectra of SN~2009dc are dominated by
  intermediate-mass elements and unburned material at early times, and
  by iron-group elements at late phases. Strong \CII\ lines are
  present until $\sim$\,2 weeks past maximum, which is unprecedented
  in thermonuclear SNe. The ejecta velocities are significantly lower
  than in normal and even subluminous SNe~Ia. No signatures of CSM
  interaction are found in the spectra. Assuming that the light curves
  are powered by radioactive decay, analytic modelling suggests that
  SN~2009dc produced $\sim$\,$1.8$ \Msun\ of \Nifs\ assuming the
  smallest possible rise time of 22\,d. Together with a derived total
  ejecta mass of $\sim$\,$2.8$ \Msun, this confirms that SN~2009dc is
  a member of the class of possible super-Chandrasekhar-mass SNe~Ia
  similar to SNe~2003fg, 2006gz and 2007if. A study of the hosts of
  SN~2009dc and other superluminous SNe~Ia reveals a tendency of these
  SNe to explode in low-mass galaxies. A low metallicity of the
  progenitor may therefore be an important pre-requisite for producing
  superluminous SNe~Ia. We discuss a number of possible explosion
  scenarios, ranging from super-Chandrasekhar-mass white-dwarf
  progenitors over dynamical white-dwarf mergers and Type
  I$\frac{1}{2}$ SNe to a core-collapse origin of the explosion. None
  of the models seem capable of explaining all properties of
  SN~2009dc, so that the true nature of this SN and its peers remains
  nebulous.
\end{abstract}


\section{Introduction}
\label{Introduction}

Owing to their remarkable homogeneity in peak luminosity and
light-curve shape, Type Ia supernovae (SNe~Ia) are considered
excellent tools to measure luminosity distances in the Universe,
suitable to constrain possible cosmologies also beyond the converged
$\Lambda$CDM model \citep[e.g.][]{Leibundgut01}. In particular, with
sufficiently precise observations of SNe~Ia at low and high redshift
the expansion history of the Universe can be reconstructed and a
possible time evolution of Dark Energy can be probed
\citep[e.g.][]{Riess07}.

Thanks to extensive observational campaigns and sophisticated
modelling efforts over the past decade, some convergence has been
achieved about the origin of `normal' SNe~Ia. There is a widespread
consensus on the single-degenerate (SD) scenario of a white dwarf (WD)
accreting matter from a non-degenerate companion until it approaches
the Chandrasekhar mass (\MCh) and ignites carbon near its centre. This
leads to a thermonuclear runaway disrupting the star
\citep[e.g.][]{Hillebrandt00}. Since the explosions always occur close
to \MCh, this scenario provides a natural explanation for the observed
homogeneity among `normal' SNe~Ia. Within this picture the ratio of
nuclear-statistical-equilibrium (NSE) material to intermediate-mass
elements (IME) in the ejecta is likely the key parameter for both the
width of the SN light curve (through the opacity generated by Fe-group
elements) and its peak luminosity (through the radioactive decay of
\Nifs; \citealt{Pinto01,Mazzali01,Mazzali07}).

However, this emerging picture has recently been challenged by the
discovery of a handful of objects whose properties cannot readily be
explained within the \MCh\ framework
\citep{Howell06,Branch06,Hicken07,Yamanaka09}. These SNe are
characterised by high peak luminosities, a factor $\sim$\,2 larger
than in all other SNe~Ia ($M_{V,\mathrm{max}}\,\sim\,-20$). At the
same time, apart from their comparatively low ejecta velocities they
share strong spectroscopic similarity with ordinary SNe~Ia. To explain
the early light curves of these events within spherical symmetry, a
\Nifs\ mass exceeding 1 \Msun\ and a total ejecta mass in excess of
1.4 \Msun\ are required \citep{Howell06}. The explosion of a
super-\MCh\ WD stabilised by strong differential rotation
\citep{Howell06,Branch06}, or a merger of two WDs, in sum again
exceeding \MCh\ \citep{Hicken07}, have been suggested as possible
scenarios. However, \citet{Hillebrandt07} and \citet{Sim07} argued
that these events could possibly be consistent with \MCh-WD explosions
if strong deviations from spherical symmetry are invoked, with the
ignition point located far off-centre and the \Nifs\ distribution being
very one-sided.  Alternatively, energy sources other than
radioactivity may have to be considered.

Understanding superluminous SNe~Ia is not only a merit by its own, but has 
important implications for cosmology. Spectroscopically similar to ordinary 
SNe~Ia, some of these events may have entered into the cosmological SN~Ia 
data sets, especially at high $z$ where they are favoured by their luminosity 
and where the data quality is mostly poor. Moreover, while they are apparently 
rare in the present Universe, without a better knowledge of their progenitors 
and explosion mechanisms it cannot be ruled out that superluminous SNe~Ia were 
more abundant in the past. Since they may not obey the light-curve width -- 
luminosity relations used to standardise SNe~Ia, they may introduce systematic 
errors in the reconstruction of $H(z)$ using SN data.

With the observations of SN~2009dc presented in this work, a comprehensive optical 
and near-IR data set of a superluminous SN~Ia becomes available, extending 
observations of the same SN presented by \citet{Yamanaka09} and \citet{Silverman10} 
in both temporal and wavelength coverage. This allows us to study the properties 
and evolution of one member of this class in unprecedented detail, and to put 
constraints on possible explosion scenarios.
The paper is organised as follows: in Section~\ref{Observations and data reduction} 
the observations are presented and the techniques applied for data reduction and 
calibration are discussed. Sections~\ref{Photometric evolution} and 
\ref{Spectroscopic evolution} are devoted to the analysis of the photometric and 
spectroscopic evolution of SN~2009dc, respectively. Important physical properties 
of SN~2009dc and possible explosion mechanisms are discussed in 
Section~\ref{Discussion}, before a brief summary of our main results is given 
in Section~\ref{Conclusions}.

\section{Observations and data reduction}
\label{Observations and data reduction}

SN~2009dc was discovered in the course of the Puckett Observatory
Supernova Search on UT 2009 April 9.31 at an unfiltered magnitude of
$16.5$, and confirmed on unfiltered exposures on April 10.42 at a
magnitude of $16.3$ \citep*{CBET1762}. No object was visible at the SN
position on images taken by T. Puckett on 2009 March 21 to a limiting
magnitude of $19.3$. A classification spectrum taken with the
Telescopio Nazionale Galileo + DOLORES on UT 2009 April 16.22 revealed
that SN~2009dc was a Type Ia supernova well before maximum light
\citep*{CBET1768}. The SN was reported to share similarity with
SN~2006gz \citep{Hicken07} at pre-maximum phases in most spectral
features, including prominent \CII\ absorption lines, but lower
expansion velocities. The presence of carbon and the resemblance to
suspected super-\MCh\ SNe~Ia such as SN~2003fg \citep{Howell06} was
confirmed by \citet*{CBET1776} based on optical and IR spectra
obtained on UT 2009 April 18 and 19. \citet{Yamanaka09} measured a
peak absolute magnitude of M$_{V,\mathrm{max}}$\,$\sim$\,$-20$ and a
slow light-curve decline of \dm15 $=0.65$, and estimated an ejected
\Nifs\ mass of at least $1.2$ \Msun. They concluded that the
spectrophotometric properties of SN~2009dc are consistent with the
explosion of a super-\MCh\ WD. In spectropolarimetric observations
presented by \citet{Tanaka09}, SN~2009dc shows moderately strong
polarisation in \SiII\ and \CaII\ lines, but small continuum
polarisation indicative of spherical symmetry on global scales. The
authors consider this as support for the explosion scenario suggested
by \citet{Yamanaka09}.

\subsection{Distance and extinction}
\label{Distance and extinction}

SN~2009dc is located in the outskirts of the S0 galaxy UGC~10064 (but see 
Section~\ref{09dc host} for an in-depth discussion on the host-galaxy 
interaction with the late-type UGC~10063, and the expected stellar population), 
at a redshift of $z = 0.0214$ (NED\footnote{NASA/IPAC Extragalactic Database\\ 
\hspace*{0.18cm} http:/$\!$/nedwww.ipac.caltech.edu/}). This is well within 
the Hubble flow, and a kinematic distance modulus of $\mu = 34.86 \pm 0.08$ mag 
is assumed, based on the average recession velocity of UGC~10064 and UGC~10063 
corrected for local flow patterns (see Table~\ref{properties} for details).

The extinction towards SN~2009dc is subject to some
uncertainty. Narrow \NaI\,D absorption lines can be discerned in the
spectra both at zero redshift and the host-galaxy rest frame,
suggesting a non-negligible amount of dust in the line of sight. The
respective equivalent widths are $0.61 \pm 0.13$ and $0.94 \pm 0.15$
\AA, measured from thirteen early-time spectra. Application of the
formula of \citet*{Turatto03} yields colour excesses of \ebv $= 0.10
\pm 0.02$ and $0.15 \pm 0.02$ mag in the Milky Way and the host
galaxy, respectively.  Alternatively, consulting the
\citet*{Schlegel98} dust maps suggests a foreground extinction of
$0.07$ mag towards UGC~10064, and we adopt this value for the Galactic
component. In `normal' SNe~Ia the colours measured at peak or during
the tail phase are used to infer the total reddening caused by dust
\citep[e.g.][]{Phillips99}. However, as will be shown in
Section~\ref{Colour evolution}, the colour evolution of SN~2009dc
differs significantly from that of ordinary SNe~Ia, so that these
methods cannot be applied with any confidence. Given the high
luminosity of SN~2009dc and its blue early-time colours already before
correcting for any host reddening, we rather consider the value
derived from the interstellar \NaI\,D lines as an upper limit for the
actual host-galaxy colour excess. In the further analysis we therefore
adopt a host-galaxy colour excess of \ebv $=0.10 \pm 0.07$ mag as our
best estimate, 50 per cent larger than the Galactic colour excess as
motivated by the ratio of the \NaI\,D lines.  Despite our
multi-wavelength coverage, the unknown intrinsic colours of the SN
also prevent us from determining the dust properties directly. We
therefore adopt a \citet*{Cardelli89} reddening law with $R_V = 3.1$,
assuming dust properties not too different from those in the
Galaxy\footnote{Adopting a lower value for the host-galaxy reddening
  of e.g. $R_V = 2.36$ as suggested by \citet{Wang09a} for normal
  SNe~Ia would reduce the inferred extinction in $V$ (and hence the
  peak luminosity calculated in Section~\ref{Bolometric light curve})
  by $\sim$\,7 per cent.}.

Details on the SN, the presumed host galaxy UGC~10064 and its interacting 
companion UGC~10063 are summarised in Table~\ref{properties}.

\begin{figure}
   \centering
   \includegraphics[width=8.4cm]{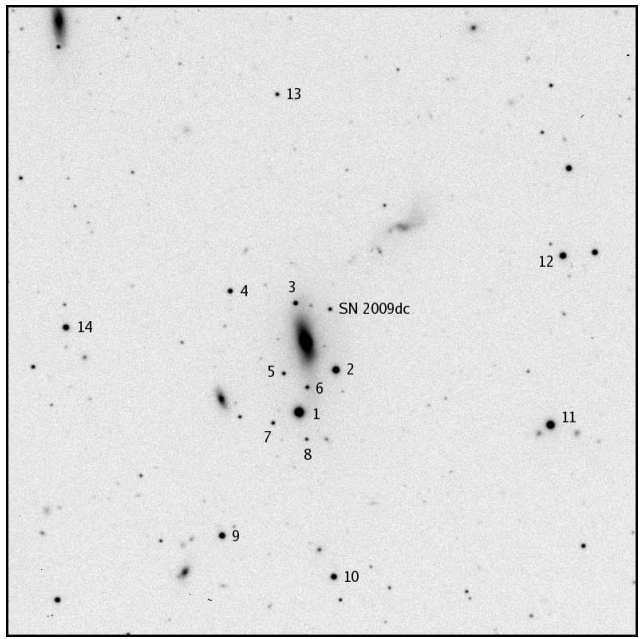}
   \caption{$R$-band image of the SN~2009dc field taken with the Calar Alto 
   2.2m Telescope + CAFOS on UT 2009 August 13. The field of view is $7.0 \times 
   7.0$ arcmin$^2$, north is up and east to the left. The local sequence stars 
   are indicated.}
   \label{fig:chart}
\end{figure}

\begin{table}
\caption{Properties of SN~2009dc and its possible host galaxies.}
\label{properties}
\begin{center}
\begin{footnotesize}
\begin{tabular}{lrr}
\hline
SN~2009dc & & \\
\hline
$\alpha$\,(J2000)                      &  15$^\mathrm{h}$51$^\mathrm{m}$12\fs10                   & 1 \\
$\delta$\,(J2000)                      &  +25\degr42\arcmin28\farcs5      & 1 \\
offset from UGC 10064 nucleus          &  15\farcs8 W, 20\farcs8 N        & 1 \\
host reddening $E(B\!-\!V)$            &  $0.10 \pm 0.07$ mag             & 2 \\
\dm15$_\mathrm{true}$                   &  $0.71 \pm 0.03$                 & 2 \\
JD$_{\mathrm{max},U}$                    &  $2\,454\,945.2 \pm 0.5$         & 2 \\
JD$_{\mathrm{max},B}$                    &  $2\,454\,947.1 \pm 0.3$         & 2 \\
JD$_{\mathrm{max},V}$                    &  $2\,454\,947.7 \pm 0.4$         & 2 \\
JD$_{\mathrm{max},R}$                    &  $2\,454\,948.2 \pm 0.5$         & 2 \\
JD$_{\mathrm{max},I}$                    &  $2\,454\,949.1 \pm 0.8$         & 2 \\
$U_\mathrm{max}$                        &  $14.604 \pm 0.038$              & 2 \\ 
$B_\mathrm{max}$                        &  $15.329 \pm 0.017$              & 2 \\ 
$V_\mathrm{max}$                        &  $15.319 \pm 0.017$              & 2 \\ 
$R_\mathrm{max}$                        &  $15.304 \pm 0.018$              & 2 \\ 
$I_\mathrm{max}$                        &  $15.385 \pm 0.018$              & 2 \\ 
M$_{U,\mathrm{max}}$                     &  $-21.06 \pm 0.34$               & 2 \\ 
M$_{B,\mathrm{max}}$                     &  $-20.22 \pm 0.30$               & 2 \\ 
M$_{V,\mathrm{max}}$                     &  $-20.07 \pm 0.23$               & 2 \\ 
M$_{R,\mathrm{max}}$                     &  $-19.99 \pm 0.20$               & 2 \\ 
M$_{I,\mathrm{max}}$                     &  $-19.79 \pm 0.16$               & 2 \\ 
\hline  
\end{tabular}
\vspace{0.3cm}
\begin{tabular}{lrr}
\hline 
UGC 10064 & & \\
\hline
$\alpha$\,(J2000)                      &  15$^\mathrm{h}$51$^\mathrm{m}$13\fs28                   & 3 \\
$\delta$\,(J2000)                      &  +25\degr42\arcmin07\farcs5      & 3 \\
redshift                               &  $0.02139 \pm 0.00007$           & 4 \\
$v_\mathrm{CMB}^a$                      &  $6508 \pm 22$ km\,s$^{-1}$      & 4 \\
$v_\mathrm{Virgo}^b$                    &  $6686 \pm 25$ km\,s$^{-1}$      & 4 \\
$v_\mathrm{Virgo\,+\,GA\,+\,Shapley}^b$ &  $7012 \pm 32$ km\,s$^{-1}$      & 4 \\
distance modulus $\mu^c$               &  $34.85 \pm 0.08$ mag            & 4 \\
extinction-corr. app. $B$ mag          &  $14.55 \pm 0.28$                & 3 \\
absolute $B$ magnitude                 &  $-20.30 \pm 0.29$               & 3,4 \\
morphological type$^d$                 &  S0, $-1.9$                      & 3 \\
Galactic reddening $E(B\!-\!V)\!\!\!$  &  $0.071$ mag                     & 5 \\
\hline
\end{tabular}
\vspace{0.3cm}
\begin{tabular}{lrr}
\hline
UGC 10063 & & \\
\hline
$\alpha$\,(J2000)                      &  15$^\mathrm{h}$51$^\mathrm{m}$08\fs51                   & 3 \\
$\delta$\,(J2000)                      &  +25\degr43\arcmin21\farcs5      & 3 \\
redshift                               &  $0.02158 \pm 0.00003$           & 4 \\
$v_\mathrm{CMB}^a$                      &  $6565 \pm 11$ km\,s$^{-1}$      & 4 \\
$v_\mathrm{Virgo}^b$                    &  $6743 \pm 16$ km\,s$^{-1}$      & 4 \\
$v_\mathrm{Virgo\,+\,GA\,+\,Shapley}^b$ &  $7068 \pm 25$ km\,s$^{-1}$      & 4 \\
distance modulus $\mu^c$               &  $34.87 \pm 0.08$ mag            & 4 \\
extinction-corr. app. vis. mag         &  $16.3$                          & 4 \\
absolute vis. magnitude                &  $-18.6$                         & 4 \\
morphological type$^d$                 &  SBd, $8.0$                      & 3 \\
Galactic reddening $E(B\!-\!V)\!\!\!$  &  $0.071$ mag                     & 5 \\
\hline
\end{tabular}
\\[1.5ex]
\flushleft
$^a$~recession velocity corrected to the CMB rest frame \citep{Fixsen96}\quad
$^b$~recession velocity corrected for Local-Group infall onto the Virgo cluster (+\,the Great 
     Attractor\,+\,the Shapley Supercluster; \citealt{Mould00})\quad
$^c$~from the average of $v_\mathrm{CMB}$, $v_\mathrm{Virgo}$ and $v_\mathrm{Virgo\,+\,GA\,+\,Shapley}$, 
     using $H_0=72\,\rmn{km}\,\rmn{s}^{-1}\rmn{Mpc}^{-1}$\quad
$^d$~numerical code according to de Vaucouleurs\\[1.8ex] 
1: \citealt{CBET1776}; 2: this work; 3: LEDA; 4: NED; 5: \citealt{Schlegel98}\\ 
\end{footnotesize}
\end{center} 
\end{table}

\subsection{Reduction of photometric data}
\label{Photometry}

The reduction of optical photometric data (bias subtraction, overscan correction and 
flat-fielding) was performed using standard routines in {\sc iraf}\footnote{{\sc 
iraf} is distributed by the National Optical Astronomy Observatories, which are 
operated by the Association of Universities for Research in Astronomy, Inc, under 
contract to the National Science Foundation.}. In the near-IR, an in-field dithering 
strategy allowed for the creation of source-free sky images, which were subtracted 
from the individual scientific images to eliminate the strong near-IR sky emission. 
Sub-exposures taken with the same filter during one night were aligned and combined 
before the photometric measurements were performed.

A sequence of field stars (Fig.~\ref{fig:chart}) was calibrated with respect to 
\citet[][for $U\!BV\!RI$]{Landolt92} and Arnica \citep[][for $JHK'$]{Hunt98} standard 
fields on several photometric nights. The calibrated magnitudes of these field stars, 
listed in Table~\ref{sequence}, were used to determine the SN magnitudes under 
non-photometric conditions. For the six local standards we have in common, our 
magnitudes are consistent with those reported by \citet{Silverman10}, with average 
systematic differences ranging from 0.008 mag ($B$ band) to 0.015 mag ($V$ band).

\begin{table*}
\caption{Magnitudes of the local sequence stars in the field of
SN~2009dc (Fig.~\ref{fig:chart}).} 
\begin{footnotesize}
\begin{center}
\begin{tabular}{@{}ccccccccc@{}}
\hline
ID & $U$ & $B$ & $V$ & $R$ & $I$ & $J$ & $H$ & $K'$\\
\hline
 1  &  $15.894\pm0.029$  &  $15.368\pm0.016$  &  $14.460\pm0.014$  &  $13.922\pm0.018$  &  $13.445\pm0.013$ &  $12.785\pm0.022$ &  $12.308\pm0.058$ &  $12.188\pm0.012$\\
 2  &  $16.939\pm0.018$  &  $16.647\pm0.016$  &  $15.850\pm0.016$  &  $15.392\pm0.016$  &  $14.969\pm0.019$ &  $14.376\pm0.015$ &  $13.967\pm0.041$ &  $13.894\pm0.010$\\
 3  &  $19.512\pm0.055$  &  $18.890\pm0.018$  &  $17.968\pm0.022$  &  $17.411\pm0.028$  &  $16.936\pm0.017$ &  $16.252\pm0.052$ &  $15.786\pm0.028$ &  $15.724\pm0.039$\\
 4  &  $18.870\pm0.036$  &  $18.649\pm0.017$  &  $17.772\pm0.019$  &  $17.229\pm0.017$  &  $16.721\pm0.018$ &  $16.057\pm0.061$ &  $15.539\pm0.019$ &  $15.426\pm0.021$\\
 5  &                    &  $20.464\pm0.034$  &  $19.156\pm0.033$  &  $18.274\pm0.019$  &  $17.464\pm0.017$ &  $16.491\pm0.066$ &  $15.849\pm0.010$ &  $15.678\pm0.072$\\
 6  &  $19.103\pm0.109$  &  $19.191\pm0.022$  &  $18.535\pm0.020$  &  $18.141\pm0.022$  &  $17.737\pm0.030$ & & &\\
 7  &  $19.199\pm0.089$  &  $19.366\pm0.040$  &  $18.742\pm0.028$  &  $18.328\pm0.031$  &  $17.913\pm0.034$ & & &\\
 8  &                    &  $21.674\pm0.077$  &  $20.015\pm0.073$  &  $18.995\pm0.050$  &  $17.214\pm0.032$ &  $15.788\pm0.010$ &  $15.230\pm0.053$ &  $14.932\pm0.012$\\
 9  &  $17.102\pm0.013$  &  $17.152\pm0.030$  &  $16.544\pm0.022$  &  $16.161\pm0.007$  &  $15.792\pm0.017$ & & &\\
10  &  $17.520\pm0.038$  &  $17.479\pm0.021$  &  $16.803\pm0.015$  &  $16.395\pm0.012$  &  $16.011\pm0.011$ & & &\\
11  &  $16.009\pm0.028$  &  $15.903\pm0.015$  &  $15.195\pm0.009$  &  $14.793\pm0.018$  &  $14.412\pm0.016$ & & &\\
12  &  $19.307\pm0.051$  &  $18.123\pm0.025$  &  $16.788\pm0.011$  &  $15.936\pm0.020$  &  $15.217\pm0.014$ & & &\\
13  &                    &  $20.073\pm0.019$  &  $18.901\pm0.021$  &  $18.125\pm0.025$  &  $17.478\pm0.020$ & & &\\
14  &  $18.492\pm0.104$  &  $17.798\pm0.033$  &  $16.839\pm0.006$  &  $16.267\pm0.006$  &  $15.794\pm0.007$ & & &\\
\hline\\[-0.7ex]
\end{tabular}
\end{center}
\end{footnotesize}
\label{sequence}
\end{table*}

\begin{figure}
   \centering
   \includegraphics[width=8.4cm]{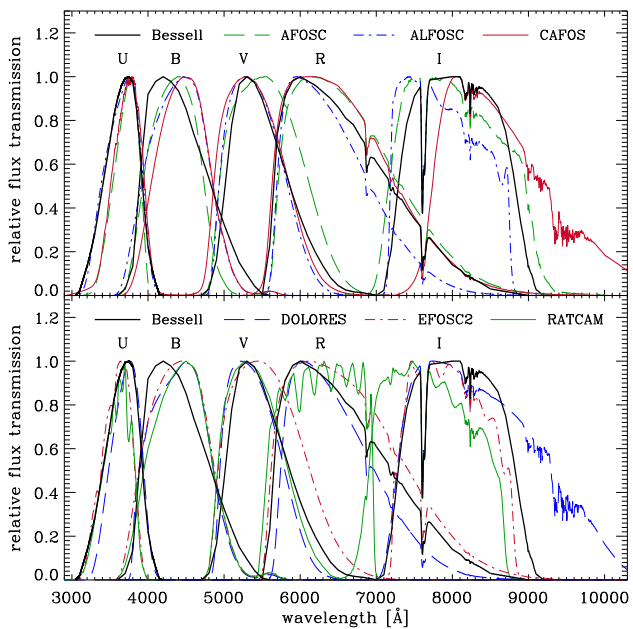}
   \caption{$U\!BV\!RI$ passbands of the instruments used during the 
   follow-up of SN~2009dc. The standard \citet{Bessell90} passbands are 
   displayed for comparison. Note that the CCD and $U\!BV\!RI$ filter set 
   of DOLORES were replaced in 2007. The curves shown here refer to 
   the new passbands, which deviate significantly from the old ones presented 
   e.g. in \citet{Taubenberger08}.}
   \label{fig:filters}
\end{figure}

All SN measurements were done with point-spread function (PSF) fitting photometry 
using the software package {\sc snoopy}, a dedicated tool for SN photometry 
written by F. Patat and implemented in {\sc iraf} by E. Cappellaro. Measurement 
errors were estimated through an artificial star experiment. Since instruments 
with very different passbands were used for the follow-up of SN~2009dc (see 
Fig.~\ref{fig:filters}), we made use of the `$S$-correction' technique 
\citep{Stritzinger02,Pignata04b} to calibrate the SN magnitudes to the standard 
photometric system of Johnson and Cousins \citep{Bessell90} in the optical bands. 
Our excellent spectroscopic coverage enabled the computation of the $S$-correction 
solely on the basis of SN~2009dc spectra. In the same manner, a $K$-correction was 
derived to compensate for the non-negligible redshift of SN~2009dc. At epochs 
where no spectra were available, the $S$- and $K$-correction terms were determined 
by linear interpolation or constant extrapolation. No $S$- and $K$-correction was 
applied in the near-IR bands, where the SN magnitudes were calibrated with 
first-order colour-term or simple zero-point corrections.

\begin{figure}
   \centering
   \includegraphics[width=8.4cm]{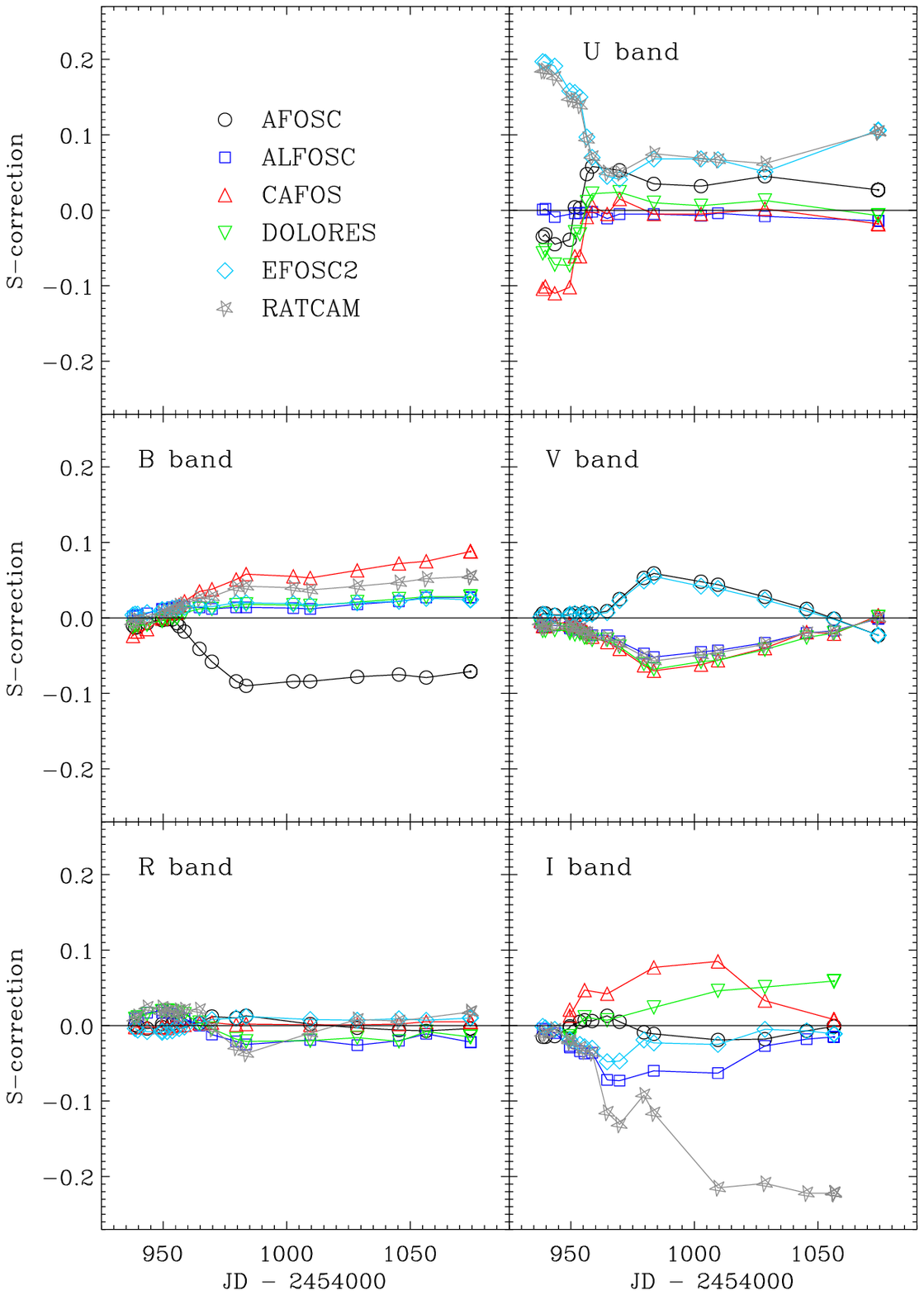}
   \caption{Temporal evolution of the $S$-correction in the $U\!BV\!RI$ bands 
   for the different instrumental configurations used for follow-up observations 
   of SN~2009dc.}
   \label{fig:Scorr}
\end{figure}

In Table~\ref{SN_mags} the fully calibrated $S$- and $K$-corrected
Bessell $U\!BV\!RI$ magnitudes of SN~2009dc are reported. The
uncertainties in brackets are the quadratic sums of the measurement
errors and the uncertainties in the photometric zero points of the
nights.  Table~\ref{S-corr} lists the $S$- and $K$-correction
separately, meant as the quantity added to the zero-point calibrated
magnitudes instead of a colour-term correction.  The temporal
evolution of the $S$-correction for the different instruments is shown
in Fig.~\ref{fig:Scorr}.  The colour-term corrected $JHK'$ magnitudes
and their uncertainties (determined analogously to the optical bands)
are shown in Table~\ref{SN_mags_IR}.

\begin{table*}
\caption{$S$- and $K$-corrected $U\!BV\!RI$ magnitudes of SN~2009dc.} 
\begin{footnotesize}
\begin{center}
\begin{tabular}{@{}rrcccccll@{}}
\hline
JD$^a$\  & Epoch$^b$    &        $U$         &        $B$         &        $V$         &        $R$         &        $I$         & Telescope & Seeing$^c$\\
\hline
 935.47\ & $-11.7$\ \ \ &                    &                    &                    &  $15.928\pm0.058$  &                    & \ \ JN    & \ \ 3.16 \\
 937.49\ &  $-9.7$\ \ \ &                    &                    &                    &  $15.608\pm0.059$  &                    & \ \ EP1   & \ \ 4.90 \\
 938.66\ &  $-8.5$\ \ \ &  $14.905\pm0.076$  &  $15.735\pm0.022$  &  $15.657\pm0.017$  &  $15.671\pm0.027$  &  $15.781\pm0.022$  & \ \ NOT   & \ \ 0.89 \\
 939.50\ &  $-7.6$\ \ \ &                    &                    &                    &  $15.443\pm0.036$  &                    & \ \ JN    & \ \ 2.96 \\
 939.69\ &  $-7.5$\ \ \ &  $14.800\pm0.040$  &  $15.639\pm0.015$  &  $15.552\pm0.023$  &  $15.603\pm0.017$  &  $15.702\pm0.027$  & \ \ NOT   & \ \ 0.70 \\
 942.44\ &  $-4.7$\ \ \ &                    &                    &                    &  $15.313\pm0.025$  &                    & \ \ JN    & \ \ 3.16 \\
 942.53\ &  $-4.6$\ \ \ &  $14.779\pm0.009$  &  $15.421\pm0.014$  &  $15.385\pm0.010$  &  $15.399\pm0.016$  &  $15.502\pm0.010$  & \ \ LT    & \ \ 1.84 \\
 943.49\ &  $-3.7$\ \ \ &  $14.648\pm0.025$  &  $15.423\pm0.027$  &  $15.378\pm0.023$  &  $15.390\pm0.027$  &  $15.520\pm0.024$  & \ \ NOT   & \ \ 2.03 \\
 944.57\ &  $-2.6$\ \ \ &  $14.767\pm0.010$  &  $15.347\pm0.016$  &  $15.325\pm0.012$  &  $15.339\pm0.011$  &  $15.439\pm0.016$  & \ \ LT    & \ \ 1.42 \\
 946.49\ &  $-0.7$\ \ \ &                    &                    &                    &  $15.247\pm0.036$  &                    & \ \ JN    & \ \ 3.16 \\
 949.53\ &   $2.4$\ \ \ &  $14.668\pm0.031$  &  $15.342\pm0.028$  &  $15.328\pm0.028$  &  $15.295\pm0.025$  &  $15.339\pm0.026$  & \ \ CA    & \ \ 2.54 \\
 949.54\ &   $2.4$\ \ \ &  $14.706\pm0.032$  &  $15.369\pm0.022$  &  $15.349\pm0.019$  &  $15.339\pm0.015$  &  $15.398\pm0.024$  & \ \ TNG   & \ \ 1.02 \\
 949.62\ &   $2.5$\ \ \ &  $14.869\pm0.026$  &  $15.348\pm0.013$  &  $15.326\pm0.012$  &  $15.297\pm0.023$  &  $15.390\pm0.016$  & \ \ LT    & \ \ 1.37 \\
 950.63\ &   $3.5$\ \ \ &  $14.943\pm0.046$  &  $15.371\pm0.012$  &  $15.350\pm0.011$  &  $15.308\pm0.011$  &  $15.387\pm0.008$  & \ \ LT    & \ \ 1.51 \\
 951.58\ &   $4.4$\ \ \ &                    &  $15.399\pm0.026$  &  $15.360\pm0.022$  &  $15.328\pm0.028$  &  $15.348\pm0.044$  & \ \ CA    & \ \ 1.70 \\
 953.68\ &   $6.5$\ \ \ &  $14.971\pm0.087$  &  $15.525\pm0.020$  &  $15.432\pm0.012$  &  $15.397\pm0.014$  &  $15.424\pm0.024$  & \ \ TNG   & \ \ 1.32 \\
 954.41\ &   $7.3$\ \ \ &                    &                    &                    &  $15.305\pm0.041$  &                    & \ \ JN    & \ \ 3.57 \\
 954.49\ &   $7.4$\ \ \ &  $15.257\pm0.018$  &  $15.542\pm0.022$  &  $15.414\pm0.009$  &  $15.350\pm0.011$  &  $15.406\pm0.017$  & \ \ LT    & \ \ 1.70 \\
 955.50\ &   $8.4$\ \ \ &  $15.128\pm0.045$  &  $15.560\pm0.041$  &  $15.428\pm0.013$  &  $15.377\pm0.022$  &  $15.411\pm0.030$  & \ \ CA    & \ \ 1.11 \\
 956.42\ &   $9.3$\ \ \ &  $15.425\pm0.037$  &  $15.628\pm0.029$  &  $15.453\pm0.019$  &  $15.410\pm0.012$  &  $15.400\pm0.007$  & \ \ LT    & \ \ 1.56 \\
 956.48\ &   $9.3$\ \ \ &  $15.258\pm0.072$  &  $15.671\pm0.022$  &  $15.455\pm0.014$  &  $15.424\pm0.032$  &  $15.442\pm0.031$  & \ \ NOT   & \ \ 0.82 \\
 958.63\ &  $11.5$\ \ \ &  $15.429\pm0.038$  &  $15.756\pm0.026$  &  $15.498\pm0.028$  &  $15.451\pm0.027$  &  $15.479\pm0.021$  & \ \ NOT   & \ \ 0.55 \\
 959.65\ &  $12.5$\ \ \ &  $15.746\pm0.036$  &  $15.809\pm0.014$  &  $15.533\pm0.022$  &  $15.438\pm0.007$  &  $15.408\pm0.020$  & \ \ LT    & \ \ 1.95 \\
 959.69\ &  $12.6$\ \ \ &  $15.547\pm0.033$  &  $15.857\pm0.016$  &  $15.527\pm0.036$  &  $15.519\pm0.024$  &  $15.473\pm0.028$  & \ \ NOT   & \ \ 1.25 \\
 963.52\ &  $16.4$\ \ \ &                    &  $16.115\pm0.018$  &  $15.672\pm0.018$  &  $15.557\pm0.019$  &  $15.429\pm0.010$  & \ \ LT    & \ \ 1.31 \\
 964.41\ &  $17.3$\ \ \ &  $16.128\pm0.049$  &  $16.202\pm0.019$  &  $15.696\pm0.017$  &  $15.576\pm0.005$  &  $15.412\pm0.029$  & \ \ LT    & \ \ 1.90 \\
 964.70\ &  $17.6$\ \ \ &  $16.001\pm0.049$  &  $16.217\pm0.024$  &  $15.684\pm0.021$  &  $15.570\pm0.023$  &  $15.398\pm0.027$  & \ \ TNG   & \ \ 1.35 \\
 966.43\ &  $19.3$\ \ \ &  $16.457\pm0.023$  &  $16.348\pm0.015$  &  $15.773\pm0.011$  &  $15.611\pm0.016$  &  $15.409\pm0.013$  & \ \ LT    & \ \ 1.37 \\
 968.56\ &  $21.4$\ \ \ &  $16.668\pm0.015$  &  $16.537\pm0.015$  &  $15.865\pm0.021$  &  $15.643\pm0.013$  &  $15.401\pm0.009$  & \ \ LT    & \ \ 1.34 \\
 969.71\ &  $22.6$\ \ \ &  $16.671\pm0.045$  &  $16.603\pm0.037$  &  $15.932\pm0.038$  &  $15.664\pm0.029$  &  $15.423\pm0.026$  & \ \ NTT   & \ \ 1.30 \\
 970.37\ &  $23.2$\ \ \ &                    &                    &                    &  $15.672\pm0.042$  &                    & \ \ JN    & \ \ 2.96 \\
 970.37\ &  $23.2$\ \ \ &                    &                    &                    &  $15.691\pm0.052$  &                    & \ \ EP2   & \ \ 3.17 \\
 972.39\ &  $25.3$\ \ \ &                    &                    &                    &  $15.705\pm0.045$  &                    & \ \ JN    & \ \ 3.16 \\
 973.47\ &  $26.3$\ \ \ &  $16.943\pm0.055$  &  $16.939\pm0.025$  &  $16.088\pm0.033$  &  $15.766\pm0.015$  &  $15.484\pm0.015$  & \ \ NOT   & \ \ 1.46 \\
 976.52\ &  $29.4$\ \ \ &  $17.237\pm0.035$  &  $17.145\pm0.025$  &  $16.179\pm0.020$  &  $15.828\pm0.035$  &  $15.494\pm0.025$  & \ \ NOT   & \ \ 0.59 \\
 977.38\ &  $30.2$\ \ \ &                    &                    &                    &  $15.845\pm0.042$  &                    & \ \ JN    & \ \ 3.16 \\
 977.43\ &  $30.3$\ \ \ &  $17.223\pm0.072$  &  $17.164\pm0.046$  &                    &                    &                    & \ \ Ekar  & \ \ 2.84 \\
 978.59\ &  $31.5$\ \ \ &                    &  $17.258\pm0.033$  &  $16.309\pm0.023$  &  $15.914\pm0.028$  &  $15.557\pm0.039$  & \ \ Ekar  & \ \ 2.65 \\
 981.44\ &  $34.3$\ \ \ &  $17.747\pm0.028$  &  $17.475\pm0.017$  &  $16.423\pm0.011$  &  $16.009\pm0.015$  &  $15.613\pm0.025$  & \ \ LT    & \ \ 1.37 \\
 982.62\ &  $35.5$\ \ \ &  $17.526\pm0.029$  &  $17.499\pm0.013$  &  $16.463\pm0.017$  &  $16.074\pm0.023$  &  $15.576\pm0.035$  & \ \ CA    & \ \ 1.75 \\
 983.41\ &  $36.3$\ \ \ &  $17.858\pm0.043$  &  $17.568\pm0.013$  &  $16.511\pm0.007$  &  $16.087\pm0.014$  &  $15.678\pm0.016$  & \ \ LT    & \ \ 1.28 \\
 987.54\ &  $40.4$\ \ \ &  $18.043\pm0.028$  &  $17.743\pm0.027$  &  $16.683\pm0.009$  &  $16.266\pm0.017$  &  $15.847\pm0.014$  & \ \ LT    & \ \ 1.06 \\
 988.53\ &  $41.4$\ \ \ &  $17.883\pm0.073$  &  $17.783\pm0.026$  &  $16.706\pm0.031$  &  $16.327\pm0.040$  &  $15.897\pm0.029$  & \ \ NOT   & \ \ 0.46 \\
 991.54\ &  $44.4$\ \ \ &  $18.129\pm0.037$  &  $17.866\pm0.017$  &  $16.809\pm0.025$  &  $16.420\pm0.016$  &  $15.973\pm0.012$  & \ \ LT    & \ \ 1.42 \\
 995.43\ &  $48.3$\ \ \ &  $18.283\pm0.050$  &  $17.930\pm0.023$  &  $16.934\pm0.024$  &  $16.555\pm0.017$  &  $16.125\pm0.031$  & \ \ LT    & \ \ 1.34 \\
 995.60\ &  $48.5$\ \ \ &  $18.014\pm0.043$  &  $17.985\pm0.020$  &  $16.945\pm0.024$  &  $16.596\pm0.017$  &  $16.187\pm0.033$  & \ \ NOT   & \ \ 0.95 \\
1006.46\ &  $59.3$\ \ \ &  $18.479\pm0.053$  &  $18.143\pm0.012$  &  $17.186\pm0.016$  &  $16.894\pm0.015$  &  $16.483\pm0.022$  & \ \ LT    & \ \ 1.17 \\
1007.48\ &  $60.3$\ \ \ &  $18.258\pm0.041$  &  $18.163\pm0.022$  &  $17.229\pm0.021$  &  $16.950\pm0.018$  &  $16.596\pm0.024$  & \ \ NOT   & \ \ 0.65 \\
1009.56\ &  $62.4$\ \ \ &  $18.528\pm0.051$  &  $18.183\pm0.014$  &  $17.248\pm0.022$  &  $16.969\pm0.016$  &  $16.575\pm0.014$  & \ \ LT    & \ \ 1.48 \\
1013.43\ &  $66.3$\ \ \ &  $18.567\pm0.052$  &  $18.255\pm0.017$  &  $17.333\pm0.017$  &  $17.085\pm0.025$  &  $16.699\pm0.014$  & \ \ LT    & \ \ 1.34 \\
1016.53\ &  $69.4$\ \ \ &  $18.265\pm0.074$  &  $18.282\pm0.018$  &  $17.374\pm0.025$  &  $17.153\pm0.018$  &  $16.766\pm0.022$  & \ \ TNG   & \ \ 1.24 \\
1021.42\ &  $74.3$\ \ \ &  $18.730\pm0.044$  &  $18.344\pm0.033$  &  $17.498\pm0.019$  &  $17.310\pm0.028$  &  $16.959\pm0.019$  & \ \ LT    & \ \ 1.40 \\
1022.43\ &  $75.3$\ \ \ &  $18.534\pm0.056$  &  $18.400\pm0.018$  &  $17.527\pm0.017$  &  $17.358\pm0.026$  &  $17.109\pm0.038$  & \ \ NOT   & \ \ 0.65 \\
1025.43\ &  $78.3$\ \ \ &  $18.854\pm0.049$  &  $18.416\pm0.013$  &  $17.582\pm0.018$  &  $17.414\pm0.011$  &  $17.099\pm0.022$  & \ \ LT    & \ \ 1.26 \\
1028.57\ &  $81.4$\ \ \ &  $18.690\pm0.064$  &  $18.443\pm0.040$  &  $17.679\pm0.026$  &  $17.507\pm0.033$  &  $17.205\pm0.028$  & \ \ NTT   & \ \ 1.92 \\
1032.42\ &  $85.3$\ \ \ &  $18.959\pm0.065$  &  $18.511\pm0.013$  &  $17.735\pm0.012$  &  $17.592\pm0.017$  &  $17.301\pm0.018$  & \ \ LT    & \ \ 1.51 \\
1033.42\ &  $86.3$\ \ \ &  $18.950\pm0.037$  &  $18.532\pm0.012$  &  $17.747\pm0.015$  &  $17.627\pm0.015$  &  $17.332\pm0.013$  & \ \ LT    & \ \ 1.87 \\
1034.42\ &  $87.3$\ \ \ &  $19.003\pm0.064$  &  $18.546\pm0.018$  &  $17.776\pm0.014$  &  $17.641\pm0.028$  &  $17.366\pm0.016$  & \ \ LT    & \ \ 1.40 \\
1037.41\ &  $90.3$\ \ \ &  $18.848\pm0.051$  &  $18.605\pm0.030$  &  $17.830\pm0.024$  &  $17.737\pm0.029$  &  $17.530\pm0.032$  & \ \ NOT   & \ \ 1.29 \\
1037.42\ &  $90.3$\ \ \ &  $19.080\pm0.067$  &  $18.586\pm0.016$  &  $17.827\pm0.013$  &  $17.726\pm0.012$  &  $17.428\pm0.019$  & \ \ LT    & \ \ 1.73 \\
1043.46\ &  $96.3$\ \ \ &  $18.951\pm0.085$  &  $18.626\pm0.029$  &  $17.912\pm0.020$  &  $17.876\pm0.033$  &  $17.624\pm0.030$  & \ \ CA    & \ \ 1.22 \\
1047.52\ & $100.4$\ \ \ &                    &  $18.748\pm0.026$  &  $18.056\pm0.027$  &  $17.987\pm0.019$  &  $17.707\pm0.021$  & \ \ LT    & \ \ 1.45 \\
1052.46\ & $105.3$\ \ \ &                    &  $18.813\pm0.040$  &  $18.141\pm0.028$  &  $18.120\pm0.016$  &  $17.834\pm0.020$  & \ \ LT    & \ \ 1.73 \\
\hline
\end{tabular}
\end{center}
\end{footnotesize} 
\label{SN_mags}
\end{table*}

\begin{table*}
\addtocounter{table}{-1}
\caption{\textit{-- continued.} $S$- and $K$-corrected $U\!BV\!RI$ magnitudes of SN~2009dc.} 
\begin{footnotesize}
\begin{center}
\begin{tabular}{@{}rrcccccll@{}}
\hline
JD$^a$\  & Epoch$^b$    &        $U$         &        $B$         &        $V$         &        $R$         &        $I$         & Telescope & Seeing$^c$\\
\hline
1056.37\ & $109.2$\ \ \ &                    &  $18.773\pm0.017$  &  $18.184\pm0.017$  &  $18.196\pm0.018$  &  $18.033\pm0.024$  & \ \ CA    & \ \ 1.54 \\
1056.40\ & $109.3$\ \ \ &  $19.231\pm0.034$  &  $18.833\pm0.023$  &  $18.187\pm0.020$  &  $18.219\pm0.024$  &  $17.947\pm0.029$  & \ \ NOT   & \ \ 0.68 \\
1061.38\ & $114.2$\ \ \ &                    &  $18.879\pm0.044$  &  $18.322\pm0.025$  &  $18.352\pm0.030$  &  $18.102\pm0.024$  & \ \ Ekar  & \ \ 1.70 \\
1063.34\ & $116.2$\ \ \ &                    &  $18.913\pm0.040$  &  $18.363\pm0.039$  &  $18.438\pm0.044$  &  $18.145\pm0.038$  & \ \ Ekar  & \ \ 1.51 \\
1063.39\ & $116.3$\ \ \ &                    &  $18.944\pm0.025$  &  $18.356\pm0.018$  &  $18.422\pm0.028$  &  $18.110\pm0.021$  & \ \ LT    & \ \ 1.59 \\
1065.38\ & $118.2$\ \ \ &  $19.623\pm0.074$  &                    &                    &                    &                    & \ \ LT    & \ \ 1.37 \\
1066.40\ & $119.3$\ \ \ &                    &  $19.002\pm0.021$  &  $18.404\pm0.014$  &  $18.481\pm0.016$  &  $18.218\pm0.018$  & \ \ LT    & \ \ 1.28 \\
1072.36\ & $125.2$\ \ \ &                    &  $19.025\pm0.045$  &  $18.532\pm0.040$  &  $18.660\pm0.048$  &  $18.490\pm0.046$  & \ \ CA    & \ \ 1.27 \\
1073.38\ & $126.2$\ \ \ &                    &  $19.070\pm0.023$  &  $18.564\pm0.017$  &  $18.675\pm0.024$  &  $18.349\pm0.029$  & \ \ LT    & \ \ 1.12 \\
1077.44\ & $130.3$\ \ \ &                    &  $19.180\pm0.030$  &  $18.643\pm0.025$  &  $18.778\pm0.020$  &  $18.422\pm0.022$  & \ \ LT    & \ \ 1.17 \\
1079.39\ & $132.3$\ \ \ &                    &  $19.248\pm0.049$  &  $18.673\pm0.051$  &  $18.801\pm0.067$  &                    & \ \ CA    & \ \ 2.70 \\
1084.39\ & $137.3$\ \ \ &                    &  $19.227\pm0.022$  &  $18.753\pm0.016$  &  $18.907\pm0.013$  &  $18.605\pm0.017$  & \ \ LT    & \ \ 1.14 \\
1086.36\ & $139.2$\ \ \ &  $20.001\pm0.096$  &                    &                    &                    &                    & \ \ LT    & \ \ 1.09 \\
1092.39\ & $145.3$\ \ \ &                    &                    &  $18.863\pm0.067$  &  $18.995\pm0.060$  &                    & \ \ LT    & \ \ 1.95 \\
1098.35\ & $151.2$\ \ \ &                    &  $19.480\pm0.021$  &  $19.007\pm0.015$  &  $19.293\pm0.028$  &  $18.897\pm0.033$  & \ \ LT    & \ \ 1.17 \\
1113.35\ & $166.2$\ \ \ &                    &  $19.684\pm0.019$  &  $19.271\pm0.015$  &  $19.664\pm0.020$  &  $19.175\pm0.029$  & \ \ LT    & \ \ 0.95 \\
1116.34\ & $169.2$\ \ \ &                    &                    &  $19.306\pm0.144$  &  $19.732\pm0.104$  &  $19.165\pm0.256$  & \ \ LT    & \ \ 1.34 \\
1209.75\ & $262.6$\ \ \ &                    &  $21.696\pm0.036$  &  $21.457\pm0.035$  &  $22.078\pm0.067$  &                    & \ \ NOT   & \ \ 0.74 \\
1247.85\ & $300.7$\ \ \ &                    &  $22.645\pm0.116$  &  $22.481\pm0.070$  &  $22.997\pm0.172$  &  $22.140\pm0.137$  & \ \ NTT   & \ \ 1.34 \\
1260.83\ & $313.7$\ \ \ &                    &  $22.941\pm0.171$  &  $22.500\pm0.160$  &  $23.338\pm0.273$  &  $22.123\pm0.123$  & \ \ NTT   & \ \ 1.25 \\
1263.90\ & $316.8$\ \ \ &                    &  $23.163\pm0.114$  &  $22.797\pm0.070$  &  $23.505\pm0.421$  &                    & \ \ NTT   & \ \ 1.03 \\
1275.69\ & $328.6$\ \ \ &                    &                    &                    &  $23.583\pm0.271$  &  $22.345\pm0.142$  & \ \ NOT   & \ \ 1.27 \\
1409.48\ & $462.3$\ \ \ &                    &                    &  $24.997\pm0.271$  &                    &  $24.329\pm0.260$  & \ \ TNG   & \ \ 0.75 \\

\hline
\end{tabular}
\\[1.5ex]
\flushleft
$^a$~JD $-$ 2\,454\,000.00\quad
$^b$~Phase in days with respect to $B$-band maximum JD $= 2\,454\,947.1 \pm 0.3$.\quad
$^c$~Stellar FWHM (arcsec).\\[1.6ex]
CA = Calar Alto 2.2\,m Telescope + CAFOS SiTe; \,http:/$\!$/www.caha.es/CA/Instruments/CAFOS/\\
TNG = 3.58\,m Telescopio Nazionale Galileo + DOLORES; \,http:/$\!$/www.tng.iac.es/instruments/lrs/\\
LT = 2.00\,m Liverpool Telescope + RATCAM; \,http:/$\!$/telescope.livjm.ac.uk/Info/TelInst/Inst/RATCam/\\
NOT = 2.56\,m Nordic Optical Telescope + ALFOSC; \,http:/$\!$/www.not.iac.es/instruments/alfosc/\\
Ekar = 1.82\,m Copernico Telescope + AFOSC; \,http:/$\!$/www.oapd.inaf.it/asiago/2000/2300/2310.html\\
NTT = 3.58\,m New Technology Telescope + EFOSC2; \,http:/$\!$/www.eso.org/sci/facilities/lasilla/instruments/efosc/\\
JN = unfiltered image by JN; 0.28\,m Celestron C11 + SBIG ST-8; \,http:/$\!$/www.astrosurf.com/snaude/\\
EP1 = unfiltered image by EP; 0.35\,m Meade LX200 GPS 14'' + SBIG ST-9XE; \,http:/$\!$/www.webalice.it/e.prosperi/\\
EP2 = unfiltered image by EP; 0.30\,m Meade LX200 12'' + SBIG ST-10XME; \,http:/$\!$/www.skylive.it/\\
\end{center}
\end{footnotesize}
\end{table*}

\begin{table*}
\caption{Colour-term\,/\,zero-point calibrated $JHK'$ magnitudes of SN~2009dc.}
\label{SN_mags_IR}
\begin{center}
\begin{footnotesize}
\begin{tabular}{rrcccll}
\hline
JD$^a$\ & Epoch$^b$  &         $J$        &        $H$       &       $K'$       & Telescope  & Seeing$^c$\\
\hline
 941.68\ &  $ -5.4$\ \ \ & $15.728\pm0.025$ & $15.826\pm0.028$ & $15.707\pm0.028$ & \ \ TNG-N  & \ \ 1.43\\
 947.81\ &  $  0.7$\ \ \ & $15.653\pm0.150$ &                  &                  & \ \ REM    & \ \ 3.24\\
 952.72\ &  $  5.6$\ \ \ & $15.668\pm0.170$ &                  &                  & \ \ REM    & \ \ 2.76\\
 954.70\ &  $  7.6$\ \ \ & $15.751\pm0.033$ & $15.680\pm0.031$ & $15.410\pm0.029$ & \ \ TNG-N  & \ \ 1.15\\
 963.55\ &  $ 16.4$\ \ \ & $16.042\pm0.032$ & $15.620\pm0.037$ & $15.388\pm0.034$ & \ \ NOT-N  & \ \ 0.68\\
 967.56\ &  $ 20.4$\ \ \ & $15.958\pm0.213$ &                  &                  & \ \ REM    & \ \ 3.36\\
 969.78\ &  $ 22.6$\ \ \ & $15.921\pm0.029$ & $15.516\pm0.029$ & $15.415\pm0.045$ & \ \ NTT-S  & \ \ 0.78\\
 993.61\ &  $ 46.5$\ \ \ & $16.119\pm0.207$ &                  &                  & \ \ REM    & \ \ 3.00\\
1031.41\ &  $ 84.3$\ \ \ & $18.332\pm0.069$ & $17.335\pm0.036$ & $17.522\pm0.074$ & \ \ TNG-N  & \ \ 0.83\\
1057.48\ &  $110.3$\ \ \ & $19.197\pm0.078$ & $18.119\pm0.064$ & $18.119\pm0.091$ & \ \ NTT-S  & \ \ 1.12\\
1212.06\ &  $264.9$\ \ \ &                  &     $\ge20.4$    &                  & \ \ LBT    & \ \ 1.55\\
1213.02\ &  $265.9$\ \ \ & $21.788\pm0.078$ &                  &                  & \ \ LBT    & \ \ 0.53\\
1305.84\ &  $358.7$\ \ \ &     $\ge19.9$    &     $\ge18.8$    &     $\ge18.6$    & \ \ NTT-S  & \ \ 1.29\\
\hline
\end{tabular}
\\[1.5ex]
\flushleft
$^a$~JD $-$ 2\,454\,000.00\quad
$^b$~Phase in days with respect to $B$-band maximum JD $= 2\,454\,947.1 \pm 0.3$.\quad
$^c$~Stellar FWHM (arcsec).\\[1.6ex]
TNG-N = 3.58\,m Telescopio Nazionale Galileo + NICS; \,http:/$\!$/www.tng.iac.es/instruments/nics/\\
REM = 0.60\,m Rapid Eye Mount + REMIR; \,http:/$\!$/www.rem.inaf.it/\\
NOT-N = 2.56\,m Nordic Optical Telescope + NOTCam; \,http:/$\!$/www.not.iac.es/instruments/notcam/\\
NTT-S = 3.58\,m New Technology Telescope + SOFI; \,http:/$\!$/www.eso.org/sci/facilities/lasilla/instruments/sofi/\\
LBT = 2$\times$8.4\,m Large Binocular Telescope + LUCIFER; \,http:/$\!$/abell.as.arizona.edu/~lbtsci/Instruments/LUCIFER/\\
\end{footnotesize}
\end{center}
\end{table*}

\subsection{Reduction of spectroscopic data}
\label{Spectroscopic data}

\begin{table*}
\caption{Log of spectroscopic observations of SN~2009dc.}
\label{spectra}
\begin{center}
\begin{footnotesize}
\begin{tabular}{rrlcclc}
\hline 
JD$^a$\ \ \ & Epoch$^b$     & Exposure time [s]     & Telescope$^c$ & Grism\,/\,Grating &  Range [\AA]  & Res.\,[\AA]$^d$ \\
\hline
 937.7\ \ \  &$ -9.4$\ \ \ \  & \ 1500                          &  TNG          & LR-B         &  3200 --  7900     & 10              \\
 938.6\ \ \  &$ -8.5$\ \ \ \  & \ 1500                          &  NOT          & gm4          &  3200 --  9100     & 13              \\
 939.7\ \ \  &$ -7.5$\ \ \ \  & \  900                          &  NOT          & gm4          &  3200 --  9100     & 13              \\
 941.7\ \ \  &$ -5.4$\ \ \ \  & \ 1800 $\times$ 2               &  TNG-N        & IJ,HK        &  8650 -- 24700     & 18,36           \\
 943.5\ \ \  &$ -3.7$\ \ \ \  & \  900                          &  NOT          & gm4          &  3200 --  9100     & 14              \\
 949.5\ \ \  &$  2.4$\ \ \ \  & \  900 $\times$ 2               &  TNG          & LR-B,LR-R    &  3200 --  9750     & 13              \\
 949.6\ \ \  &$  2.5$\ \ \ \  & \ 1800 $\times$ 2               &  CA           & b200,r200    &  3300 -- 10200     & 10              \\
 951.6\ \ \  &$  4.5$\ \ \ \  & \ 2400                          &  CA           & b200         &  3200 --  8750     & 10              \\
 953.7\ \ \  &$  6.5$\ \ \ \  & \  900 $\times$ 2               &  TNG          & LR-B,LR-R    &  3250 --  9750     & 10              \\
 954.7\ \ \  &$  7.5$\ \ \ \  & \ 1800 $\times$ 2               &  TNG-N        & IJ,HK        &  8650 -- 24700     & 18,36           \\
 955.5\ \ \  &$  8.4$\ \ \ \  & \ 1800 $\times$ 2               &  CA           & b200,r200    &  3250 -- 10300     &  9              \\
 956.5\ \ \  &$  9.3$\ \ \ \  & \  900                          &  NOT          & gm4          &  3250 --  9100     & 14              \\
 958.6\ \ \  &$ 11.5$\ \ \ \  & \  900                          &  NOT          & gm4          &  3250 --  9100     & 14              \\
 959.6\ \ \  &$ 12.5$\ \ \ \  & \  600 $\times$ 3               &  NOT          & gm4          &  3300 --  9100     & 13              \\
 964.7\ \ \  &$ 17.5$\ \ \ \  & \  900 $\times$ 2               &  TNG          & LR-B,LR-R    &  3200 --  9600     & 10              \\
 969.7\ \ \  &$ 22.5$\ \ \ \  & \  900 $\times$ 2               &  NTT          & gm11,gm16    &  3350 --  9500     & 14              \\
 970.8\ \ \  &$ 23.6$\ \ \ \  & \ 3240 + 3000                &  NTT-S        & GB,GR        &  9350 -- 25000     & 23,32           \\
 979.5\ \ \  &$ 32.4$\ \ \ \  & \ 2400 + 2000                &  Ekar         & gm4,gm2      &  3600 --  9200     & 24,34           \\
 983.6\ \ \  &$ 36.4$\ \ \ \  & \ 2100 $\times$ 2               &  CA           & b200,r200    &  3300 -- 10300     & 10              \\
1002.6\ \ \  &$ 55.5$\ \ \ \  & \ 2100                          &  TNG          & LR-B         &  3400 --  7900     & 14              \\
1009.5\ \ \  &$ 62.4$\ \ \ \  & \ 2400 $\times$ 2               &  CA           & b200,r200    &  3400 -- 10300     & 10              \\
1028.5\ \ \  &$ 81.4$\ \ \ \  & \ 1800 $\times$ 2               &  NTT          & gm11,gm16    &  3400 --  9500     & 21              \\
1031.5\ \ \  &$ 84.3$\ \ \ \  & \ 3600 + 5760                &  TNG-N        & IJ,HK        &  8650 -- 24700     & 18,36           \\
1045.4\ \ \  &$ 98.3$\ \ \ \  & \ 3600 $\times$ 2               &  CA           & b200,r200    &  3600 -- 10000     & 10              \\
1056.5\ \ \  &$109.3$\ \ \ \  & \ 4800\,/\,2700\,$\times$\,2 &  NOT\,/\,CA   & gm4\,/\,r200 &  3650 -- 10000     & 17,11           \\
1074.4\ \ \  &$127.2$\ \ \ \  & \ 2700 $\times$ 4               &  CA           & b200         &  3400 --  8750     & 12              \\
1112.3\ \ \  &$165.1$\ \ \ \  & \ 3600 $\times$ 2               &  CA           & b200         &  3500 --  8750     & 12              \\
\hline
\end{tabular}
\\[1.5ex]
\flushleft
$^a$~JD $-$ 2\,454\,000.0\quad
$^b$~Phase in days with respect to $B$-band maximum JD $= 2\,454\,947.1 \pm 0.3$.\quad
$^c$~See Tables~\ref{SN_mags} and \ref{SN_mags_IR} for details.\quad
$^d$~Full-width at half maximum (FWHM) of isolated, unblended night-sky lines.\\
\end{footnotesize}
\end{center}
\end{table*}

An overview of our spectroscopic observations of SN~2009dc is given in
Table~\ref{spectra}. All spectra were taken with the slit along the
parallactic angle to avoid differential flux losses
\citep{Filippenko82}. The reduction of the optical data followed
standard procedures. The two-dimensional spectroscopic frames were
debiased and flat-fielded, before an optimal, variance-weighted
extraction of the spectra \citep*{Horne86} was performed using the
{\sc iraf} routine {\sc apall}. Wavelength calibration was
accomplished with the help of arc-lamp exposures and checked against
isolated night-sky lines. Second-order contamination in the spectra
taken with NOT\,+\,ALFOSC was eliminated following the method of
\citet{Stanishev07}. The instrumental response functions were
determined from observations of spectrophotometric standard stars
\citep{Oke90,Hamuy92,Hamuy94}. An atmospheric extinction correction
was applied using tabulated extinction coefficients for each
observatory. Telluric features were identified in the smooth spectra
of the spectrophotometric standard stars and removed from the SN
spectra. To check the flux calibration of the spectra, synthetic
photometry was computed using \citet{Bessell90} passbands. If
necessary, the spectral fluxes were adjusted to match the
contemporaneous photometry. Finally, spectra obtained during the same
night (at late phases also in subsequent nights) were combined to
increase the signal-to-noise ratio (S/N); if the wavelength range of
these spectra was different, they were averaged in their overlap
region.

The reduction of our near-IR spectra mostly followed the procedure
described for the optical bands, with a few noticeable differences.
The total integration time in the near IR was split into several
sub-exposures, with the target off-set along the slit. Subsequent
exposures could thus be subtracted from each other to remove the sky
emission. After extraction and wavelength-calibration the SN spectra
were divided by those of a telluric A0 standard star taken at similar
airmass to remove telluric absorptions, and multiplied by a Vega
spectrum to provide a relative flux calibration.  Proper absolute flux
calibration was achieved by comparison with contemporaneous $JHK'$
photometry.

\section{Photometric evolution}
\label{Photometric evolution}

\begin{figure*}
   \centering
   \includegraphics[width=14cm]{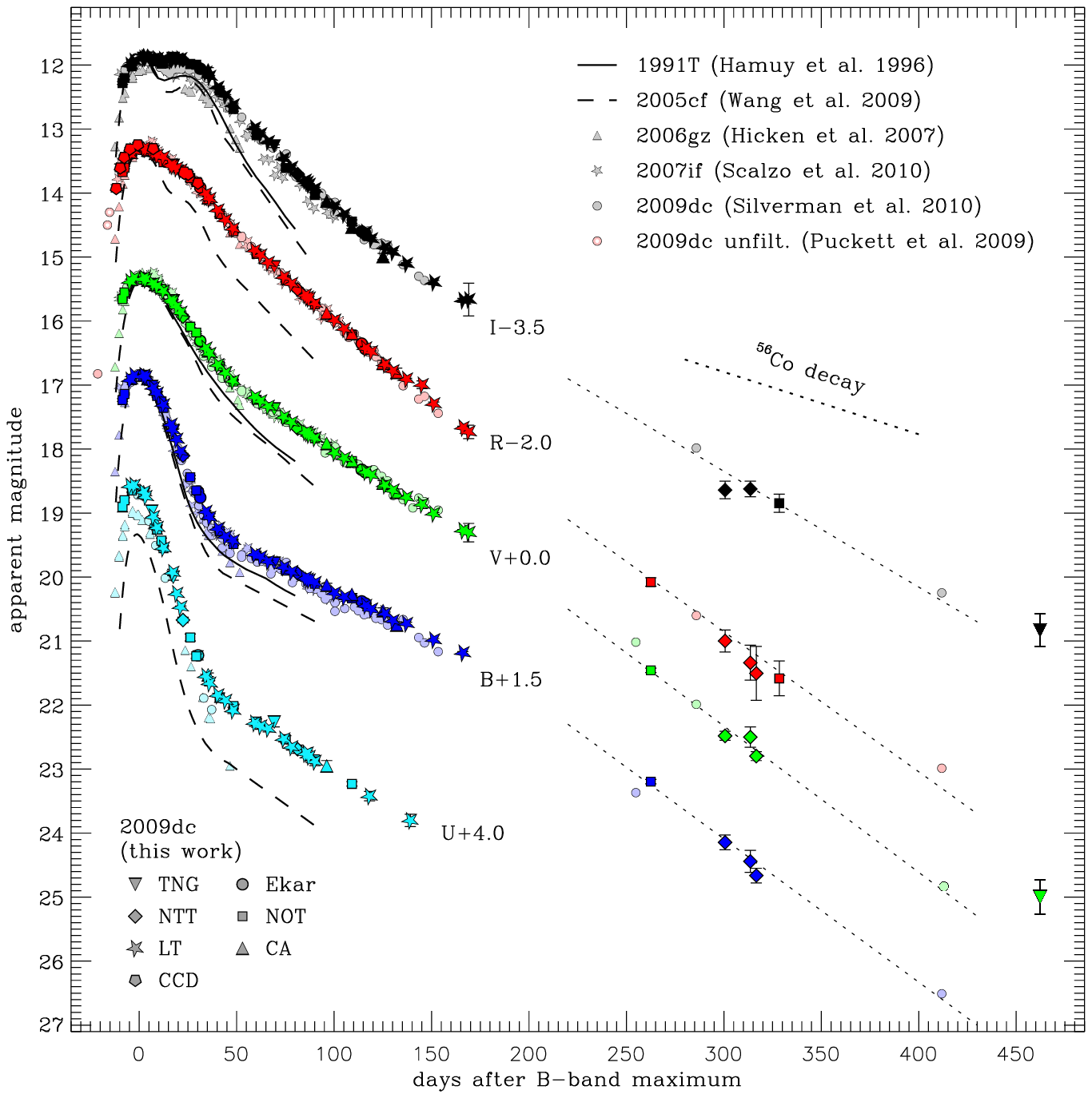}
   \caption{Bessell $U\!BV\!RI$ light curves of SN~2009dc (coloured
     symbols). The data are $S$- and $K$-corrected with the exception
     of the unfiltered amateur data, which were calibrated to the $R$
     band employing colour-term corrections (Table~\ref{SN_mags}). The
     unfiltered discovery magnitudes reported by \citet[][open red
     pentagons]{CBET1762} are plotted with the same vertical shift as
     our $R$ magnitudes. In this Figure, a constant shift of $-0.2$
     mag has been applied to the Liverpool-Telescope (LT) $U$-band
     photometry to eliminate a systematic offset with respect to the
     $U$-band data of other telescopes. The SN~2009dc light
     curves of \citet{Silverman10}, the light curves of the
     superluminous SNe~2006gz \citep{Hicken07} and 2007if
     \citep{Scalzo10}, and light-curve templates of the luminous SN~Ia
     1991T \citep{Hamuy96} and the normal SN~Ia 2005cf \citep{Wang09}
     are shown for comparison, shifted in time and magnitude to match
     SN~2009dc at maximum light in $B$.}
   \label{fig:UBVRI}
\end{figure*}

Figure~\ref{fig:UBVRI} shows that our photometry of SN~2009dc is
generally in good agreement with that published by
\citet{Silverman10}. There are some systematic discrepancies in the
$U$ band and -- at more advanced epochs -- in the $B$ band, which we
attribute to missing $S$- and $K$-corrections in the Silverman et
al. data.  With absolute peak magnitudes between $-19.8$ and $-21.1$
in the optical bands, SN~2009dc is roughly a factor 2 more luminous
than the bulk of SNe~Ia, and also slightly more luminous than the
possible super-\MCh\ SNe~2003fg \citep{Howell06} and 2006gz
\citep{Hicken07}.

Morphologically, the light curves of SN~2009dc resemble those of other
SNe~Ia.  In the $I$ and probably also the $JHK'$ bands, a secondary
light-curve maximum exists, delayed by $\sim$\,25\,d with respect to
the peak in $B$. The immediate post-maximum decline is faster in the
bluer bands, accompanied by a strong evolution towards redder colours
within the first month after peak.

However, compared to other SNe~Ia the evolution of the light curves is
very slow, both during the rise and the decline. With an unfiltered
discovery magnitude of 16.5 \citep{CBET1762} on April 9.3, SN~2009dc
was only $\sim$\,1.2 mag below peak more than 16\,d before $B$-band
maximum light. \citet{Silverman10} reported a detection of SN~2009dc
at 3.5 mag below peak in an even earlier, unfiltered image taken with
KAIT on April 4, 21.2\,d before $B$-band maximum. This is clearly
longer than the canonical rise time of SNe~Ia of 17--20\,d
\citep{Riess99,Conley06,Strovink07,Hayden10}. \citet{Silverman10} also
mentioned a non-detection in an unfiltered image taken on March 28,
28\,d before $B$-band maximum, to a limiting magnitude of
$\sim$\,19.3. The rise time of SN~2009dc is therefore at least 22\,d,
but probably not more than 30\,d.

With a polynomial fit to the $S$-corrected $B$-band light curve of
SN~2009dc, a \dm15 of $0.69\pm0.03$ is inferred. Correcting for the
effect of reddening \citep{Phillips99}, this turns into
\dm15$_\mathrm{true}=0.71\pm0.03$.  \citet{Yamanaka09} reported a
\dm15\ of $0.65 \pm 0.03$, marginally consistent with our result
within the uncertainties, whereas \citet{Silverman10} derived \dm15\
$= 0.72 \pm 0.03$, in excellent agreement with our value. This decline
rate is among the lowest ever measured for SNe~Ia, similar to those
of SNe~2006gz (0.69; \citealt{Hicken07}) and 2007if (0.71;
\citealt{Scalzo10}). 

The decline of the $R$-band light curve is remarkably linear for more
than 300\,d after maximum, with only a slight shoulder after one
month. The first and second maximum in the $I$ band are almost equally
bright. However, the minimum in between is not very pronounced, and
with $\sim$\,25\,d the offset of the two maxima is not particularly
large. This is similar to the situation in SN~2006gz, but contrary to
the trend of more pronounced and delayed secondary maxima that is
observed in other SNe~Ia with small \dm15\ \citep{Hamuy96}.
Moreover, the first $I$-band maximum does not precede those in the
bluer band as in other luminous SNe~Ia, but is slightly delayed.  This
behaviour is reminiscent of subluminous SNe~Ia. However, in the latter
it appears to be a consequence of the small \Nifs\ mass \citep{Kasen06},
which cannot be the reason in SN~2009dc. Alternatively,  \Nifs\ could
be more strongly mixed \citep{Kasen06}, but this is disfavoured by the
long rise time and the weak line blanketing in the UV part of early
spectra, indicative of little Fe-group material in the outer shells
(c.f. Section~\ref{Optical spectra}).

The late-time photometric behaviour of SN~2009dc deserves special
attention after the lesson taught by SN~2006gz \citep{Maeda09}: the
slow decline of the light curves of SN~2006gz soon after peak must
have been followed by a rapid drop at later times, when the SN was too
close to the Sun to be observed. An attempt to recover the SN
photometrically one year after maximum failed, and the derived upper
limits indicate a faster average post-maximum decline than observed in
normal SNe~Ia \citep{Maeda09}. In SN~2009dc, no accelerated decline is
seen until 180\,d after peak. The decline rates between 60 and 170\,d,
ranging from $1.42\pm0.02$ mag (100\,d)$^{-1}$ in $B$ to $2.62\pm0.02$
mag (100\,d)$^{-1}$ in $R$ (c.f. Table~\ref{slopes}), are comparable
to those of normal SNe~Ia. However, thereafter the situation changes
completely. While the $R$-band light curve continues its linear
decline from earlier phases, the bluer bands now show a much more
rapid fading than before, with decline rates very similar to that in
$R$ (Table~\ref{slopes}). This behaviour is at odds with both that of
normal SNe~Ia and the expectation that around 300\,d the (bolometric)
decline rate should slow down and approach the \Cofs\ decay rate of
0.98 mag (100\,d)$^{-1}$, since the ejecta are then fully transparent
to $\gamma$-rays, but still mostly opaque to positrons. There are
hints that the decline might eventually slow down more than 400\,d
after the explosion. However, this very late decline rate is very
uncertain, since our last photometric measurement might be
contaminated by host-galaxy light or background sources. The
discrepancy between expectation and observation is addressed again
when studying the pseudo-bolometric light curve in
Section~\ref{Bolometric light curve}, and possible reasons are
discussed in Section~\ref{Late-time decline}.

\begin{table}
\caption{Light-curve tail decline rates of SN~2009dc.}
\label{slopes}
\begin{center}
\begin{footnotesize}
\begin{tabular}{rlllll}
\hline 
Interval$^a$ &  $\Delta U^b$ &  $\Delta B^b$ &  $\Delta V^b$ &
$\Delta R^b$ &  $\Delta I^b$ \\
\hline
 60--170\,d  &  1.94  &  1.42  &  1.95  &  2.62  &  2.59  \\
150--330\,d  &        &  2.19  &  2.28  &  2.48  &  2.04  \\
260--330\,d  &        &  2.59  &  2.30  &  2.40  &        \\
260--460\,d  &        &        &  1.69  &        &  1.42  \\
\hline
\end{tabular}
\\[1.5ex]
\flushleft
$^a$~With respect to $B$-band maximum.\quad
$^b$~In mag (100\,d)$^{-1}$.\\
\end{footnotesize}
\end{center}
\end{table}

\begin{figure}
   \centering
   \includegraphics[width=8.4cm]{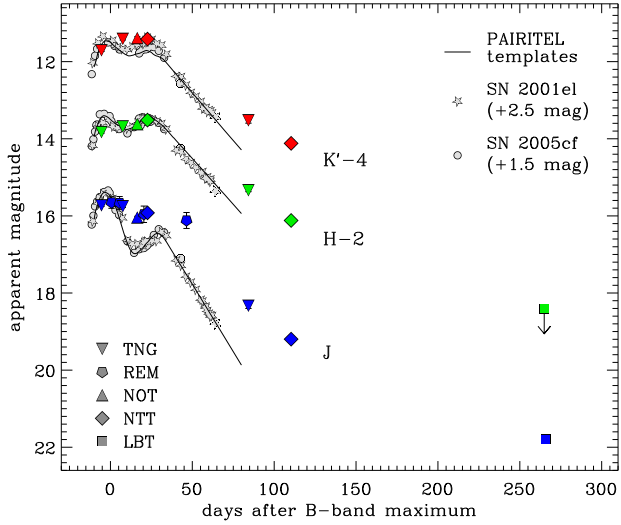}
   \caption{$JHK'$ light curves of SN~2009dc. The phase is in days with respect 
   to $B$-band maximum light on JD $= 2\,454\,947.1 \pm 0.3$. The NIR light 
   curves of SNe~2001el \citep{Krisciunas03} and 2005cf \citep{Pastorello07a,Wang09}, 
   and the NIR template light curves constructed from PAIRITEL observations 
   \citep{WoodVasey08} are shown for comparison, shifted to line up with SN~2009dc 
   at maximum.}
   \label{fig:JHK}
\end{figure}

In the $JHK'$ bands (Fig.~\ref{fig:JHK}), the paucity of observations
makes an analysis of the light curves less robust. Nevertheless, in
all three bands the secondary maximum seems to be particularly bright,
whereas the first maximum is weak ($J$) or degraded to a shoulder
($HK'$). The steep decline in $J$ characteristic for normal SNe~Ia
after the first peak is almost entirely absent in SN~2009dc. Since the
majority of the flux in the near IR is caused by fluorescence
\citep[e.g.][]{Kromer09}, the lack of emission compared to ordinary
SNe~Ia around the time of $B$-band maximum suggests less flux
re-distribution from the blue and UV part of the spectrum to redder
wavelengths. This is in agreement with a high UV flux until
$\sim$\,10\,d after maximum light as also reported by
\citet{Silverman10}, and a strong fading in the UV thereafter. After
the secondary peak, the $JHK'$ light curves of SN~2009dc decline more
slowly than those of ordinary SNe~Ia, and it was possible to recover
the SN in deep $J$-band images taken at the LBT 266\,d after $B$-band
maximum light. Note that with peak absolute magnitudes of about
$-19.3$ to $-19.4$, SN~2009dc is overluminous also in $J$, $H$ and
$K'$. It does not follow the behaviour of normal SNe~Ia, which are
nearly standard candles in the near IR \citep[with
M$_{JHK'}$\,$\sim$\,$-18.1$ to $-18.3$;][]{Krisciunas04,WoodVasey08}.

\subsection{Colour evolution}
\label{Colour evolution}

\begin{figure}   
   \centering
   \includegraphics[width=8.4cm]{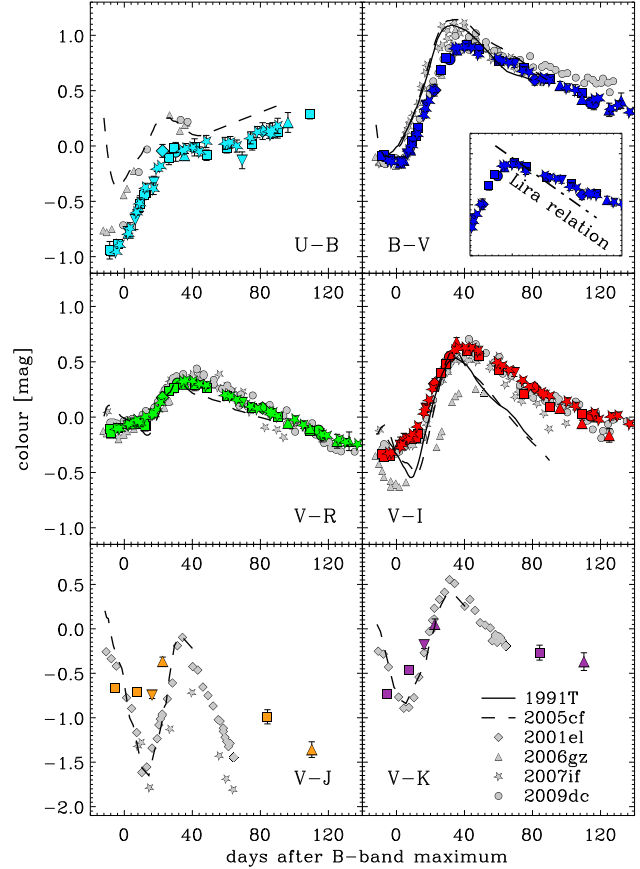}
   \caption{Time-evolution of various colour indices of SN~2009dc,
     reddening-corrected adopting a \citet{Cardelli89} extinction law
     with $E(B-V)=0.17$ mag and $R_V=3.1$ (coloured symbols). A
     constant shift of $-0.2$ mag has been applied to the
     Liverpool-Telescope (LT) $U$-band photometry to eliminate a
     systematic offset with respect to other telescopes. The 
     SN~2009dc colour curves of \citet{Silverman10}, the
     colour curves of the superluminous SNe~2006gz \citep{Hicken07} and
     2007if \citep{Scalzo10}, and those of the Type Ia SNe~1991T
     \citep{Hamuy96}, 2001el \citep{Krisciunas03,Krisciunas07} and
     2005cf \citep{Pastorello07a,Wang09} -- all of them
     reddening-corrected -- are shown for comparison. The
     inset in the top-right panel illustrates the deviation of the
     $B-V$ curve of SN~2009dc from the Lira relation
     \citep{Lira95,Phillips99}.}
   \label{fig:colours}
\end{figure}
 
Figure~\ref{fig:colours} presents the time-evolution of the $U-B$,
$B-V$, $V-R$ and $V-I$ colours of SN~2009dc. The basic behaviour of
these curves appears typical of a SN~Ia, starting with relatively blue
colours before and around maximum light, turning redder until
$\sim$\,40\,d thereafter, and then again evolving towards bluer
colours (except for $U-B$ which monotonically becomes
redder). However, at a higher level of detail differences become
evident.  The early $U-B$ colour of SN~2009dc is unusually blue
because of little UV line blanketing and the weakness of \CaII\ H\&K
(c.f. Section~\ref{Optical spectra}). In normal SNe~Ia, the $V-R$ and
$V-I$ colours are bluest around and immediately after maximum light,
whereas in SN~2009dc they monotonically turn redder from the start of
our observations until 40\,d after maximum light. Even more
importantly, SN~2009dc does not conform with the uniform $B-V$ colour
evolution of other SNe~Ia between 30 and 90\,d after maximum, known as
the Lira relation \citep{Lira95,Phillips99}. Instead, its $B-V$ colour
evolves much more slowly, and is redder than predicted by the Lira
relation by almost 0.3 mag at 90\,d.  Hence, for SN~2009dc, none of
the colour criteria usually used to estimate the reddening in SNe~Ia
can be employed confidently.

Note that the colour evolution of SN~2009dc shares more similarities
with those of SNe~2006gz and 2007if than with the bulk of SNe~Ia
(Fig.~\ref{fig:colours}), suggesting that some of the observed
colour characteristics may be generic for superluminous SNe~Ia.

\subsection{Bolometric light curve}
\label{Bolometric light curve}

\begin{figure*}   
   \centering
   \includegraphics[width=13.5cm]{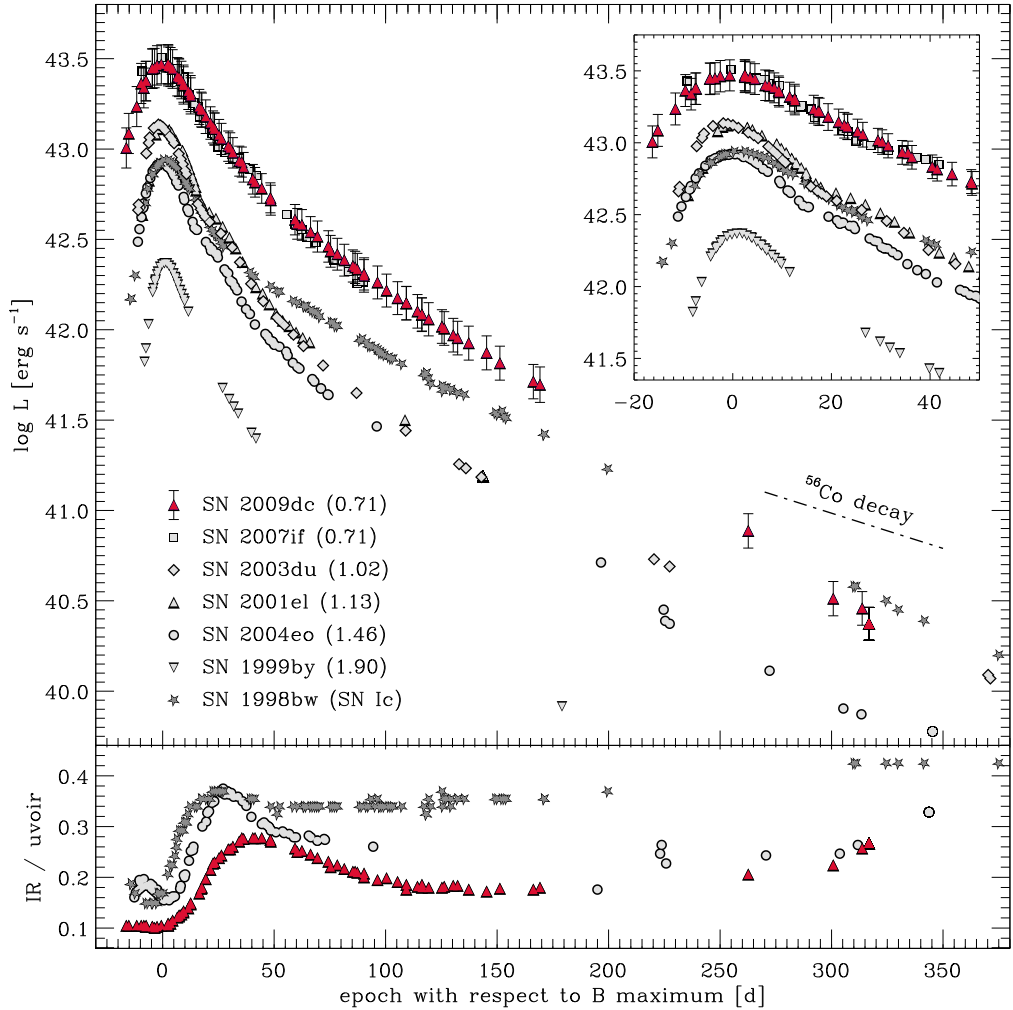}
   \caption{Top panel: Pseudo-bolometric light curves of SN~2009dc,
     the superluminous SN~Ia 2007if, the normal Type Ia SNe~2003du,
     2001el and 2004eo, and the subluminous SN~Ia 1999by. The
     exceptionally energetic and luminous Type Ic SN~1998bw is shown
     for comparison. All light curves were obtained by integrating the
     $U$-through-$K'$-band fluxes.  Error bars are shown for SN~2009dc
     only, and account for uncertainties in the photometric calibration, 
     distance and extinction estimates. The $\Delta$m$_{15}(B)_\mathrm{true}$ 
     of all SNe~Ia is given in parentheses. Bottom panel: NIR contribution to the
     pseudo-bolometric light curves of SNe~2009dc, 2004eo and 1998bw.}
   \label{fig:bolom}
\end{figure*}

A pseudo-bolometric light curve \citep*[see, e.g.,][]{Nomoto90} of
SN~2009dc was constructed from our filtered photometry. To this aim,
the $U$-through-$K'$-band magnitudes were first converted into
monochromatic fluxes. The spectral energy distribution (SED) was then
interpolated linearly and integrated over wavelength, assuming zero
flux at the blue edge of the $U$ band and the red edge of the $K'$
band. In case of missing $JHK'$ data, a NIR correction to the
$U$-through-$I$ light curve was estimated by interpolation between
adjacent epochs with NIR data. At very late phases, constant $I-H$-
and $I-K'$ colours were assumed. We find a NIR contribution to the
pseudo-bolometric light curve of $\sim$\,10 per cent around maximum
light, $\sim$\,28 per cent 40\,d after maximum, and $\sim$\,20 per
cent 100\,d after maximum. Compared to ordinary SNe~Ia and
core-collapse SNe, this NIR contribution in SN~2009dc is quite low
(Fig.~\ref{fig:bolom}, bottom panel). Wavelength regions other than
the optical and near-IR are believed to contribute very little to the
total bolometric flux in SNe~Ia \citep*{Contardo00}, and are neglected
here.\footnote{Actually, owing to the relatively high UV flux in
  SN~2009dc indicated by its very blue early-time spectra, the mid and
  far UV could have a non-negligible effect on the pre-maximum
  bolometric light curve. Unfortunately, the SWIFT UVOT photometry
  presented by \citet{Silverman10} only starts at maximum light, when
  the contribution of these bands has dropped below the 10 per cent
  level.}  The resulting quasi-bolometric light curve of SN~2009dc is
shown in Fig.~\ref{fig:bolom} (top panel), along with those of the
normal Type Ia SNe~2003du \citep[][\dm15 = 1.02]{Stanishev07a}, 2001el
\citep[][\dm15 = 1.13]{Krisciunas03} and 2004eo \citep[][\dm15 =
1.46]{Pastorello07}, the subluminous SN~Ia 1999by \citep[][\dm15 =
1.90]{Garnavich04}, the superluminous SN~Ia 2007if \citep[][\dm15 =
0.71]{Scalzo10} and the Type Ic hypernova 1998bw
\citep{Galama98,McKenzie99,Sollerman00,Patat01}. Neither SN~1998bw,
which is one of the most luminous non-interacting core-collapse SNe
known to date (with about 0.4 \Msun\ of \Nifs\ synthesised;
\citealt{Maeda06}), nor any of the classical SNe~Ia can compete with
SN~2009dc in terms of luminosity. With $\log L = 43.47 \pm 0.11$ [erg
s$^{-1}$], SN~2009dc outshines ordinary SNe~Ia by a factor $\sim$\,$2$
at peak\footnote{For the same assumption on the host-galaxy reddening
  [\ebv$_\mathrm{host}$ = 0.10 mag], \citet{Silverman10} derive a peak
  bolometric luminosity of $\log L = 43.52 \pm 0.08$ [erg s$^{-1}$].
  Based on a slightly larger host reddening [\ebv$_\mathrm{host}$ =
  0.14 mag] \citet{Yamanaka09} obtain $\log L = 43.52 \pm 0.12$ [erg
  s$^{-1}$]. Both results are in agreement with our estimate.}, and
the difference increases during the first $\sim$\,150\,d past
maximum. Only SN~2007if keeps up with SN~2009dc, showing a strikingly
similar bolometric light curve. After the peak phase, SNe generally
have a decline rate which is significantly faster than the \Cofs\
decay. The reason for this behaviour is an increasing escape fraction
of $\gamma$-rays due to decreasing opacities. SN~2009dc also shows
this behaviour, but fades more slowly than normal SNe~Ia until
$\sim$\,150\,d past maximum. This is an indication for a larger
$\gamma$-ray opacity, caused by the low ejecta velocities or by a
larger ejecta mass. SN~1998bw has very massive ejecta
(\citealt{Maeda06} estimated M$_\mathrm{ej} \sim 10$ \Msun) and thus
shows an even slower late-time decline.

However, as already noted in Section~\ref{Photometric evolution}, the
behaviour of SN~2009dc changes substantially 200--250\,d after
maximum. The decline of the pseudo-bolometric light curve becomes
steeper than it was the months before. In fact, one year after the
explosion, SN~2009dc is no longer more luminous than the normal SN~Ia
2003du. Unfortunately, we do not have full wavelength coverage at
those late phases (the pseudo-bolometric light curve is based on
$B$-through-$J$-band photometry), so that this statement must be taken
with a grain of salt. Nevertheless, there is a clear flux deficit in
the optical regime (out to the $J$ band) with respect to expectations
based on an extrapolation from earlier epochs. It is unlikely that an
increased $H$- and $K'$-band flux could make up for this.

\section{Spectroscopic evolution}
\label{Spectroscopic evolution}

 \subsection{Optical spectra}
 \label{Optical spectra}

\begin{figure*}   
   \centering
   \includegraphics[width=13cm]{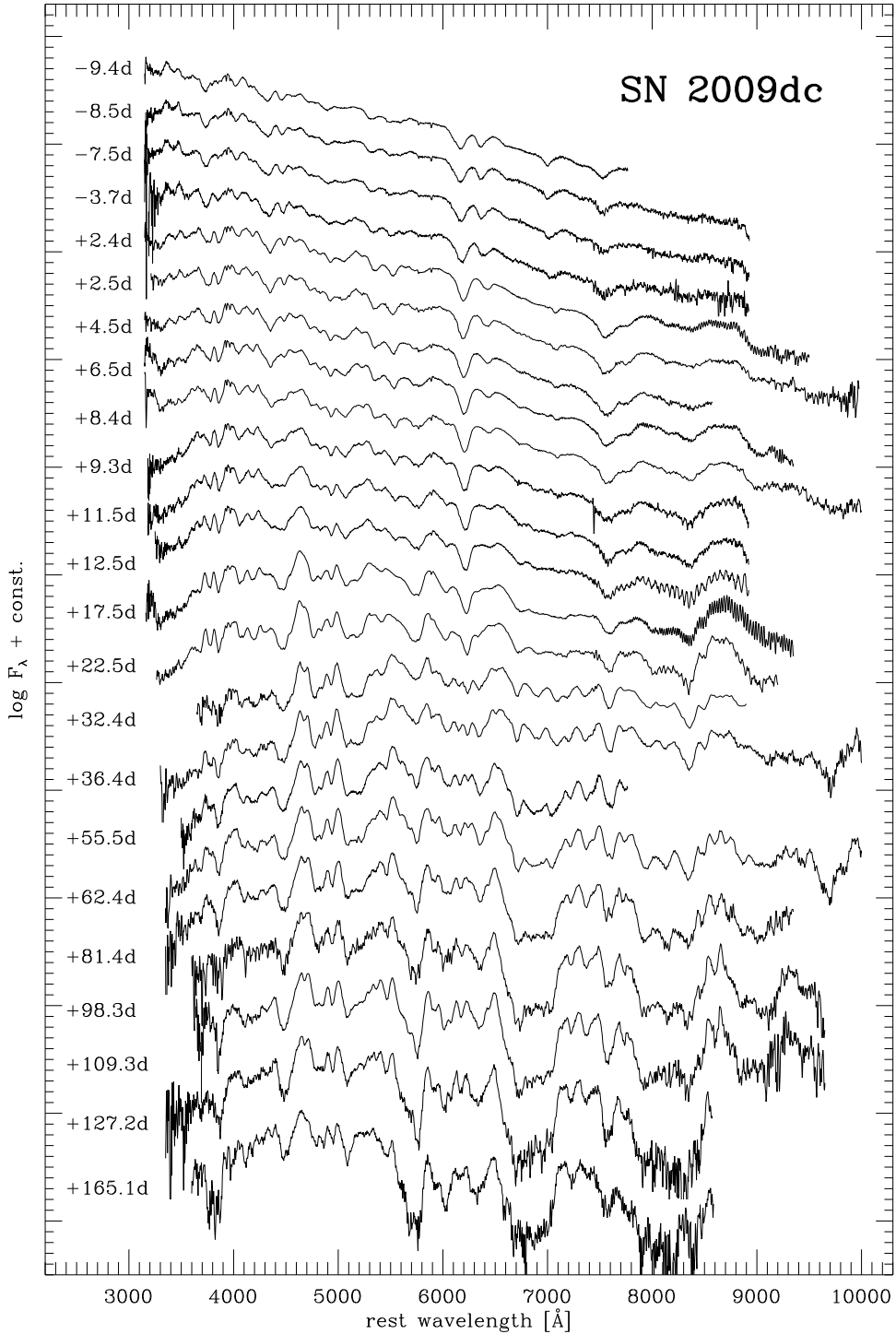}
   \caption{Time sequence of SN~2009dc spectra in the SN rest frame. The 
   phases reported next to each spectrum are with respect to $B$-band 
   maximum light (JD $= 2\,454\,947.1 \pm 0.3$). The +165.1\,d spectrum 
   has been smoothed with a kernel of 2500 \kms\ for presentation purposes.}
   \label{fig:spectra}
\end{figure*}

The optical spectra of SN~2009dc, presented in Fig.~\ref{fig:spectra}
show a transition from a blue pseudo-continuum with superimposed
P-Cygni features at early phases to line-dominated emission after a
few months. Throughout its evolution, SN~2009dc is characterised by
comparatively narrow spectral features, indicating a small extension
(in velocity space) of the line-formation zone, and resulting in
reduced line blending compared to ordinary SNe~Ia. In this respect,
SN~2009dc resembles 02cx-like SNe~Ia \citep{Li03}, which have
comparably low ejecta velocities. However, given that the latter are
at least moderately subluminous, no direct conclusions on similar
progenitors or explosion mechanisms can be drawn. In what follows, the
spectra of SN~2009dc at several characteristic epochs will be
described in more detail and compared with those of other normal and
peculiar SNe~Ia.

\subsubsection{One week before maximum}

\begin{figure}   
   \centering
   \includegraphics[width=8.4cm]{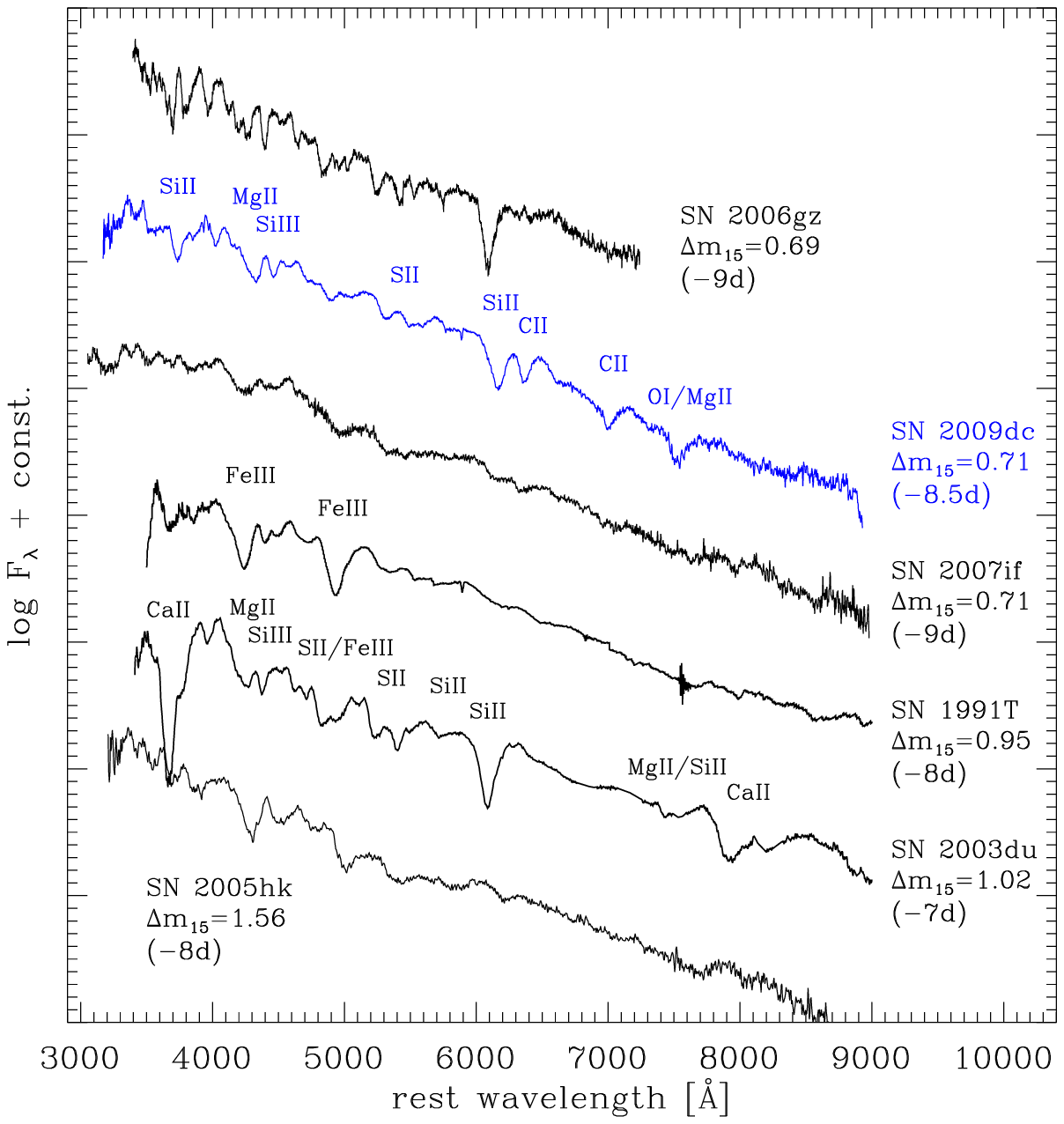}
   \caption{Comparison of pre-maximum spectra of SN~2009dc with those of 
   other superluminous SNe~Ia (2006gz and 2007if; \citealt{Hicken07,Scalzo10}), 
   the classical luminous SN~1991T \citep*{Filippenko92,RuizLapuente92,Gomez96}, 
   the normal SN~Ia 2003du \citep{Stanishev07a} and the peculiar, 02cx-like 
   SN~2005hk \citep{Phillips07,Stanishev07b}. The spectra of SNe~2006gz and 2007if 
   have been slightly smoothed for clarity. An identification of the major 
   absorption lines has been attempted.} 
   \label{fig:comp_minus08}
\end{figure}

In the $-8.5$ day spectrum of SN~2009dc (Fig.~\ref{fig:comp_minus08}),
lines of \SiII, \SiIII, \SII, \CII, \OI\ and possibly \MgII\ can be
identified. Compared to normal SNe~Ia, represented by SN~2003du in
Fig.~\ref{fig:comp_minus08}, all features due to intermediate-mass
elements appear shallower. In particular, the \CaII\ lines, very
prominent in SN~2003du, are weak.  The \CaII\ NIR triplet cannot be
identified, and \CaII\ H\&K form merely a shoulder to the red of
\SiII\ $\lambda3859$.  The equivalent width (EW) of \SiII\
$\lambda6355$ is only 44\,\AA, compared to 71\,\AA\ in SN~2003du. On
the contrary, lines due to unburned material are very
pronounced. While in most normal SNe~Ia at best a hint of \OI\
$\lambda7774$ can be discerned at early phases (probably blended with
\MgII), the line is one of the strongest features in SN~2009dc. The
same is true for \CII\ features ($\lambda6580$, $\lambda7234$), which
are only occasionally identified in very early SN~Ia spectra
\citep[e.g.][]{Thomas07,Tanaka08,Taubenberger08}, but are present at
unprecedented strength in the $-8.5$ day spectrum of SN~2009dc. The
overall continuum shape is rather blue, and down to the blue cut-off
of the spectrum at $\sim$\,3200\,\AA\ no severe flux depression due to
blanketing of Fe-group lines is observed. In summary, lines from
unburned material are strong in SN~2009dc at early phases, whereas all
lines from burning products are relatively weak. Of course, line
formation does not only depend on the composition, but also on the
ionisation and excitation conditions, and indeed the weakness of lines
from singly ionised IMEs and Fe-group elements could be explained by a
high ionisation (supported also by the relative strength of \SiIII\
$\lambda4563$). However, the simultaneous strength of \OI\
$\lambda7774$ and the lack of detectable \FeIII\ lines do not favour
this scenario. Instead, we believe that the composition of the
line-formation zone at day $-8.5$ is indeed dominated by unburned
material.

With its spectral  properties, SN~2009dc deviates from any of the
established  sub-classes of peculiar SNe~Ia. The classical very
luminous SN~Ia 1991T is characterised by very high ionisation, the
blue spectrum being dominated by prominent \FeIII\ lines and little
else. In particular, lines from IMEs are very weak, and no features of
\CII\ or \OI\ are found. Contrary to the low-velocity SN~2009dc,
the lines in SN~1991T are rather broad, and their blueshift is
similar to that in normal SNe~Ia. 02cx-like SNe like SN~2005hk show
the same low ejecta velocities as SN~2009dc, but apart from that their
early-time spectra closely resemble those of 91T-like objects.

A comparison with other superluminous SNe~Ia reveals that SN~2007if
has similar velocities as SN~2009dc, but apparently a higher
ionisation. This results in much weaker lines from IMEs and unburned
material and the presence of \FeIII\ lines. However, \FeIII\ in
SN~2007if is not as strong as in SN~1991T, indicating that the
ionisation is not quite as high, or the Fe-group-element abundance
above the photosphere is lower. In SN~2006gz, on the contrary, the
same features as in SN~2009dc are present (though with weaker \CII\
lines). Line blending seems to be even lower, and all features are
sharp and pronounced. However, contrary to common sense these narrow
lines do not correspond to particularly low velocities of the ejecta
as determined from the blueshift of the P-Cygni minima (which is
comparable to that in normal SNe~Ia and significantly larger than in
SN~2009dc).

\begin{figure}   
   \centering
   \includegraphics[width=8.4cm]{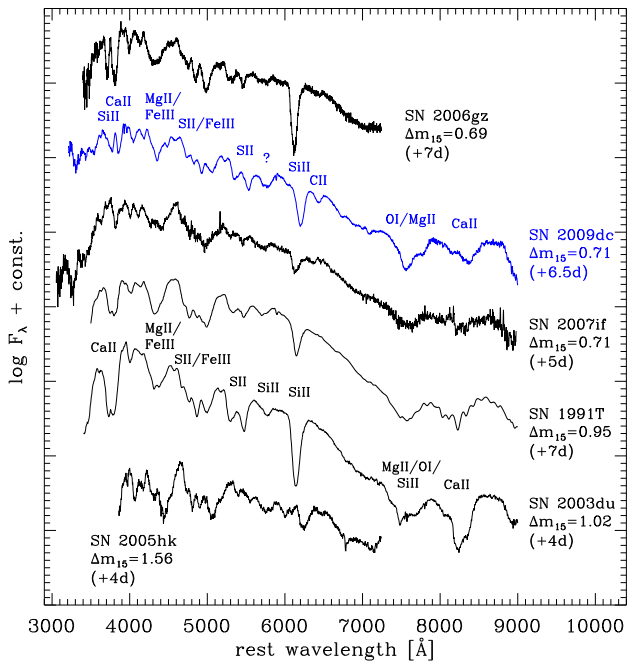}
   \caption{The same as Fig.~\ref{fig:comp_minus08}, but for spectra taken about 
   a week after maximum light.} 
   \label{fig:comp_plus06}
\end{figure}

\subsubsection{One week past maximum}

By a week after maximum (Fig.~\ref{fig:comp_plus06}), the spectrum of
SN~2009dc has evolved significantly, but its main characteristics have
not changed too much. There is still a blue continuum, spectral
features are not particularly deep, and the line velocities are even
lower than before. The UV flux has diminished compared to the first
observations, suggesting stronger UV line blanketing due to Fe-group
elements.  Also the lines from IMEs (\SiII, \SII, \MgII) have become
more pronounced, and the \CaII\ H\&K and NIR triplet lines are now
clearly visible (though still weak compared to normal SNe~Ia). \OI\
$\lambda7774$ is very pronounced, although the shape of the feature
indicates blending with another line, most likely \MgII\
$\lambda7890$. The \CII\ lines are weaker than before maximum light,
but still clearly detected. This is remarkable given that in no other
SN~Ia have such strong \CII\ lines been found at such a relatively
late epoch. The identification of an absorption feature at
$\sim$\,5760\,\AA\ remains unclear: compared to \SiII\ $\lambda6355$
it is at too blue a wavelength to be identified with \SiII\
$\lambda5972$. \NaI\,D, possibly blended with \SiII\ $\lambda5972$,
could be an explanation. 

Apart from the lower velocities and the \CII\ lines, the spectrum of
SN~2009dc is now fairly similar to that of SN~1991T. A normal SN~Ia
like 2003du has deeper \SiII, \SII\ and \CaII\ lines, whereas in
SN~2005hk these IME features are shallower than in SN~2009dc and at
even lower velocity. SN~2009dc's superluminous colleagues again reveal 
some variation in line velocities and strengths within this subgroup:
both SN~2006gz and 2007if show higher velocities than SN~2009dc at
this phase, but while the absorption lines in SN~2006gz are as deep
as in a normal SN~Ia, those in SN~2007if are even less pronounced than
those in SN~2009dc. 

\begin{figure}   
   \centering
   \includegraphics[width=8.4cm]{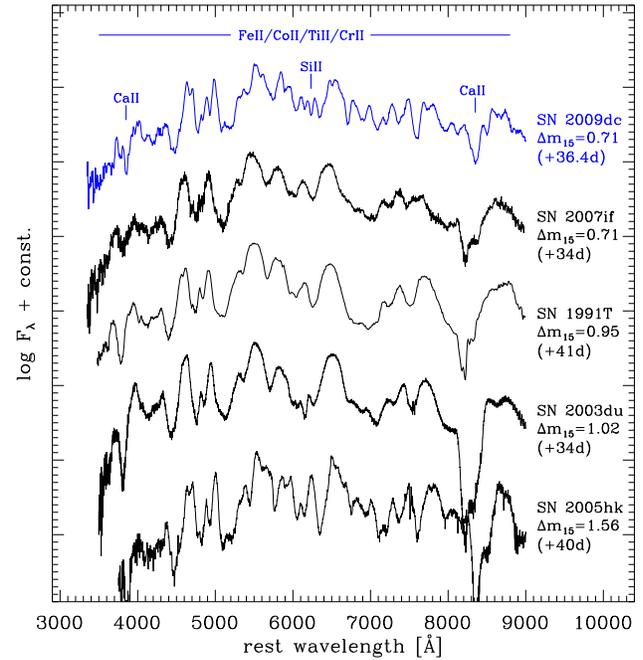}
   \caption{The same as Fig.~\ref{fig:comp_minus08}, but 5--6 weeks after 
   maximum light. The spectrum of SN~2007if has been slightly smoothed for 
   better clarity.}
   \label{fig:comp_plus36}
\end{figure}

\subsubsection{Five weeks past maximum}

Five to six weeks after maximum the spectra of all SNe~Ia are
dominated by lines of Fe-group elements (\FeII, \CoII, \TiII, \CrII)
and \CaII, and SN~2009dc is no exception in this respect
(Fig.~\ref{fig:comp_plus36}). SN~2003du, SN~1991T and the
superluminous SN~2007if are very similar at this epoch, showing fairly
broad and blended spectral features. This suggests that at least the
inner ejecta of SN~2007if have velocities comparable to normal
SNe~Ia. SN~2009dc, on the contrary, has much narrower and less blended
lines. There is now a remarkable similarity between SN~2009dc and the
02cx-like SN~Ia 2005hk in terms of velocities and line blending, but
also in the relative strengths of different lines. Exceptions are
\CaII\ (which is weaker in SN~2009dc) and \SiII\ $\lambda6355$ (which
can still be identified in SN~2009dc but is absent in SN~2005hk).

\begin{figure}   
   \centering
   \includegraphics[width=8.4cm]{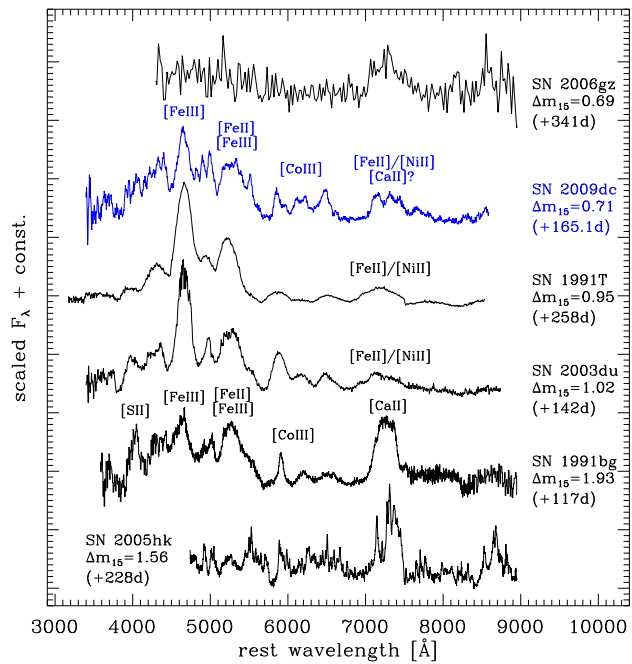}
   \caption{Comparison of SN~2009dc with SNe~2006gz \citep{Maeda09},
     1991T, 2003du, 2005hk \citep{Sahu08} and the subluminous 1991bg
     \citep{Turatto96} during the nebular phase. The +165.1\,d
     spectrum of SN~2009dc has been smoothed with a kernel of 2500
     \kms\ for presentation purposes. Prominent emission features have
     been identified.}
   \label{fig:comp_nebular}
\end{figure}

\subsubsection{Nebular phase}

Figure~\ref{fig:comp_nebular} presents late-time spectra of
SNe~2009dc, 2006gz, 1991T, 2003du, 2005hk and the prototypical
subluminous Type Ia SN~1991bg. Several of the spectra are too early to
be considered fully nebular, but in all of them the (forbidden)
emission lines dominate over residual photospheric flux. SNe~2003du
and 1991T have very similar spectra with prominent [\FeII], [\FeIII]
and [\CoIII] emission lines, the reduced strength of the [\CoIII]
lines in SN~1991T being a consequence of the later epoch of the
spectrum with more Co already decayed to Fe. The lines in SN~1991T are
also broader than those in SN~2003du, indicating more Fe-group
material at higher velocities, in agreement with the larger Ni mass
and the detection of Fe-group material in the outer layers in early
spectra \citep{RuizLapuente92,Mazzali95}. SN~2009dc once more shows
narrower and better resolved emission lines owing to the low ejecta
velocities. However, there are also more subtle differences. The
emission feature at $\sim$\,4650\,\AA, mostly a blend of [\FeIII]
lines, is much less pronounced in SN~2009dc, indicating a lower
\FeIII-to-\FeII\ ratio than normal. This lower ionisation is most
likely explained by enhanced recombination as a consequence of the
relatively low velocities\,/\,high densities in the inner ejecta of
SN~2009dc. In this particular aspect, SN~2009dc resembles subluminous
SNe~Ia such as SN~1991bg, which also show suppressed [\FeIII] lines in
their nebular spectra because of low velocities and temperatures
\citep{Mazzali97}. However, 91bg-like SNe also show a prominent
emission feature at $\sim$\,7300\,\AA, identified as [\CaII]
$\lambda\lambda7291,7323$ by Mazzali et al., which is significantly
weaker and probably dominated by [\FeII] and [\NiII] in SN~2009dc.

The nebular spectra of SNe~2006gz \citep{Maeda09} and 2005hk
\citep{Sahu08} deserve some comments. The SN 2005hk spectrum is
characterised by a multitude of extremely narrow features (much
narrower now than those of SN~2009dc) and by the complete absence of
prominent [\FeII] and [\FeIII] lines. In fact, despite the relatively
advanced epoch, this spectrum is not yet nebular in most parts, which
is probably a consequence of the low velocities and the corresponding
high densities. The spectrum of SN~2006gz also is highly peculiar:
there is a strong feature at $\sim$\,7300\,\AA, but like in SN~2005hk
there is almost no emission in the blue (though this region suffers
from very low S/N).  In SN~2006gz this unusual spectral appearance
comes along with an unexpectedly low luminosity for the given epoch,
seemingly inconsistent with the large \Nifs\ mass derived from
early-time data. Given that we find an accelerated light-curve decline
also in SN~2009dc past 200\,d (Section~\ref{Photometric evolution}),
one could speculate that at a sufficiently late epoch (more than a
year after the explosion) the spectra of SN~2009dc might look similar
to the nebular spectrum of SN~2006gz.

\subsubsection{\rm C\,\textsc{ii} \it lines}

\begin{figure}
   \centering
   \includegraphics[width=8.4cm]{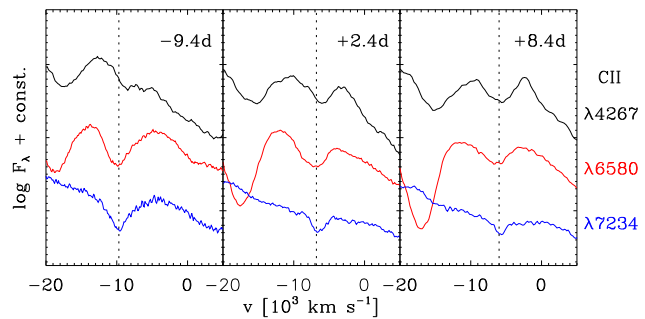}
   \caption{\CII\ lines in SN~2009dc. The profiles of \CII\
     $\lambda4267$ (top black line; possibly \CrII\ instead of \CII,
     see text), \CII\ $\lambda6580$ (middle red line) and \CII\
     $\lambda7234$ (bottom blue line) are compared at three different
     epochs. The vertical dotted lines correspond to velocities of
     9700, 6800 and 6000 km\,s$^{-1}$ at $-9.4$, $+2.4$ and $+8.4$\,d,
     respectively.}
   \label{fig:carbon}
\end{figure}

\begin{figure*}
   \centering
   \includegraphics[width=15cm]{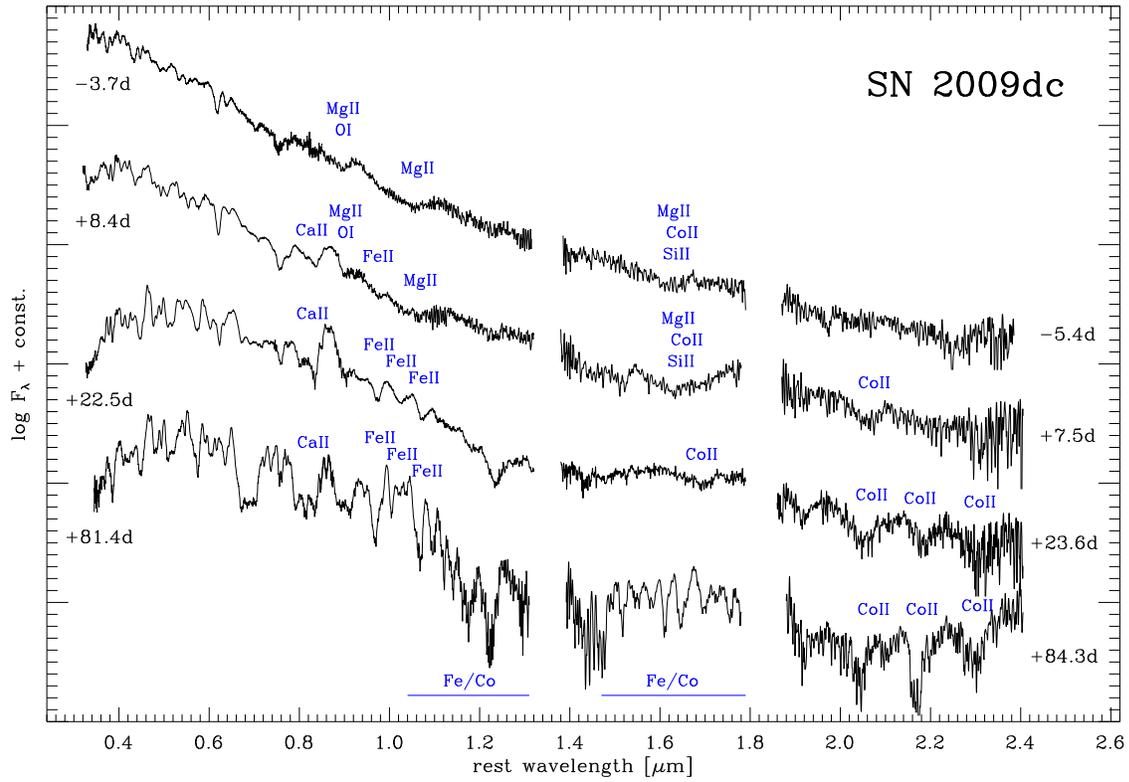}
   \caption{Combined optical and NIR spectra of SN~2009dc. The NIR spectra have 
   always been combined with the optical spectrum closest in time. The $-5.4$ and 
   $+84.3$\,d spectra of SN~2009dc have been smoothed 
   for better clarity. Line identifications have been adopted from \citet{Marion09}.}
   \label{fig:IR_opt}
\end{figure*}

As mentioned earlier, SN~2009dc is the Type Ia SN with the strongest
and most persistent \CII\ lines ever observed. \CII\ can be identified
from the earliest spectra to about two weeks after maximum
light. Previously, \CII\ features were only reported in very early
spectra of normal SNe~Ia \citep[e.g.][]{Thomas07}, in pre-maximum
spectra of the superluminous SN~Ia 2006gz \citep{Hicken07} and -- more
tentatively -- around maximum light in some 02cx-like SNe
\citep[e.g.,][]{Sahu08} and the superluminous SNe~2003fg
\citep{Howell06} and 2007if \citep{Scalzo10}. In SN~2009dc there is
little doubt that the \CII\ identification is correct, since not only the
$\lambda6580$ line, but also the $\lambda7234$ line is clearly
detected (Fig.~\ref{fig:carbon}). There may even be \CII\
$\lambda4267$ in form of a notch around 4150\,\AA, but
\citet{Scalzo10} suggested \CrII\ as an alternative explanation for
this feature in SN~2007if. Given that this line strengthens while
\CII\ $\lambda6580$ and $\lambda7234$ fade with time we tend to agree
with this interpretation.

\begin{figure*}
   \centering
   \includegraphics[width=11cm]{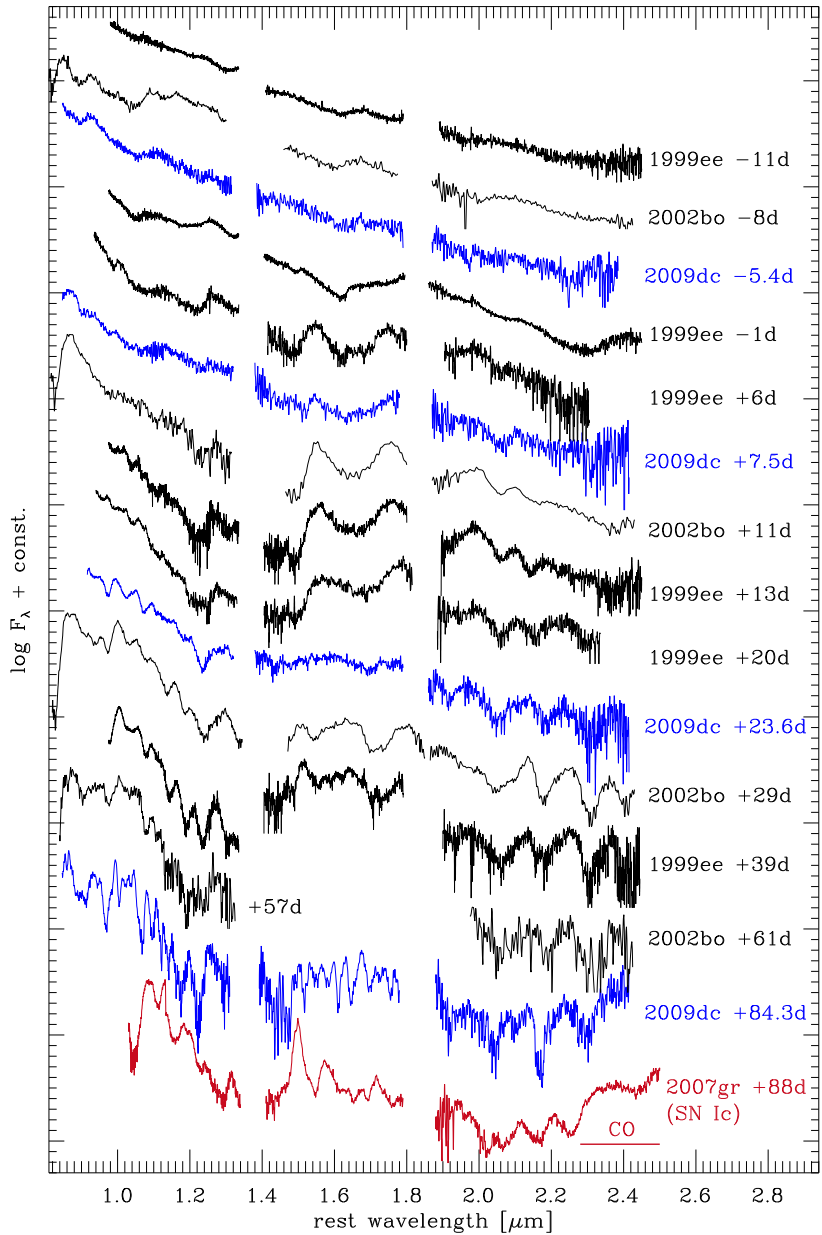}
   \caption{Comparison of NIR spectra of SN~2009dc with those of the
     normal Type Ia SNe~1999ee \citep{Hamuy02} and 2002bo
     \citep{Benetti04}. The $-5.4$ and $+84.3$\,d spectra of SN~2009dc
     have been smoothed
     for presentation purposes. A late-phase NIR spectrum of the Type
     Ic SN~2007gr is shown to illustrate the possibility of CO
     molecular-band emission in the $+84.3$\,d spectrum of
     SN~2009dc.}
   \label{fig:comp_IR}
\end{figure*}

\subsection{NIR spectra}

At four epochs, our optical spectra of SN~2009dc are complemented by
NIR spectra, so that combined spectra can be constructed which cover a
wavelength interval from $\sim$\,0.35 to $\sim$\,2.45\,$\mu$m (with
gaps in the NIR where the earth's atmosphere is opaque). These spectra
are presented in Fig.~\ref{fig:IR_opt}.  Starting with a relatively
smooth, blue continuum before maximum light, the NIR spectra of
SN~2009dc show increasing structure as the SN evolves, and develop
some flux deficit in the $J$ band. In the $+23.6$\,d spectrum, \CoII\
lines start to be visible in the $K$-band region, which is a typical
feature of SNe~Ia at these epochs \citep{Marion09}. The +84.3\,d
spectrum, finally, shows a large number of narrow absorptions and
emissions in the $J$- and $H$-band regions, most of them due to \CoII,
\FeII\ and other Fe-group elements \citep{Marion09}. In spite of the
prominent \CII\ features in the optical part of the spectrum, there is
no evidence at any time for \CI\ lines in the NIR. On the contrary, in
the carbon-rich Type Ic SN~2007gr, \CI\ was unambiguously detected in
the NIR early on \citep{Valenti08}, indicating a quite different level
of ionisation. Similarly, we do not find any features which could be
attributed to the NIR \HeI\ 1.083 and 2.058\,$\mu$m lines. Being the
strongest \HeI\ lines in the entire optical and NIR regime, these
lines are sensitive indicators of the amount of He present in the SN
ejecta. Their absence requires nearly He-free conditions, which could
be a challenge for some of the potential explosion models discussed in
Section~\ref{Explosion models}.

Figure~\ref{fig:comp_IR} presents a comparison of the NIR spectra of
SN~2009dc with those of the well-sampled normal SNe~Ia 1999ee
\citep{Hamuy02} and 2002bo \citep{Benetti04}. While SNe~1999ee and
2002bo are very consistent with each other in their spectral
evolution, SN~2009dc shows both similarities and differences. The
overall spectral energy distribution of SN~2009dc and its time
evolution resemble those of the other objects, but the flux deficit
between 1.2 and 1.5 $\mu$m in the +23.6\,d spectrum of SN~2009dc is
less pronounced compared to SNe~1999ee and 2002bo. This is reflected
by the weaker post-maximum drop in the $J$ band seen in the light
curves (Fig.~\ref{fig:JHK}). 

With its fairly smooth, blue continuum, the $-5.4$\,d spectrum of
SN~2009dc resembles the $-8$\,d spectrum of SN~2002bo. At $+7.5$\,d
the difference to normal SNe~Ia is most pronounced in the region
around $1.5$\,$\mu$m, where SN~2009dc shows only a hint of the strong
P-Cygni feature seen in other objects and normally attributed to
Fe-group elements \citep{Marion09}. On day $+23.6$, SN~2009dc shows
the typical blends of \CoII\ lines in the $K$-band region, which all
SNe~Ia develop a few weeks after maximum light. Also the $J$-band
region strongly resembles e.g. that of SN~2002bo on day $+29$. At the
same time, however, the $H$-band region in SN~2009dc shows less
structure than in ordinary SNe~Ia.

Interestingly, as a consequence of the decreasing velocities the same
region is resolved into a large number of individual, narrow lines by
day $+84.3$, most of them attributed to Fe-group elements (\CoII\ and
\FeII\ in particular). Probably the same lines dominated this region
already at earlier epochs, but were so strongly blended that they
mimicked a continuous emission. The $J$-band region of SN~2009dc on
day $+84.3$ is characterised by the same features as that of SN~2002bo
on day $+61$ (\FeII\ and other Fe-group lines), but again all lines
are narrower and better resolved. Only the \CoII\ lines in the
$K$-band region are still strongly blended. Accordingly, little
evolution is seen in that region compared to the $+23.6$\,d spectrum,
except for a relative increase in the flux beyond 2.3\,$\mu$m. The
latter might be caused by emission of the first overtone band of CO,
as observed at similar epochs in some stripped-envelope CC-SNe
\citep[e.g. in SN~2007gr,][see Fig.~\ref{fig:comp_IR}]{Hunter09}.
Given that SN~2009dc had similarly strong C and O features at early
phases and comparably low ejecta velocities as SN~2007gr, molecule
formation at comparable epochs in these two objects appears possible,
although it has never been observed in ordinary SNe~Ia even at much
later phases \citep{Bowers97,Spyromilio04}.

\subsection{Ejecta velocities}
\label{Ejecta velocities}

\begin{figure}
   \centering
   \includegraphics[width=8.4cm]{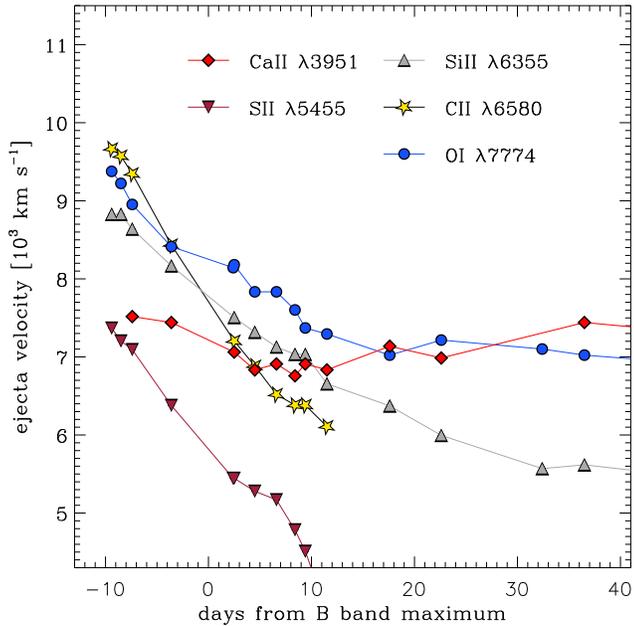}
   \caption{Expansion velocities of \CaII\ $\lambda3951$, \SII\ $\lambda5455$, 
   \SiII\ $\lambda6355$, \CII\ $\lambda6580$ and \OI\ $\lambda7774$, 
   determined from the minima of P-Cygni absorption profiles.}
   \label{fig:velocities}
\end{figure}

One of the most characteristic properties of SN~2009dc throughout its
spectroscopic evolution is the low expansion velocity of the ejecta
deduced from spectral lines. As shown in Fig.~\ref{fig:velocities},
even 10\,d before maximum light the velocities do not exceed 10\,000
\kms, regardles of which element is considered. The highest velocities
are measured in carbon and oxygen, in agreement with the idea that
unburned material should predominantly be found in the outer
layers. IMEs are at somewhat lower velocities: \SiII\ $\lambda6355$
yields a velocity estimate of $\sim$\,9000 \kms, whereas the \SII\
blend with an effective wavelength of 5455\,\AA\ yields just 7500
\kms. Given that the sulphur lines are intrinsically quite weak, they
are often considered to be a better tracer of the photosphere than
\SiII\ $\lambda6355$.

During the following 20--30\,d, the measured line velocities generally
decrease, levelling at $\sim$\,7000 \kms\ in the case of oxygen, and
at $\sim$\,5500 \kms\ in the case of silicon. \CII\ $\lambda6580$ has
a steeper velocity gradient than \SiII\ $\lambda6355$, and by the time
the carbon features disappear (about two weeks after maximum) they
form in deeper layers than \SiII\ $\lambda6355$. This suggests the
presence of a zone at $\sim$\,$6000$--$9000$ \kms\ where IMEs and
unburned material are mixed. Global asphericities could in principle
lead to a similar behaviour, but are disfavoured for the respective
layers by the spectropolarimetry presented by \citet{Tanaka09}. \SII\
$\lambda5455$ always has the lowest velocities of all lines studied
here, reaching 5000 \kms\ already a week after maximum light.
Thereafter, the identification of this feature is not certain; in fact
the accelerated slow-down seen in Fig.~\ref{fig:velocities} is an
indication that blending with other lines becomes important at that
epoch. 

An exception to the trend of decreasing velocities is made by
\CaII. The \CaII\ H\&K lines are hardly detected in the earliest spectra,
but then remain at a fairly constant velocity of 7000--7500 \kms\
throughout the photospheric phase. In particular, no high-velocity
features are seen in the \CaII\ lines, which are otherwise ubiquitous
in early spectra of SNe~Ia \citep{Mazzali05}. The constantly low velocity
constrains the \CaII\ abundance in the layers above $\sim$\,8000 \kms\
to be very low, since already small amounts of \CaII\ should be
visible given the intrinsic strength of the H\&K lines.

\begin{figure}
   \centering
   \includegraphics[width=8.4cm]{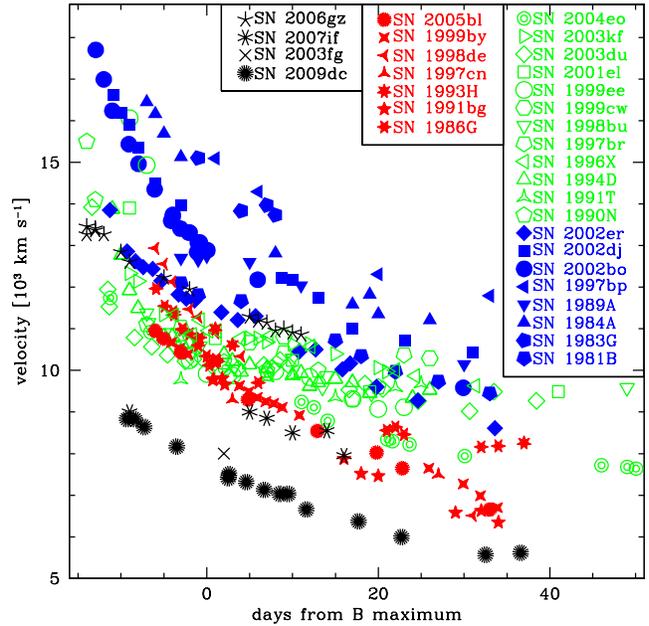}
   \caption{\SiII\ $\lambda6355$ velocity evolution of SN~2009dc compared to 
   that of other superluminous SNe~Ia (SNe~2003fg, 2006gz and 2007if) as well as 
   normal and subluminous SNe~Ia. Open green symbols stand for LVG SNe, filled 
   blue symbols for HVG SNe, and red starred symbols correspond to the FAINT SNe, 
   as defined by \citet{Benetti05}. Superluminous SNe~Ia are shown as black 
   symbols.}
   \label{fig:vel_vs_ph}
\end{figure}

In Fig.~\ref{fig:vel_vs_ph}, where the \SiII\ $\lambda6355$ velocity
evolution of a large number of SNe~Ia is compared, the exceptional
position of SN~2009dc is evident. At all times, SN~2009dc has
significantly lower velocities than all other SNe~Ia, including
subluminous, 91bg-like events. Even most other superluminous SNe~Ia
(with the exception of SN~2003fg, for which only one epoch of
spectroscopy is available) show significantly higher velocities than
SN~2009dc, SN~2007if by $\sim$\,2000 \kms\ and SN~2006gz by
$\sim$\,4000 \kms.

\begin{figure}
   \centering
   \includegraphics[width=8.4cm]{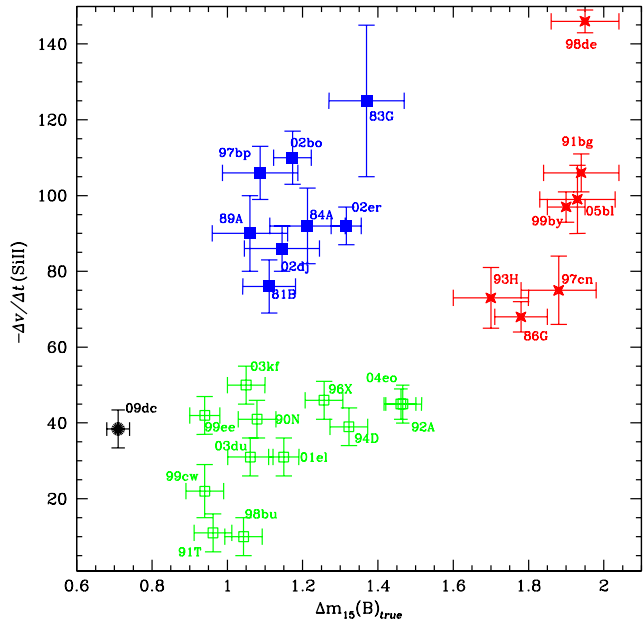}
   \caption{\SiII\ $\lambda6355$ post-maximum velocity gradient $\dot v$ versus 
   $\Delta$m$_{15}(B)$ (cf. \citealt{Benetti05} for a precise definition of 
   $\dot v$). Filled blue squares are HVG SNe, open green squares LVG SNe, and 
   filled red stars FAINT SNe. The velocity gradient of SN~2009dc is comparable 
   to those of LVG SNe.}
   \label{fig:dv_vs_dm15}
\end{figure}

In a hierarchical cluster analysis, \citet{Benetti05} arranged all
SNe~Ia in three groups (LVG, HVG and FAINT) based on their
post-maximum velocity gradient and their \dm15. With a velocity
gradient of $\dot v = 38 \pm 5$ \kms\,(100\,d)$^{-1}$, and \dm15 $=
0.71 \pm 0.03$, SN~2009dc extends the group of LVG SNe to smaller
\dm15\ (Fig.~\ref{fig:dv_vs_dm15}). However, given the peculiarities
of SN~2009dc, we hesitate to blindly assign it to the LVG group.
\citet{Branch06a} proposed an alternative scheme, based on the EWs of
\SiII\ $\lambda6355$ and $\lambda5972$ at maximum light.  With
50--60\,\AA\ and 5--8\,\AA, respectively, SN~2009dc would end up in
the shallow-silicon group, close to the boundary to core-normal
SNe. However, again the overall spectrophotometric similarity of
SN~2009dc to other members of this group (e.g. SN~1991T) is limited.

\section{Discussion}
\label{Discussion}

\subsection{Explosion parameters}
\label{Arnett}

In order to estimate the most relevant explosion parameters of
SN~2009dc, we have employed the analytic description of
\citet{Arnett82}. This model allows to derive the mass of \Nifs\
produced in the explosion, the ejecta mass and the total explosion
energy on the basis of the early-time bolometric luminosity evolution
(c.f.  Section~\ref{Bolometric light curve}). The model makes the
simplifying assumptions of spherical symmetry, homologously expanding
ejecta, no mixing of \Nifs, a constant mean optical opacity
$\kappa_\mathrm{opt}$, radiation-pressure dominance, and the
applicability of the diffusion approximation for photons, restricting
its application to early phases when the ejecta are optically thick.

In the Arnett model the peak of the bolometric light curve occurs when
the radiative losses equal the energy release by radioactivity.
Numerical simulations \citep{Hoeflich96} have shown that this equality
holds at least approximately, and that the deviations to either side
are typically not larger than 20 per cent in a broad set of models
studied in detail. Lacking information on the nature of SN~2009dc, we
make use of the original Arnett relation, but add in quadrature an
additional 20 per cent to the error of the derived \Nifs\ masses. The SN
luminosity at peak is thus given by
\begin{equation}
L(t_\mathrm{max}) = \varepsilon_\mathrm{Ni}\, M_\mathrm{Ni}(t_\mathrm{max}) + 
\varepsilon_\mathrm{Co}\, M_\mathrm{Co}(t_\mathrm{max}),
\end{equation}
where $\varepsilon_\mathrm{Ni/Co} =
\frac{Q_\mathrm{Ni/Co}}{m_\mathrm{Ni/Co}\,
  \tau_\mathrm{Ni/Co}}$. $Q_\mathrm{Ni/Co}$ is the energy release per
decay (1.73 MeV and 3.57 MeV for \Nifs\ and \Cofs, respectively, not
taking into account the energy in neutrinos, for which the ejecta are
optically thin).  $\tau_\mathrm{Ni/Co}$ is the decay time (8.77\,d and
111.4\,d, respectively) and $m_\mathrm{Ni/Co}$ the atomic mass of
\Nifs/\Cofs. For a given $t_\mathrm{max}$, the masses of \Nifs\ and
\Cofs\ at peak can be calculated from the initial \Nifs\ mass
($M_\mathrm{Ni}^0$) through the decay chain \Nifs\ $\rightarrow$
\Cofs\ $\rightarrow$ \Fefs\ (assuming that $M_\mathrm{Co}^0 = 0$).

A complication in the case of SN~2009dc is the unknown rise
time. Since the explosion was different from a normal SN~Ia,
there is a priori no reason to believe that the rise time should be
the same. A lower limit to the rise time is provided by a very early
detection 21.2\,d before maximum light in a pre-discovery image, as
reported by \citet{Silverman10}, when the SN was about 3.5 mag below
peak. The last non-detection down to a limiting magnitude of 19.3
dates to one week before this detection, i.e. 28\,d before maximum
light \citep{Silverman10}. We have therefore carried out the
calculations twice, once for a rise time of 22\,d and once for a rise
time of 28\,d. With $\log L(t_\mathrm{max}) = 43.47\pm0.11$ [erg
s$^{-1}$] and $t_\mathrm{max} = 22$\,d, a \Nifs\ mass of
$M_\mathrm{Ni}^0 = 1.78\pm0.58$ \Msun\ is derived. For $t_\mathrm{max}
= 28$\,d, a \Nifs\ mass of $M_\mathrm{Ni}^0 = 2.21\pm0.75$ \Msun\
would be required to power the light curve. In both cases the quoted
errors include uncertainties in the distance and reddening estimates,
and a 20 per cent error intrinsic to the analytic model. These numbers
are consistent with those obtained by \citet{Yamanaka09} and
\citet{Silverman10} using Arnett's law and assuming a similar host
reddening, i.e., $M_\mathrm{Ni}^0 = 1.8\pm0.5$ \Msun\ [for
\ebv$_\mathrm{host}$ = 0.14 mag and $t_\mathrm{max}=20$\,d] and
$1.7\pm0.4$ \Msun\ [for \ebv$_\mathrm{host}$ = 0.10 mag and
$t_\mathrm{max}=23$\,d], respectively.  Based on our calculation, the
lowest \Nifs\ mass for SN~2009dc, as derived for a short rise time,
small distance, low reddening and a peak luminosity exceeding the
instantaneous energy deposition by 20 per cent, is 1.20 \Msun. This is
just below \MCh, which may be important to reconcile SN~2009dc with
thermonuclear explosion scenarios (see Section~\ref{Explosion
  models}).

The \citet{Arnett82} description also allows to obtain estimates of
the ejecta mass $M_\mathrm{ej}$ of SN~2009dc, using the relation
\begin{equation}
\label{tau}
\tau_\mathrm{m} = \left(\frac{2}{\beta c} \frac{\kappa_\mathrm{opt} 
M_\mathrm{ej}}{v_\mathrm{sc}} \right)^{1/2}\! \propto \
\kappa_\mathrm{opt}^{1/2}\, M_\mathrm{ej}^{3/4}\, E_\mathrm{kin}^{-1/4}.
\end{equation}
In this formula, $\tau_\mathrm{m}$ is the effective photon diffusion
time, $v_\mathrm{sc}$ the velocity scale of the homologous expansion,
and $\beta$ an integration constant. We have calculated all quantities
relative to SN~2003du [\dm15\ = 1.02], assuming that the mean optical
opacity $\kappa_\mathrm{opt}$ is approximately the same in these two
SNe.\footnote{Given the large abundance of Fe-group material in
  SN~2009dc, the $\kappa_\mathrm{opt}$ of a Type Ia SN appears to be a
  better match than that of a core-collapse SN, no matter what the
  nature of SN~2009dc actually is.}  The ratio of the effective
diffusion times in SNe~2009dc and 2003du has been estimated measuring
the widths of the bolometric light-curve peaks [between points where
$L(t) = e^{-1}\ L(t_\mathrm{max})$], yelding $\tau_\mathrm{m,09dc}
\approx 1.65\ \tau_\mathrm{m,03du}$. Similarly, the ratio of velocity
scales has been estimated from the velocities measured in \SiII\
$\lambda6355$ at maximum light, yielding $v_\mathrm{sc,09dc} \approx
0.75\ v_\mathrm{sc,03du}$. With Eq.~\ref{tau} we obtain
$M_\mathrm{ej,09dc} \approx 2.04\,M_\mathrm{ej,03du}$ = 2.84\,\Msun\
assuming that SN~2003du was a \MCh-explosion. $E_\mathrm{kin} \propto
M_\mathrm{ej}\, v_\mathrm{sc}^2$ then yields $E_\mathrm{kin,09dc}
\approx 1.15\,E_\mathrm{kin,03du} \approx 1.61$ foe.\\

These numbers confirm the conclusion of \citet{Yamanaka09} and
\citet{Silverman10}, i.e. that it is impossible to explain SN~2009dc
by the explosion of a \MCh-WD. In fact, already the \Nifs\ mass --
whose estimate is more robust than that of the total ejecta mass --
probably exceeds 1.4 \Msun. A total ejecta mass of $\sim$\,2.8 \Msun\
is furthermore an utmost challenge for all scenarios that invoke
thermonuclear explosions of WDs.
It is noteworthy that the total explosion energy is just $\sim$\,15
per cent larger than in ordinary SNe~Ia. At the same time, the energy
production through nuclear burning probably exceeds that in normal
SNe~Ia by at least a factor of $2$. This indicates that whatever the
progenitor is, it should have a high binding energy per unit mass
unless we underestimate the amount of low-opacity material (and hence
also the total ejecta mass).

Of course, there is the caveat that in the case of SN~2009dc Arnett's
law might simply not be applicable, if some of the assumptions that
enter into this model are not fulfilled. Strong deviations from
spherical symmetry could for instance produce significant deviations
from the Arnett model.  However, the degree of asphericity is limited
by the negligible continuum polarisation found by \citet{Tanaka09} in
polarisation spectra of SN~2009dc.

\subsection{Enhanced late-time decline}
\label{Late-time decline}

As shown in Sections~\ref{Photometric evolution} and \ref{Bolometric
  light curve}, starting at 200--250\,d after maximum the late-time
light curves of SN~2009dc decline much more rapidly than before. This
behaviour is most pronounced in the $B$ and $V$ filters
(Fig.~\ref{fig:UBVRI}), which are characterised by forbidden Fe
emission lines (Section~\ref{Optical spectra}). Since the $B$ and $V$
bands dominate the optical emission at late times, also the
pseudo-bolometric light curve, constructed from $B$- to $J$-band data
with estimated corrections for the $U$, $H$ and $K'$ contributions,
reflects this trend (Fig.~\ref{fig:bolom}). The enhanced late-time
fading of SN~2009dc was also noted by \citet{Silverman10}, who showed
that an estimate based on their photometry taken 403\,d after maximum
would yield a \Nifs\ mass of only $\sim$\,0.4 \Msun. In the following
we discuss possible explanations for this behaviour, although no final
conclusion can be achieved with the available data.

First of all, it should be stressed that an increased luminosity
decline after some point cannot be explained by other radioactive
species starting to dominate over \Cofs. In order to dominate at
\textit{late} times, such a nucleus needs to have a longer half life
than \Cofs, which would result in a slow-down of the decline rather
than an acceleration once it starts to dominate the energy deposition.

Depending on the perspective, the situation encountered in SN~2009dc
can be interpreted in two ways: as a late-time flux deficit (at least
in the optical regime) or as a flux excess during the first
200--250\,d after the explosion. If interpreted as late-time deficit,
there are again two scenarios: a decrease in the true bolometric
luminosity, or a re-distribution of flux into wavelength regions not
covered by our observations.

A drop in the true bolometric luminosity would require a change in the
energy deposition rate, caused by an increased $\gamma$-ray or
positron escape fraction. While this happens in all SNe as the ejecta
expand, it is normally a gradual process, and it may be doubted
whether a situation can be constructed that leads to a rather sudden
drop of the opacity after some point in time. The alternative is a
re-distribution of the emission into other wavelength regions, most
likely the IR \citep{Silverman10}. Since we have no constraints on the
late-time IR luminosity of SN~2009dc beyond the $J$ band, this is,
however, pure speculation. A flux re-distribution into the IR could be
achieved by an unexpectedly early IR catastrophe \citep{Axelrod80},
where \FeII\ recombines to \FeI\ which emits predominantly at IR
wavelengths. Alternatively, dust may form within the ejecta, leading
to an absorption of optical light and re-emission in form of a thermal
continuum determined by the temperature of the dust grains. Dust
formation has never been observed in SNe~Ia so far. However, from the
early spectra we have indications that the ejecta of SN~2009dc
contained more carbon than those of other SNe~Ia, which opens the
possibility of graphite condensation once the ejecta have cooled down
sufficiently. In fact, if the (highly uncertain) detection of CO
emission in the +84.3\,d NIR spectrum was correct, the additional
cooling through molecular bands might create favourable conditions for
dust formation at a later moment \citep{Fassia01}. The scenario of an
IR catastrophe, on the other hand, seems to suffer from more
shortcomings. First of all, even in very late observations of SNe~Ia
no IR catastrophe has ever been observed, although theoretical models
predict it to occur between one and two years after the explosion
\citep[see, e.g.,][]{Leloudas09}. An IR catastrophe requires low
densities; the densities in the slowly expanding SN~2009dc after
200\,d, however, are much higher than those in a two or three times
older normal SN~Ia.

Seen from a different perspective, it might well be that the reason
for the strange light curve of SN~2009dc must not be searched for in
its late-time, but rather in its early-time behaviour, in a sense that
during the first 200--250\,d there was some extra emission that
stopped after that time. Such a flux excess could arise from
ejecta-CSM interaction. This scenario would help to turn SN~2009dc
into a more `normal' object since much less \Nifs\ would be required
to power the light curve. Moreover, it would partially explain the low
velocities by kinetic energy transfer to swept-up material, and -- if
the extra emission was in the form of an underlying continuum -- the
fairly shallow spectral features at early epochs could be understood
\citep{Hamuy03}.  What argues against this interaction scenario is the
lack of direct interaction signatures such as high-velocity spectral
features or narrow emission lines. At least if the CSM was
hydrogen-rich, narrow H$\alpha$ emission should have shown up.  On the
other hand, a sufficiently massive, hydrogen-free CSM is not
straightforward to obtain. Eventually, a lot of fine-tuning would be
necessary for the interaction contribution to mimic the temporal
evolution of a SN light curve, especially if the interaction is
supposed to boost the SN luminosity by a factor 2 or more (in the end,
the colour evolution of SN~2009dc is not too different from that of a
normal SN~Ia, and the light-curve decline after the peak phase follows
the slope of objects powered by \Cofs\ decay).

Whatever the truth behind the late-time light-curve dimming is, it was
very likely also the reason for the highly peculiar late-time
behaviour of SN~2006gz \citep{Maeda09}. The mechanism may have worked
more efficiently in SN~2006gz -- a spectrum taken a year after maximum
shows essentially no emission in the blue part, and a photometric
recovery of the SN almost failed -- but by extrapolation one may guess
that SN~2009dc is on a good way to catch up.

\subsection{Explosion models}
\label{Explosion models}

The high luminosity of SN~2009dc, its low ejecta velocities, and the
chemical composition of the ejecta with a lot of IMEs, C and O, but
without any trace of H or He, pose a challenge to any explosion model
one can think of, be it thermonuclear or core collapse. In particular,
as demonstrated by \citet{Yamanaka09} and \citet{Silverman10} and as
is evident from the derived explosion parameters
(Section~\ref{Arnett}), in the absence of CSM interaction there is no
way for SN~2009dc to be the explosion of a regular \MCh-WD. In what
follows, we go through possible progenitors and explosion channels for
SN~2009dc, some of them already scrutinised in the literature, others
being new suggestions to widen the discussion. A successful model has
to explain the large production of \Nifs, the ejection of $\geq 2.5$
\Msun\ of material with low kinetic energy per mass, the presence of C
and O in all layers down to 6000 \kms, the absence of H and He, and
the peculiar evolution of the late-time SN luminosity. This discussion
is based on SN~2009dc and the explosion parameters derived for
it. However, given the similarities within the group of superluminous
SNe~Ia, most arguments are also valid for other members of this class.

\subsubsection{Rotating `super-\MCh' white dwarfs}

Rapidly rotating WDs were first suggested by \citet{Howell06} as a
progenitor for the superluminous SN~Ia 2003fg. Indeed, as shown by
\citet{Yoon05}, differentially rotating WDs can be stabilised by
centrifugal force and exceed the Chandrasekhar mass of non-rotating
WDs by quite a margin. However, this model faces some problems.

At least within the single-degenerate scenario it is difficult for a
WD to accrete all the material and gain all the angular momentum
needed to grow to $\geq2$ \Msun. CO WDs are typically born with masses
of $\sim$\,0.6 \Msun\ \citep*{Koester79,Weidemann83,Liebert05} and
with slow rotation \citep[e.g.,][]{Spruit98}. Even if the WD was
initially as massive as 1 \Msun, it would still have to accrete at
least another solar mass of material, provided by the binary
companion. This large mass requires the companion itself to be quite
massive, but its mass is limited by the fact that its ZAMS mass must
have been lower than that of the primary, since it is less
evolved. This narrows down the possible parameter space for primary
and secondary significantly. In fact, \citet{Langer00} estimated that
no CO WD can grow much beyond 2 \Msun\ by accretion from a
non-degenerate companion, which is probably not enough to explain the
ejecta mass of $\sim$\,2.8 \Msun\ derived for SN~2009dc
(Section~\ref{Arnett}).

Some of the problems concerning the accretion and spin-up could be
avoided if the secondary was also a CO WD, disrupted in the course of
a merger, with its material being accreted steadily onto the primary
\citep{Hicken07}. However, in the case of a merger of two massive WDs
with a mass ratio close to one, a dynamical merger with a detonation
triggered already during the act of merging \citep{Pakmor10} may be
the more likely outcome. This scenario is discussed in the next
section. 

A problem of all rotating models, be it a single WD or the result of a
WD merger, may be the kinetic energy of the ejecta. Synthesising at
least 1.2 \Msun\ of \Nifs\ plus some IMEs, SN~2009dc produced 50 to
100 per cent more energy than a normal SN~Ia.  If this energy is distributed
to 2 \Msun\ of ejecta, it will result in high ejecta velocities unless
the specific binding energy is significantly larger than in a
\MCh-WD. \citet{Yoon05} indeed see an increase in the specific binding
energy by a factor $\sim$\,1.7 when going from a non-rotating \MCh-WD
to a rotating WD with 2 \Msun. However, given that the binding energy
is only a fraction of the total explosion energy this does not affect
the specific kinetic energy by too much. Accordingly,
\citet*{Pfannes10} find that the ejecta velocities of a detonating 2.1
\Msun\ WD are very similar to those encountered in normal SNe~Ia, and
too high to be compatible with SN~2009dc.

Finally, rapidly rotating massive WDs are significantly deformed
\citep{Yoon05}, and the oblate geometry of the WD would likely result
in an aspherical SN explosion. However, spectropolarimetry suggests
that SN~2009dc was overall spherical \citep{Tanaka09}.

\subsubsection{Dynamical white-dwarf mergers}

\citet{Pakmor10} presented WD merger calculations in which a
detonation was triggered dynamically in a hot spot during the process
of merging. The computation was performed for the merger of two 0.9
\Msun\ WDs, which produced $\sim$\,0.1 \Msun\ of \Nifs. However, the
prerequisite for a dynamical explosion is not only the absolute mass
of the WDs, but also a mass ratio close to one \citep{Pakmor10}. The
central densities of WDs are a steep function of their mass. Hence, it
may be expected that e.g. a pair of 1.2 \Msun\ WDs produces much more
\Nifs\ during a dynamical merger, although it may not be sufficient
for SN~2009dc. The total ejecta mass of such an event would be 2.4
\Msun, which is at least closer to the value deduced for SN~2009dc
than rotating models can probably get. Of course, CO WDs with $\geq
1.2$ \Msun\ are rare \citep[e.g.][]{Liebert05} and so are binary
systems of them, but since the rate of superluminous SNe~Ia is low
(and not well known) this would per se not exclude this scenario.

Beyond 1.2 \Msun, essentially all WDs which have not grown through
accretion or mergers are ONe rather than CO WDs
\citep{Koester79,Weidemann83}. From the detection of carbon in the
early spectra one can conclude that SN~2009dc cannot have been a
merger of two ONe WDs, but a merger of an ONe WD and a similarly
massive CO WD could be feasible. Because of its higher mean atomic
mass, an ONe WD releases less energy when burned to \Nifs\ than a CO
WD, which could help in keeping the ejecta velocities low. A problem
might be the susceptibility of ONe WDs to electron captures on
$^{20}$Ne and $^{24}$Mg. In fact, numerical simulations have shown
that ONe WDs do not explode, but collapse to a neutron star when
approaching \MCh\ \citep{Miyaji80,Saio85}. This, of course, does not
have to be true for a dynamical merger, where the detonation triggered
by an `external event' and propagating supersonically might leave the
ONe WD no time to collapse.

\subsubsection{Core collapse in an envelope-stripped massive star}

Explaining SN~2009dc as a core-collapse explosion of a stripped star
has the obvious advantage that no strict limits on the ejecta mass
apply. Core-collapse events whose ejecta exceed \MCh\ are common.
Unfortunately, other hallmarks of core-collapse explosions are not met
by SN~2009dc. Typical core-collapse SNe produce of the order of 0.1
\Msun\ of \Nifs, and this number goes up to $\sim$\,0.5 \Msun\ for
some $\gamma$-ray burst related hypernovae like SNe~1998bw
\citep{Galama98,Maeda06} or 2003dh \citep{Hjorth03} which --
contrary to SN~2009dc -- are characterised by a very large kinetic
energy and high ejecta velocities.
Beyond this, \citet{Umeda08} recently constructed a series of
core-collapse models from very massive (20--100 \Msun) progenitors,
assuming little fall-back of synthesised \Nifs\ onto the newly formed
black hole. They found ejected \Nifs\ masses up to several \Msun\
\citep[see also][]{Moriya10}. However, given the very large total
ejecta masses, the composition of these objects -- as of all
core-collapse SNe -- cannot be considered Ni-rich. SN~2009dc is
different in this respect, with probably more than half of its ejecta
being made up of \Nifs. More generally, the overall abundance pattern
as suggested by the spectral time series of SN~2009dc is not typical
of core-collapse SNe. SN~2009dc is characterised by prominent IME
lines, most notably those of \SiII\ and \SII, which are weak
\citep[\SiII\ $\lambda6355$, see e.g.][]{Branch06a} or absent (\SII\
lines) in stripped-envelope core-collapse SNe. The latter events, on
the other hand, always eject a large amount of oxygen, giving rise to
a prominent \OIa\ $\lambda6300,6364$ emission in the nebular
phase. SN~2009dc does show \OI\ lines in early-time spectra, but no
hint of \OIa\ $\lambda6300,6364$ at late times, contrary to all
objects with an undisputed classification as stripped-envelope
core-collapse SNe known to date. This indicates a lack of mixing of O
and \Nifs, which is difficult to achieve in core-collapse explosions.

\subsubsection{Core collapse with activity of the central remnant}

Within the core-collapse scenario a possible way to reduce the mass of
\Nifs\ needed to power the light curve of SN~2009dc could be heating
from an active compact remnant. Indeed, the birth of a magnetar was
proposed by \citet{Maeda07} to explain the unusual light-curve
evolution of the Type Ib SN~2005bf, characterised by a broad, delayed
peak and a deep luminosity drop thereafter. An energy source different
from radioactivity allows more freedom in the light-curve shape, as an
exponential tail is not a necessary consequence. Accordingly, within
such a scenario the late-time luminosity drop seen in SN~2009dc could
be explained by the termination of the heating from the central
object. Compared to CSM interaction as an alternative
non-radioactivity-related energy source, heating from a compact
remnant has the advantage that the energy is injected at the centre
and released on photon diffusion time scales, which allows the light
curve to rise smoothly to peak \citep{Maeda07}. Moreover, since the
photons propagate through the entire ejecta, the normal processes of
spectrum formation take place, resulting in a fairly normal SN
spectrum \citep[as was the case for
SN~2005bf;][]{Tominaga05,Folatelli06}. Narrow emission lines are not
expected. 

What remains a problem, however, is the SN~Ia-like abundance pattern
derived from the spectra of SN~2009dc, as detailed above. SN~2005bf
was spectroscopically undoubtedly a stripped-envelope core-collapse
SN, and showed the hallmark \OIa\ and \CaIIa, but only weak Fe
emission lines in its nebular spectra. All of this is not the case for
SN~2009dc.

\subsubsection{Type I$\frac{1}{2}$ SNe: thermonuclear explosions of AGB-star cores}

Developed back in the 1970s (\citealt{Ergma74,Couch75,Iben82}, but see
\citealt{Iben83}), the idea of exploding AGB star cores has not found
much resonance during the past two decades. While stars with ZAMS
masses between $\sim$\,8 and 10 \Msun\ are predicted to form ONeMg
cores and possibly undergo electron-capture core collapse, those
between 0.5 and 8 \Msun\ end up with degenerate CO cores after
core He burning.  The subsequent He shell burning or AGB phase is
characterised by strong mass loss, and is usually terminated by the
ejection of the envelope as planetary nebula and the formation of a CO
WD that cools by radiative processes.  However, in the most massive
AGB stars the degenerate CO core might grow to 1.4 \Msun\ before
mass-loss processes can reduce the total mass of the star below that
value. In this case, carbon would be ignited in the core and a
thermonuclear runaway would drive a flame through the entire star and
disrupt it. \citet{Iben83} proposed the term `Type I$\frac{1}{2}$ SN'
for such an explosion, based on its thermonuclear origin but the
likely presence of H lines in the spectrum.

To make Type I$\frac{1}{2}$ SNe a suitable model for SN~2009dc,
several conditions have to be fulfilled. At the time of explosion, the
AGB star needs to have a total mass between 2.5 and 3 \Msun, which is
probably feasible. It must have lost its H envelope through stellar
winds or binary interaction to prevent H lines, most notably
H$\alpha$, from appearing in the SN spectra. The mass of the
degenerate CO core (i.e., 1.4 \Msun) is an upper limit for the amount
of \Nifs\ that can be produced in such an explosion, but if the core
is burned entirely to Fe-group material this might still be consistent
with SN~2009dc within the large uncertainties
(Section~\ref{Arnett}). Such a complete burning of the core
is only possible if the thermonuclear flame propagates as a
detonation, and not as a deflagration that gives the star time to
pre-expand. The He envelope would then have to be burned to carbon,
oxygen and IMEs, in order to match the abundance pattern observed in
the ejecta of SN~2009dc. This could be problematic, since too much
energy might be released by the He burning to be consistent with the
low ejecta velocities. If, on the other hand, a part of the He is left
unburned, it might show up in the spectra, in particular in the near
IR. If such Type I$\frac{1}{2}$ SNe exist, they are probably rare, but
small rates are not a problem as long as only superluminous SNe~Ia
shall be explained. In the end, detailed numerical simulations of
hydrodynamics and radiative transfer are required to assess the
feasibility of this model.

\subsection{The host galaxy of SN~2009dc}
\label{09dc host}

As mentioned in Section~\ref{Distance and extinction}, SN~2009dc is
located in the outskirts of the S0 galaxy UGC~10064. Although
lenticular galaxies are by no means dead stellar systems
\citep[e.g.][]{Emsellem07,Kawabata09}, such a host galaxy might still
be considered as an indication for a predominantly old stellar
population.  This is consistent with the lack of \HII-region emission
lines in the spectra of SN~2009dc, suggesting no strong star formation
activity at the SN location.

\begin{figure}   
   \centering
   \includegraphics[width=8.4cm]{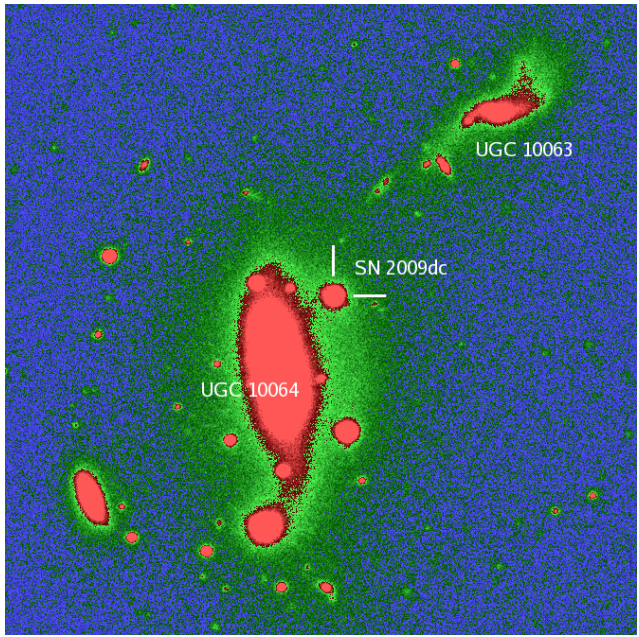}
   \caption{Stacked, contrast-enhanced $r$-band image of the environment of 
   SN~2009dc. The two potential host galaxies are labelled, and the tidal 
   bridge between them can be discerned. The field of view is $3 \times 3$ 
   arcmin$^2$, north is up and east to the left.}
   \label{fig:environment}
\end{figure}

However, UGC~10064 is not an isolated, unperturbed system. At the same
redshift, about 98.1 arcsec to its north-west (corresponding to a
projected distance of 44.6 kpc), there is the irregularly shaped,
strongly distorted galaxy UGC~10063.  Almost an order of magnitude
less luminous than UGC~10064, UGC~10063 is characterised by a
distinctly blue colour, indicative of luminous, young stars.  It shows
a tidal tail bridging the gap to its larger neighbour, and, as already
mentioned by \citet{Silverman10}, SN~2009dc is located at the
approximate endpoint of this tail (see Fig.~\ref{fig:environment}).

\begin{figure}   
   \centering
   \includegraphics[width=8.4cm]{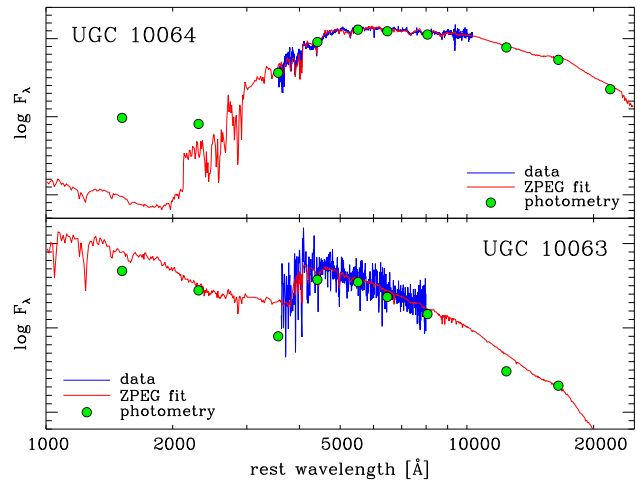}
   \caption{Optical spectra of UGC~10064 (upper panel) and UGC~10063
     (lower panel).The fluxes derived from UV (GALEX), optical (NOT)
     and NIR (2MASS) photometry, and the best-fit {\sc zpeg} template
     spectra are overplotted. Compared to the template, a UV excess in
     UGC~10064 is evident.}
   \label{fig:host_spectra}
\end{figure}

Coincident to our observations of SN~2009dc, we obtained deep
photometry of UGC~10064 and UGC~10063 at the NOT on 2009 August 12,
spectroscopy of UGC~10064 with DOLORES at the TNG on 2009 July 4, and
spectroscopy of UGC~10063 with CAFOS at the Calar Alto 2.2m on 2009
May 4. In addition to our own optical photometry, UV photometry from
GALEX \citep{galex} and NIR photometry from 2MASS \citep{twomass} were
available. We used {\sc sextractor} \citep{sextractor} in common
aperture mode, using the $B$ band to define the apertures, to derive
magnitudes from all imaging data. We then used {\sc zpeg} \citep{zpeg}
to match our photometry to galaxy evolutionary scenarios. Using the
standard set of nine evolutionary scenarios included with this code
and the internal presciptions for treating extinction by dust, we
match the photometry of UGC~10064 to a galaxy with a mass
$\log(M_*/M_\odot)=10.68^{+0.07}_{-0.06}$ that underwent a starburst
4~Gyr ago. The photometry of UGC~10063 matches an Sd galaxy of mass
$\log(M_*/M_\odot)=9.52^{+0.01}_{-0.08}$. Our spectra, the fluxes
inferred from our photometry, and the best-fit {\sc zpeg} template for
each galaxy are shown in Figure~\ref{fig:host_spectra}.

Interestingly, there appears to be an excess of UV flux for UGC~10064
compared to the best fitting {\sc zpeg} template, that of a starburst
of 4.0~Gyr age.  Inspection of the GALEX images indeed shows a visible
UV flux concentrated at the core of UGC~10064. Using the methods
described by \citet{Childress10}, we measure the emission-line fluxes
of UGC~10064 by subtracting a best-fit stellar template. Though weak
emission lines are detected, their flux ratios indicate ionisation by
an active galactic nucleus according to the BPT \citep*{bpt81} diagram
boundaries defined by \citet{kewley06}. This indicates that the UV
excess and weak-emission-line flux in UGC~10064 are likely to be
caused by an active nucleus, and that no significant star-formation
has occurred in the galaxy for at least several Gyr. Thus, if SN~2009dc
was born from the stellar population of UGC~10064, it is likely to
have a long delay time.

In contrast to the old stellar population of UGC~10064 is the much
younger population of UGC~10063. The best fitting {\sc zpeg} template
is that of an Sd galaxy whose luminosity-weighted stellar age is
1.5~Gyr. The spectrum of UGC~10063 shows no emission from recent star
formation, and is dominated by strong Balmer absorption features
indicating the galaxy to be in a post-starburst phase \citep[as was
also noted by][]{Silverman10}. This implies that UGC~10063 underwent a
strong burst of star formation a few hundred Myr before the explosion
of SN~2009dc, perhaps induced by a close encounter with
UGC~10064. Intriguigingly, this is a time scale similar to that
reported by \citet{Childress10} for the parent population of
SN~2007if, and could be indicative of a similar time scale for the
explosion of SN~2009dc.

A more in-depth study of the stellar population at the SN site has to
be postponed to a moment when the SN will have faded below
detectability. For the time being, it is not possible to decide if the
SN descended from an old, low-mass stellar population as expected for
its S0 host, or from the significantly younger population that
dominates the tidal tail of the companion galaxy.

\subsection{Host galaxies of superluminous SNe~Ia}
\label{hosts}

\begin{table*}
\caption{Properties of the host galaxies of superluminous SNe~Ia.}
\label{superluminous}
\begin{footnotesize}
\begin{tabular}{@{}lcccccrr@{}}
\hline
SN            & SN~$\Delta m_{15}(B)$ & Survey$^a$  & Host galaxy  & Morphology$^b$ & Redshift    & Host M$_{B/g}^c$     &  Host $\log(M_*/M_\odot)$  \\
\hline
SN 2003fg       & $0.84\pm0.15$ & SNLS              & anonymous    & Irr            & 0.244       &  ?\quad \quad\       &  $8.93^{+0.81}_{-0.50}\,^d$\ \quad\ \\
SN 2004gu       & $0.80\pm0.04$ & Texas             & FGC 175A     & spiral         & 0.046       &  $-18.9$\quad\       & $10.45^{+0.04}_{-0.20}$\quad\quad\ \\ 
SN 2006gz       & $0.69\pm0.04$ & Puckett\,/\,LOSS  & IC 1277      & Scd            & 0.024       & $-21.0$\quad\        & $10.28^{+0.01}_{-0.14}$\quad\quad\ \\
SN 2007if       & $0.71\pm0.06$ & ROTSE\,/\,SNF     & anonymous    &  ?             & 0.074       & $-14.1$\quad\        & $7.30^{+0.31}_{-0.31}\,^e$\ \quad\ \\
SNF20080723-012 &       ?       & SNF               & anonymous    & spiral         & 0.075       & $-17.1$\quad\        & $8.52^{+0.04}_{-0.07}$\quad\quad\ \\
SN 2009dc       & $0.71\pm0.03$ & Puckett           & UGC 10064    & S0             & 0.021       & $-20.3$\quad\        & $10.68^{+0.07}_{-0.06}$\quad\quad\ \\
                &               &                   & UGC 10063    & SBd            & 0.021       & $-18.8$\quad\        & $9.52^{+0.01}_{-0.08}$\quad\quad\ \\
SN 2009dr       &       ?       & PTF               & anonymous    &  ?             & $\sim$\,0.1 & $\geq -15.6$\quad\   & $\leq 8.3$\quad\quad\quad\ \\
\hline
\end{tabular}
\\[1.5ex]
\flushleft $^a$~SNLS = Supernova Legacy Survey, Texas = Texas
Supernova Search, Puckett = Puckett Supernova Search, LOSS = Lick
Observatory Supernova Search, ROTSE = ROTSE Supernova Verification
Project, SNF = Nearby Supernova Factory, PTF = Palomar Transient
Factory.\ \
$^b$~From LEDA, NED.\ \
$^c$~From LEDA, NED and own measurements in archival images.\ \
$^d$~\citet{Howell06}.\ \
$^e$~\citet{Childress10}.
\end{footnotesize}
\end{table*}

The question of host galaxies and parent populations is crucial for
narrowing down the range of possible explosion scenarios for
superluminous SNe~Ia. In particular, it might help discriminating
between a thermonuclear or core-collapse origin, since the stellar
populations involved are significantly different. To date, at least
seven SNe~Ia are known to be superluminous in the sense that their
\Nifs\ yields are incompatible with the explosions of \MCh-WDs. These
SNe do not form a homogeneous group: differences in their early
(ionisation state, line velocities) and late spectra (line widths,
strength of Fe lines; Taubenberger et al. in prep.) are evident. Their
hosts also exhibit some heterogeneity, but we will show below that the
ensemble of superluminous SN~Ia hosts have on average lower masses,
and presumably lower metallicities, than those of full samples of SNe
Ia from untargeted surveys.  In Table~\ref{superluminous}, some basic
properties of the host galaxies of superluminous SNe Ia are summarised.

We derived the host masses by collecting previously published host
masses where available, and by applying the code {\sc zpeg} to
multi-band photometry of the other hosts, available from public
sources (SDSS, GALEX, 2MASS) or our own observations. Previously
published masses for host galaxies of superluminous SNe~Ia include
that of the host of SN~2003fg at
$\log(M_*/M_\odot)=8.93^{+0.81}_{-0.50}$ \citep{Howell06} and that of
the host of SN~2007if at $\log(M_*/M_\odot)=7.30\pm0.31$
\citep{Childress10}. For FGC~175A, the host of SN~2004gu
\citep{Contreras10}, optical $ugriz$ photometry was obtained from SDSS
DR7 \citep{sdssdr7}, UV data from GALEX, and $JHK$ photometry from
2MASS. Archival $BVR$ photometry from Subaru was available for IC~1277
\citep{Maeda09}, the host of SN~2006gz, along with 2MASS data.  We
used our own optical VLT + FORS2 $BVRI$ photometry for the host of
SNF20080723-012 to derive its mass. Our own NOT images of 2009 August
12 provided $UBVRI$ photometry of both UGC~10064 and UGC~10063, while
GALEX UV and 2MASS IR were also available. Optical $ugriz$ photometry
from SDSS is available at the site of SN~2009dr \citep{sn2009dr}, but
the host is not detected.

We used {\sc sextractor} in common aperture mode, using either the $g$
band or $B$ band to define the apertures, to derive magnitudes from
all imaging data. We then used {\sc zpeg} to derive host masses from
our photometry, finding $\log(M_*/M_\odot)=10.45^{+0.04}_{-0.20}$ for
FGC~175A, $\log(M_*/M_\odot)=10.28^{+0.01}_{-0.14}$ for IC~1277,
$\log(M_*/M_\odot)=8.52^{+0.04}_{-0.07}$ for the host of
SNF20080723-012, $\log(M_*/M_\odot)=10.68^{+0.07}_{-0.06}$ for
UGC~10064, and $\log(M_*/M_\odot)=9.52^{+0.01}_{-0.08}$ for UGC~10063.
The host of SN~2009dr is undetected in SDSS data, so using the
$3\sigma$ SDSS $g$-band limiting magnitude of $m_g\,\approx\,22.6$
\citep{stoughton02} at the redshift of this object
\citep[$z\,\approx\,0.10$][]{sn2009dr} we estimate that the host must
be fainter than $M_g\,\approx\,-15.6$. With a (conservatively large)
solar mass-to-light ratio from \citet{blanton03}, this places an upper
mass limit for its host at $\log(M_*/M_\odot)\,\lesssim\,8.3$.

The mass-metallicity (MZ) relation \citep{Tremonti04} supports the
interpretation of this result as a prevalence of low-metallicity
environments for the preferred birthplace of superluminous SNe~Ia.  To
confirm this interpretation, we collected spectroscopic data for the
hosts of SN~2004gu, SN~2006gz, SNF20080723-012 and SN~2009dc.  Note
that contrary to the case of SN~2007if \citep{Childress10}, whose host
was a dwarf galaxy and therefore chemically well mixed, a metallicity
gradient is expected for these more massive galaxies. The derived core
metallicities are hence upper limits for the actual metallicities in
the immediate SN environments.

The emission-line fluxes from the SDSS spectrum of the host of
SN~2004gu were derived by \citet{Tremonti04}, while we measured
emission-line fluxes for the host of SN~2006gz from an archival Subaru
+ FOCAS spectrum \citep{Maeda09}, and for the host of SNF20080723-012
from our own nebular-phase observations of the SN using FORS2 on the
VLT. We then employed the O3N2 method of \citet{pp04} to derive
gas-phase oxygen abundances of
$12+\log(\mathrm{O/H})_\mathrm{04gu,PP04} = 8.74\pm0.05$,
$12+\log(\mathrm{O/H})_\mathrm{06gz,PP04} = 8.35\pm0.03$, and
$12+\log(\mathrm{O/H})_\mathrm{SNF,PP04} = 8.36\pm0.15$.  We note that
the error bar for the metallicity of the host of SNF20080723-012 is
likely underestimated, as contamination from the nebular SN prevents
fitting of the stellar continuum to obtain absorption corrections for
Balmer line fluxes. Using the conversion formulae of \citet{ke08}, we
converted these metallicity values to the scale of \citet{Tremonti04}
to place them on a common scale with the fiducial MZ relation, and
compared these to the solar oxygen abundance of
\citet{delahaye10}. This yields final metallicities of
$\log(Z/Z_\odot)_\mathrm{04gu} = 0.16\pm0.05$,
$\log(Z/Z_\odot)_\mathrm{06gz} = -0.28\pm0.04$, and
$\log(Z/Z_\odot)_\mathrm{SNF} = -0.27\pm0.15$.  The metallicity of the
host of SN~2007if was determined in precisely the same manner by
\citet{Childress10}, who found $\log(Z/Z_\odot)_\mathrm{07if} =
-1.15\pm0.14$.  The metallicity of UGC~10064 was previously measured
by \citet{wegrog08} using stellar absorption features. They found
$\log(Z/Z_\odot)_\mathrm{UGC~10064} = 0.45\pm0.07$. Though systematic
discrepancies between galaxy metallicities measured from stellar
features and those from nebular emission are known, this metallicity
for UGC~10064 is consistent with its host mass and the MZ relation.

\begin{figure}   
   \centering
   \includegraphics[width=8.3cm]{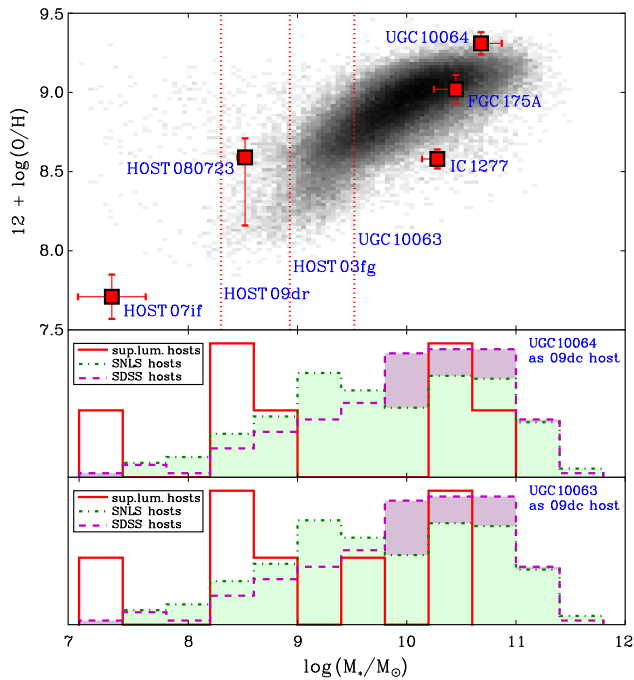}
   \caption{Top panel: masses and metallicities of the hosts of
     superluminous SNe~Ia, compared to SDSS galaxies. Whenever no
     spectroscopic metallicity was available, a vertical line was
     drawn. The mass reported for the SN 2009dr host is an upper
     limit, following the non-detection in SDSS images. Middle and
     bottom panels: binned mass distribution of superluminous SN~Ia
     hosts, compared to 162 SN~Ia hosts from SDSS and 231 SN~Ia hosts
     from SNLS. In the middle panel UGC~10064 is considered to be the
     host of SN~2009dc, in the lower panel UGC~10063. The
     distributions are scaled by arbitrary amounts to enable a
     comparison by eye.}
   \label{fig:MZ}
\end{figure}

In Fig.~\ref{fig:MZ} (top panel) we compare the derived masses and
metallicities of the hosts of superluminous SNe~Ia with those of SDSS
galaxies. For the hosts of SNe~2003fg and 2009dr, as well as for
UGC~10063, no metallicities could be determined, so that these
galaxies are included as vertical lines; for the SN~2009dr host the
upper mass limit is shown. There seems to be a tendency of the hosts
of superluminous SNe~Ia to have on average lower masses than the SDSS
galaxies. This trend also holds in the histograms in the lower two
panels, where the host-mass distibution of superluminous SNe~Ia is
compared to those of SNe~Ia from the non-targeted SDSS and SNLS
\citep{Sullivan10} surveys. In the middle panel UGC~10064 has been
assumed to be the host of SN~2009dc, in the bottom panel UGC~10063. 

The host-mass distribution of superluminous SNe~Ia has a mean and
dispersion of $\log(M_*/M_\odot) = 9.2 \pm 1.3$ or
$\log(M_*/M_\odot) = 9.0 \pm 1.1$, depending on whether UGC~10064
or UGC~10063 is considered as the host of SN~2009dc. These numbers are
conspicuously lower than the $\log(M_*/M_\odot) = 9.8 \pm 1.0$ and
$\log(M_*/M_\odot) = 10.0 \pm 0.9$ obtained for all SNLS and SDSS SN~Ia
hosts, respectively.

To verify whether the observed distributions differ to a statistically
significant degree, we ran a Kolmogorov-Smirnov test, using the SDSS
and SNLS host-mass distributions as a reference, and assuming in our
null hypothesis that the hosts of superluminous SNe~Ia have been drawn
from the same distributions. At a customary significance level of
$\alpha=0.05$, this null hypothesis is not rejected for both reference
distributions. If, however, the significance is relaxed to
$\alpha=0.10$, the null hypothesis is rejeceted for the SDSS reference
distribution (but not yet for the SNLS reference distribution). This
outcome is independent of which galaxy is adopted as the host of
SN~2009dc, since it is driven by the high frequency of low-mass dwarf
galaxies among the hosts of superluminous SNe~Ia. Of course, this is
all low-number ($n=7$) statistics, and one should note that the
addition of a single event might change the result
considerably. Nevertheless, we tentatively claim weak evidence for an
excess of low-mass galaxies as hosts of superluminous SNe~Ia.

\subsection{Concerns for cosmology}
\label{Cosmology}

The group of superluminous SNe~Ia deserves particular attention for
their potential role as troublemakers in SN cosmology.
Spectroscopically fairly similar to ordinary SNe~Ia (of the
core-normal or shallow-silicon variety; \citealt{Branch06a}), these
objects will be identified as SNe~Ia in surveys, especially at high
$z$ where the S/N of spectra is usually poor.

\begin{figure}
   \centering
   \includegraphics[width=8.4cm]{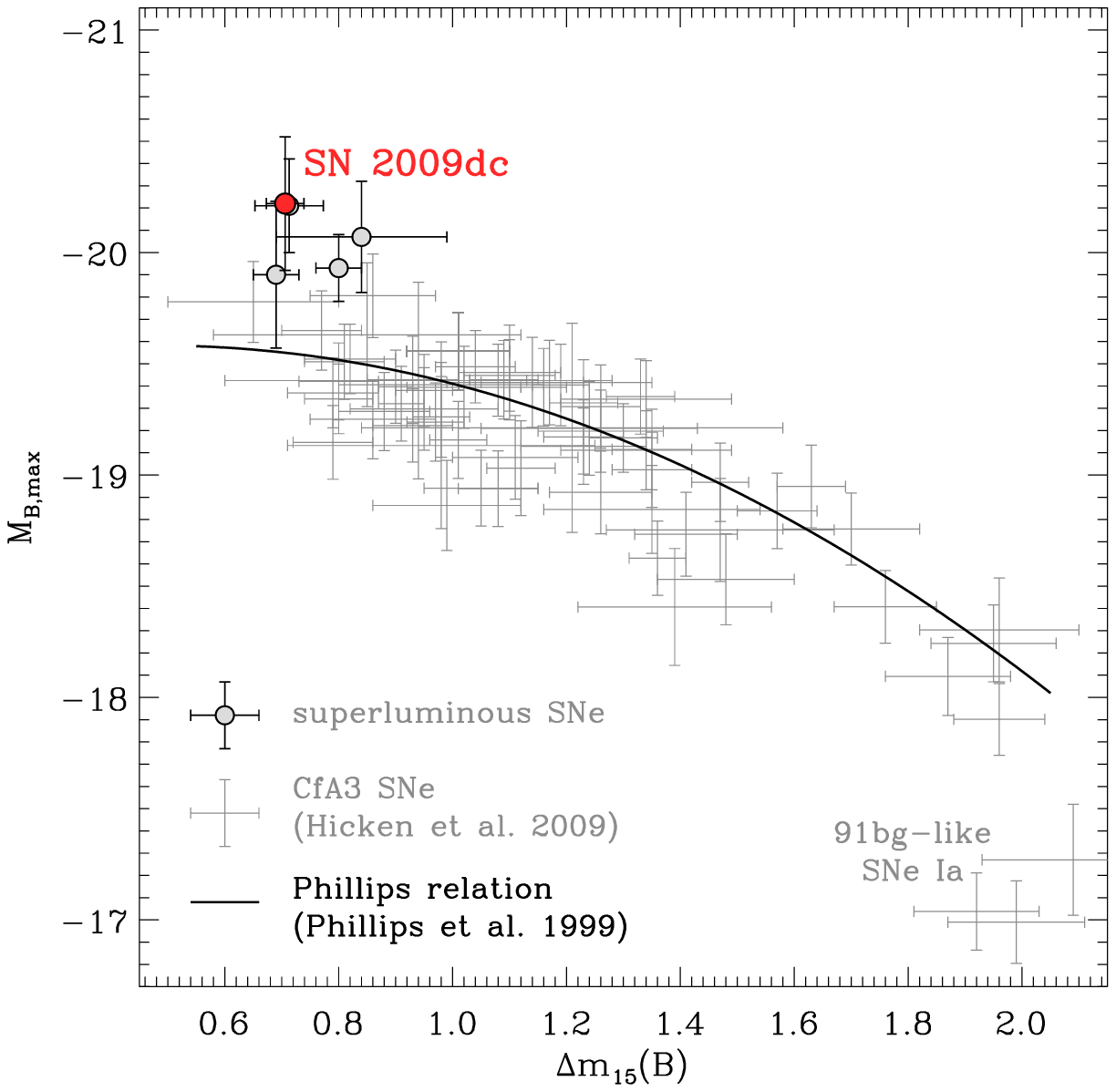}
   \caption{Absolute-magnitude [M$_{B,\mathrm{max}}$] vs. decline-rate
     [\dm15] diagram for SN~2009dc, other superluminous SNe~Ia and a
     subsample of the CfA3 SNe~Ia \citep{Hicken09}.  SN~2009dc is
     clearly more luminous than normal SNe~Ia with similar decline
     rates, and a factor $\sim$\,1.8 more luminous than predicted
     by the \citet{Phillips99} relation for the given \dm15. In this
     context, the Phillips relation is representative for all
     calibration methods used in present-day surveys; the basic result
     should not be sensitive to which light-curve fitter is actually
     employed.}
   \label{fig:Phillips}
\end{figure}

However, superluminous SNe~Ia are no good standardisable candles.  This
is demonstrated in Fig.~\ref{fig:Phillips}, where SN~2009dc is plotted
in an absolute-magnitude vs. decline-rate diagram. It is clearly
brighter than all other SNe~Ia (from the CfA3 data set;
\citealt{Hicken09}), and also $\sim$\,0.65 mag brighter than
anticipated by the \citet{Phillips99} relation for its \dm15\ of 0.69.

It is hence likely that in high-$z$ SN surveys SN~2009dc would be
regarded as less luminous than it actually is, and that too small a
distance would be inferred. At the same time, if the redshift was to
be determined from a cross-correlation with other SN~Ia spectra, it
would probably be slightly overestimated due to the lower expansion
velocities. In sum, these effects would move SN~2009dc towards the
lower-right in a Hubble diagram where the distance is plotted as a
function of redshift.

Having a small number of outliers in the Hubble diagram is by itself
not sufficient to really affect SN cosmology. Instead, an evolution of
the average SN~Ia properties with look-back time would be
required. However, such an evolutionary effect could exactly be
generated by superluminous SNe~Ia. First of all, while apparently rare
in the local Universe, their tentative tendency to be affiliated with
low-metallicity dwarf galaxies (c.f. Section~\ref{hosts}) might
suggest an intrinsically larger abundance at high $z$. Even more
importantly, a Malmquist bias is expected, making it more likely for
cosmological SN~Ia data sets to contain superluminous SNe~Ia at the
highest redshifts, where they are favoured by their sheer
luminosity. Systematic errors in the reconstruction of $H(z)$ using SN
Ia data may thus be introduced by superluminous SNe~Ia unless they are
recognised and treated correctly.

\section{Conclusions}
\label{Conclusions}

Optical and NIR observations of SN~2009dc have revealed an overall
similarity with SNe~Ia in light-curve shape and spectral evolution,
but also highlighted a number of important differences.  Most
evidently, SN~2009dc is by a factor $\geq 2$ more luminous than
ordinary SNe~Ia over the first $\sim$\,200\,d after the explosion
(M$_{V,\mathrm{max}} = -20.07\pm0.23$, $\log L_\mathrm{peak} =
43.47\pm0.11$), and is characterised by a slow decline from the
light-curve peak (\dm15$_\mathrm{true}=0.71\pm0.03$). The $IJHK'$
bands show secondary maxima, delayed by $\sim$\,25\,d with respect to
the peak in $B$. The minima between the first and secondary maxima,
however, are not very pronounced.  The spectra of SN~2009dc are
characterised by a blue pseudo-continuum with lines of IMEs, carbon
and oxygen early on, but dominated by Fe-group material at later
phases, when the photosphere recedes into \Nifs-rich zones. In particular,
the nebular spectra show mostly [\FeII] emission, indicative of a low
level of ionisation. At early epochs most spectral lines are
relatively shallow, and the ionisation is typical of normal to bright
SNe~Ia. The ejecta velocities are $\sim$\,30 per cent lower than in
typical LVG SNe~Ia.

A curiosity of SN~2009dc is an enhanced fading of its optical light
curves after $\sim$\,200\,d, at a phase when the decline in normal
SNe~Ia slows down. Lacking IR information, we cannot verify whether
this results from a re-distribution of radiation into the IR regime
(through dust formation or an IR catastrophe). Nevertheless, we are
inclined to believe this, since the only alternative seems to be CSM
interaction that contributes to the earlier light curve and suddenly
ends at $\sim$\,200\,d -- a scenario that generates more problems than
it solves (see below). The late-time behaviour of SN~2009dc reminds of
what was observed in SN~2006gz, and we speculate that both SNe shared
the same fate.

Using the analytic light-curve model of \citet{Arnett82}, the \Nifs\
mass synthesised in the explosion was estimated. The biggest
uncertainty in this exercise is the unknown rise time. For the minimum
possible rise time of 22\,d we obtained $M_\mathrm{Ni}^0 =
1.78\pm0.58$ \Msun, for a rise time of 28\,d $M_\mathrm{Ni}^0 =
2.21\pm0.75$ \Msun. Using the velocity information from spectra,
the model also allowed us to estimate a total ejected mass of
$\sim$\,2.8 \Msun, and a kinetic energy of $\sim$\,1.6 foe. These
numbers confirm that -- in the absence of CSM interaction -- the
progenitor of SN~2009dc cannot have been a \MCh-WD.

Finding an explosion scenario that explains SN~2009dc with all its
peculiarities turns out to be difficult. For all models based on WDs
an ejecta mass of 2.8 \Msun\ is either the very limit of what could
potentially be achieved (for a merger of two \MCh-WDs) or already
beyond (for a super-\MCh-WD grown by accretion from a non-degenerate
companion). Thermonuclear explosions in hydrogen-stripped stellar
cores (Type I$\frac{1}{2}$ SNe) might be an alternative, but detailed
numerical calculations which show whether this model may work have not
yet been performed. In particular, it is unclear whether H- and
He-free SNe can be achieved in this way, and whether the ejecta
velocities would match the observations. Core collapse of an
envelope-stripped progenitor, finally, might be another option. Large
ejecta masses are easy to achieve in this way, but the \Nifs- and
IME-rich composition of the SN~2009dc ejecta and the lack of nebular
\OIa\ $\lambda\lambda6300,6364$ emission disfavour this scenario.

A completely different approach is to assume that SN~2009dc was a
relatively normal (probably thermonuclear) explosion, and that the
observed peculiarities arise from interaction with a dense CSM. Such a
scenario might explain both the high luminosity and the low ejecta
velocities by conversion of kinetic energy into radiation, and the
light-curve fading after $\sim$\,200\,d by the end of the CSM
interaction. However, we have argued that it would require a lot of
fine-tuning to have a temporal evolution of the interaction strength
that mimics a radioactivity-driven light curve over more than half a
year. Moreover, an interaction strong enough to boost the SN
luminosity by a factor 2 or more should leave a direct spectroscopic
imprint in the form of narrow emission lines (which are not
observed). At least a H-rich CSM can thus be excluded.

Whatever mechanism produces objects like SN~2009dc, it should be
consistent with the finding that these events show a tendency to
explode in low-mass hosts (and probably low-metallicity
environments). While this trend for the entire group is fairly robust,
little can be said about the stellar population giving rise to
SN~2009dc in particular, since it is not clear whether the SN
`belongs' to the outskirts of an S0 galaxy (UGC~10064) or to a tidal
tail from an interacting less luminous, blue companion (UGC~10063).

A final warning is addressed to the high-$z$ SN~Ia surveys.
Superluminous SNe~Ia deviate systematically from the light-curve width
vs. luminosity relations used to standardise SNe~Ia for cosmological
distance determination. For this reason, care should be taken to
exclude superluminous SNe from the cosmological SN~Ia data sets, which
could be achieved e.g. by disregarding SNe with \dm15\ $\leq 0.9$
(Fig.~\ref{fig:Phillips}). This is even more important as their high
intrinsic luminosity and their association with (probably metal-poor)
low-mass host galaxies should result in a bias that favours their
detection at high redshift. If not accounted for, this could lead to
systematic errors in the reconstruction of the expansion history of
the Universe.

\section*{Acknowledgments}

This work is based on observations collected at the 2.2\,m Telescope
of the Centro Astron\'omico Hispano Alem\'an (Calar Alto, Spain), the
Italian 3.58\,m Telescopio Nazionale Galileo, the 2.56\,m Nordic
Optical Telescope and the 2.0\,m Liverpool Telescope (La Palma,
Spain), the 3.58\,m New Technology Telescope and 0.60\,m Rapid Eye
Mount (La Silla, Chile), the 1.82\,m Copernico Telescope on Cima Ekar
(Asiago, Italy), and the 2\,$\times$\,8.2\,m Large Binocular Telescope
(Arizona, United States).  The Telescopio Nazionale Galileo is
operated by the Fundaci\'on Galileo Galilei of the INAF (Instituto
Nazionale di Astrofisica) at the Spanish Observatorio del Roque de los
Muchachos of the Instituto de Astrofisica de Canarias.  ESO
observations have been performed under programmes 083.D-0728,
083.D-0970 and 184.D-1140.  We thank the support astronomers at the
Telescopio Nazionale Galileo, the 2.2\,m Telescope at Calar Alto, the
Nordic Optical Telescope and the Large Binocular Telescope for
performing the follow-up observations of SN~2009dc.

This research made use of the NASA/IPAC Extragalactic Database (NED),
operated by the Jet Propulsion Laboratory, California
Institute of Technology, under contract with the National Aeronautics
and Space Administration; the Lyon-Meudon Extragalactic Database
(LEDA), supplied by the LEDA team at the Centre de Recherche
Astronomique de Lyon, Observatoire de Lyon; the Online Supernova
Spectrum Archive (SUSPECT), initiated and maintained at the Homer
L. Dodge Departent of Physics and Astronomy, University of Oklahoma;
and the SMOKA archive, operated by the Astronomy Data Center,
National Astronomical Observatory of Japan. Some data used in this
paper were obtained from the Sloan Digital Sky Survey (SDSS).  Funding
for the SDSS and SDSS-II has been provided by the Alfred P. Sloan
Foundation, the Participating Institutions, the National Science
Foundation, the U.S. Department of Energy, the National Aeronautics
and Space Administration, the Japanese Monbukagakusho, the Max Planck
Society, and the Higher Education Funding Council for England. The
SDSS Web Site is http:/$\!$/www.sdss.org/.  We also benefited greatly
from the information provided by the Bright Supernova web pages
(maintained by D. Bishop) as part of the Rochester Academy of Sciences
(http:/$\!$/www.RochesterAstronomy.org/snimages).

The authors are indebted to the referee, D. Branch, for his constructive 
comments. Our thanks go to F.~K. R\"opke, S.~A. Sim, I.~R. Seitenzahl,
A.~J. Ruiter, M. Fink, I. Maurer, K. Nomoto and K. Maeda for inspiring
discussions, to M. Fink and S. Ben\'itez Herrera for assistance with
observations, and to K. Maeda and K. Kawabata for images and spectra
of SN~2006gz obtained with the Subaru telescope.  ST acknowledges
support by the Transregional Collaborative Research Centre TRR 33 `The
Dark Universe' of the German Research Foundation (DFG). MC is
supported by the Director, Office of Science, Office of High Energy
Physics, of the U.S. Department of Energy under Contract
No. DE-AC02-05CH11231 and by a grant from the Gordon \& Betty Moore
Foundation. SB, FB, PAM and MT are partially supported by the
PRIN-INAF 2009 with the project `Supernovae Variety and
Nucleosynthesis Yields'. VS acknowledges financial support from
Funda\c{c}\~{a}o para a Ci\^{e}ncia e a Tecnologia under program
Ci\^{e}ncia 2008. This research has benefited from the European
supernova collaboration led by SB.

\addcontentsline{toc}{chapter}{Bibliography}
\markboth{Bibliography}{Bibliography}
\bibliographystyle{mn2e}

\begin{thebibliography}{}
%
\bibitem[\protect\citeauthoryear{Abazajian et~al.}{Abazajian et~al.}{2009}]{sdssdr7}
Abazajian K.\,N., et~al., 2009, ApJS, 182, 543
%
\bibitem[\protect\citeauthoryear{Arnett}{Arnett}{1982}]{Arnett82}
Arnett W.\,D., 1982, ApJ, 253, 785
%
\bibitem[\protect\citeauthoryear{Axelrod}{Axelrod}{1980}]{Axelrod80}
Axelrod T.\,S., 1980, PhD thesis, Univ. California
%
\bibitem[\protect\citeauthoryear{Baldwin, Phillips \& Terlevich}{Baldwin et~al.}{1981}]{bpt81}
Baldwin J.\,A., Phillips M.\,M., Terlevich R., 1981, PASP, 93, 5 
%
\bibitem[\protect\citeauthoryear{Benetti et~al.}{Benetti et~al.}{2004}]{Benetti04}
Benetti S., et~al., 2004, MNRAS, 348, 261
%
\bibitem[\protect\citeauthoryear{Benetti et~al.}{Benetti et~al.}{2005}]{Benetti05}
Benetti S., et~al., 2005, ApJ, 623, 1011
%
\bibitem[\protect\citeauthoryear{Bertin \& Arnouts}{Bertin \& Arnouts}{1996}]{sextractor}
Bertin E., Arnouts S., 1996, A\&AS, 117, 393
%
\bibitem[\protect\citeauthoryear{Bessell}{Bessell}{1990}]{Bessell90}
Bessell M.\,S., 1990, PASP, 102, 1181
%
\bibitem[\protect\citeauthoryear{Blanton et~al.}{Blanton et~al.}{2003}]{blanton03}
Blanton M.\,R., et~al., 2003, ApJ, 592, 819
%
\bibitem[\protect\citeauthoryear{Bowers et~al.}{Bowers et~al.}{1997}]{Bowers97}
Bowers E.\,J.\,C., Meikle W.\,P.\,S., Geballe T.\,R., Walton N.\,A., Pinto P.\,A., Dhillon V.\,S., 
Howell S.\,B., Harrop-Allin M.\,K., 1997, MNRAS, 290, 663
%
\bibitem[\protect\citeauthoryear{Branch}{Branch}{2006}]{Branch06}
Branch D., 2006, Nature, 443, 283
%
\bibitem[\protect\citeauthoryear{Branch et~al.}{Branch et~al.}{2006}]{Branch06a}
Branch D., et~al., 2006, PASP, 118, 560
%
\bibitem[\protect\citeauthoryear{Cardelli, Clayton \& Mathis}{Cardelli
  et~al.}{1989}]{Cardelli89}
Cardelli J.\,A., Clayton G.\,C., Mathis J.\,S., 1989, ApJ, 345, 245
%
\bibitem[\protect\citeauthoryear{Childress et~al.}{Childress et~al.}{2010}]{Childress10}
Childress M., et~al., 2010, ApJ submitted
%
\bibitem[\protect\citeauthoryear{Conley et~al.}{Conley et~al.}{2006}]{Conley06}
Conley A., et~al., 2006, AJ, 132, 1707
%
\bibitem[\protect\citeauthoryear{Contardo, Leibundgut \& Vacca}{Contardo et~al.}{2000}]{Contardo00}
Contardo G., Leibundgut B., Vacca W.\,D., 2000, A\&A, 359, 876
%
\bibitem[\protect\citeauthoryear{Contreras et~al.}{Contreras et~al.}{2010}]{Contreras10}
Contreras C., et~al., AJ, 139, 519
%
\bibitem[\protect\citeauthoryear{Couch \& Arnett}{Couch \& Arnett}{1975}]{Couch75}
Couch R.\,G., Arnett W.\,D., 1975, ApJ, 196, 791
%
\bibitem[\protect\citeauthoryear{Delahaye et~al.}{Delahaye et~al.}{2010}]{delahaye10}
Delahaye F., Pinsonneault M.\,H., Pinsonneault L., Zeippen C.\,J., 2010,
APJL submitted (arXiv:1005.0423)
%
\bibitem[\protect\citeauthoryear{Emsellem et~al.,}{Emsellem et~al.}{2007}]{Emsellem07}
Emsellem E., et~al., 2007, MNRAS, 379, 401
%
\bibitem[\protect\citeauthoryear{Ergma \& Paczy\'nski}{Ergma \& Paczy\'nski}{1974}]{Ergma74}
Ergma E., Paczy\'nski B., 1974, Acta Astronomica, 24, 1
%
\bibitem[\protect\citeauthoryear{Fassia et~al.}{Fassia et~al.}{2001}]{Fassia01}
Fassia A., et~al., 2001, MNRAS, 325, 907
%
\bibitem[\protect\citeauthoryear{Filippenko}{Filippenko}{1982}]{Filippenko82}
Filippenko A.\,V., 1982, PASP, 94, 715
%
\bibitem[\protect\citeauthoryear{Filippenko et~al.}{Filippenko et~al.}{1992}]{Filippenko92}
Filippenko A.\,V., et~al., 1992, ApJ, 384, L15
%
\bibitem[\protect\citeauthoryear{Fixsen et~al.}{Fixsen et~al.}{1996}]{Fixsen96}
Fixsen D.\,J., Cheng E.\,S., Gales J.\,M., Mather J.\,C., Shafer R.\,A., Wright E.\,L., 1996, ApJ, 473, 576
%
\bibitem[\protect\citeauthoryear{Folatelli et~al.}{Folatelli et~al.}{2006}]{Folatelli06}
Folatelli G., et~al., 2006, ApJ, 641, 1039
%
\bibitem[\protect\citeauthoryear{Galama et~al.}{Galama et~al.}{1998}]{Galama98}
Galama T.\,J., et~al., 1998, Nature, 395, 670
%
\bibitem[\protect\citeauthoryear{Garnavich et~al.}{Garnavich et~al.}{2004}]{Garnavich04}
Garnavich P.\,M., et~al., 2004, ApJ, 613, 1120
%
\bibitem[\protect\citeauthoryear{G\'omez, L\'opez \& S\'anchez}{G\'omez et~al.}{1996}]{Gomez96}
G\'omez G., L\'opez R., S\'anchez F., 1996, AJ, 112, 2094
%
\bibitem[\protect\citeauthoryear{Hamuy et~al.}{Hamuy et~al.}{1992}]{Hamuy92}
Hamuy M., Walker A.\,R., Suntzeff N.\,B., Gigoux P., Heathcote S.\,R., Phillips M.\,M., 1992, PASP, 104, 533
%
\bibitem[\protect\citeauthoryear{Hamuy et~al.}{Hamuy et~al.}{1994}]{Hamuy94}
Hamuy M., Suntzeff N.\,B., Heathcote S.\,R., Walker A.\,R., Gigoux P., Phillips M.\,M., 1994, PASP, 106, 566
%
\bibitem[\protect\citeauthoryear{Hamuy et~al.}{Hamuy et~al.}{1996}]{Hamuy96}
Hamuy M., Phillips M.\,M., Suntzeff N.\,B., Schommer R.\,A., Maza J., Smith R.\,C., Lira P., Avil\'es R., 1996, 
AJ, 112, 2438
%
\bibitem[\protect\citeauthoryear{Hamuy et~al.}{Hamuy et~al.}{2002}]{Hamuy02}
Hamuy M., et~al., 2002, AJ, 124, 417
%
\bibitem[\protect\citeauthoryear{Hamuy et~al.}{Hamuy et~al.}{2003}]{Hamuy03}
Hamuy M., et~al., 2003, Nature, 424, 651
%
\bibitem[\protect\citeauthoryear{Harutyunyan, Elias-Rosa \& Benetti}{Harutyunyan et~al.}{2009}]{CBET1768}
Harutyunyan A., Elias-Rosa N., Benetti S., 2009, CBET, 1768
%
\bibitem[\protect\citeauthoryear{Hayden et~al.}{Hayden et~al.}{2010}]{Hayden10}
Hayden B.\,T., et~al., 2010, ApJ, 712, 350
%
\bibitem[\protect\citeauthoryear{Hicken et~al.}{Hicken et~al.}{2007}]{Hicken07}
Hicken M., Garnavich P.\,M., Prieto J.\,L., Blondin S., DePoy D.\,L., Kirshner R.\,P., Parrent J., 2007, ApJ, 669, L17
%
\bibitem[\protect\citeauthoryear{Hicken et~al.}{Hicken et~al.}{2009}]{Hicken09}
Hicken M., et~al., 2009, ApJ, 700, 331
%
\bibitem[\protect\citeauthoryear{Hillebrandt \& Niemeyer}{Hillebrandt \& Niemeyer}{2000}]{Hillebrandt00}
Hillebrandt W., Niemeyer J., 2000, ARA\&A, 38, 191
%
\bibitem[\protect\citeauthoryear{Hillebrandt, Sim \& R\"opke}{Hillebrandt et~al.}{2007}]{Hillebrandt07}
Hillebrandt W., Sim S.\,A., R\"opke F.\,K., 2007, A\&A, 465, L17
%
\bibitem[\protect\citeauthoryear{Hjorth et~al.}{Hjorth et~al.}{2003}]{Hjorth03}
Hjorth J., et~al., 2003, Nature, 423, 847
%
\bibitem[\protect\citeauthoryear{H\"oflich \& Khokhlov}{H\"oflich \& Khokhlov}{1996}]{Hoeflich96}
H\"oflich P., Khokhlov A., 1996, ApJ, 457, 500
%
\bibitem[\protect\citeauthoryear{Horne}{Horne}{1986}]{Horne86}
Horne K., 1986, PASP, 98, 609
%
\bibitem[\protect\citeauthoryear{Howell et~al.}{Howell et~al.}{2006}]{Howell06}
Howell D.\,A., et~al., 2006, Nature, 443, 308
%
\bibitem[\protect\citeauthoryear{Hunt et~al.}{Hunt et~al.}{1998}]{Hunt98}
Hunt L.\,K., Mannucci F., Testi L., Migliorini S., Stanga R.\,M., Baffa C., Lisi F., Vanzi L., 1998, AJ, 115, 2594
%
\bibitem[\protect\citeauthoryear{Hunter et~al.}{Hunter et~al.}{2009}]{Hunter09}
Hunter D., et~al., 2009, A\&A, 508, 371
%
\bibitem[\protect\citeauthoryear{Iben}{Iben}{1982}]{Iben82}
Iben I. Jr., 1982, ApJ, 253, 248
%
\bibitem[\protect\citeauthoryear{Iben \& Renzini}{Iben \& Renzini}{1983}]{Iben83}
Iben I. Jr., Renzini A., 1983, ARAA, 21, 271
%
\bibitem[\protect\citeauthoryear{Kasen}{Kasen}{2006}]{Kasen06}
Kasen D., 2006, ApJ, 649, 939
%
\bibitem[\protect\citeauthoryear{Kawabata et~al.}{Kawabata et~al.}{2009}]{Kawabata09}
Kawabata K.\,S., Maeda K., Nomoto K., Taubenberger S., Tanaka M., Hattori T., Itagaki K., 2010, Nature, 465, 326
%
\bibitem[\protect\citeauthoryear{Kewley \& Ellison}{Kewley \& Ellison}{2008}]{ke08}
Kewley, L.\,J., Ellison S.\,L., 2008, ApJ, 681, 1183
%
\bibitem[\protect\citeauthoryear{Kewley et~al.}{Kewley et~al.}{2006}]{kewley06}
Kewley L.\,J., Groves B., Kauffmann G., Heckman T., 2006, MNRAS, 372, 961
%
\bibitem[\protect\citeauthoryear{Koester, Schulz \& Weidemann}{Koester et~al.}{1979}]{Koester79}
Koester D., Schulz H., Weidemann V., 1979, A\&A, 76, 262
%
\bibitem[\protect\citeauthoryear{Krisciunas et~al.}{Krisciunas et~al.}{2003}]{Krisciunas03}
Krisciunas K., et~al., 2003, AJ, 125, 166
%
\bibitem[\protect\citeauthoryear{Krisciunas, Phillips \& Suntzeff}{Krisciunas et~al.}{2004}]{Krisciunas04}
Krisciunas K., Phillips M.\,M., Suntzeff N.\,B., 2004, ApJ, 602, L81
%
\bibitem[\protect\citeauthoryear{Krisciunas, Phillips \& Suntzeff}{Krisciunas et~al.}{2007}]{Krisciunas07}
Krisciunas K., et~al., 2007, AJ, 133, 58
%
\bibitem[\protect\citeauthoryear{Kromer \& Sim}{Kromer \& Sim}{2009}]{Kromer09}
Kromer M., Sim S.\,A., 2009, MNRAS, 398, 1809
%
\bibitem[\protect\citeauthoryear{Landolt}{Landolt}{1992}]{Landolt92}
Landolt A.\,U., 1992, AJ, 104, 340
%
\bibitem[\protect\citeauthoryear{Langer et~al.}{Langer et~al.}{2000}]{Langer00}
Langer N., Deutschmann A., Wellstein S., H\"oflich P., 2000, A\&A,
362, 1046
%
\bibitem[\protect\citeauthoryear{Le Borgne \& Rocca-Volmerange}{Le Borgne \& Rocca-Volmerange}{2002}]{zpeg}
Le\,Borgne D., Rocca-Volmerange B., 2002, A\&A, 386, 446
%
\bibitem[\protect\citeauthoryear{Leibundgut}{Leibundgut}{2001}]{Leibundgut01}
Leibundgut B., 2001, ARA\&A, 39, 67
%
\bibitem[\protect\citeauthoryear{Leloudas et~al.}{Leloudas et~al.}{2009}]{Leloudas09}
Leloudas G., et~al., 2009, A\&A, 505, 265
%
\bibitem[\protect\citeauthoryear{Li et~al.}{Li et~al.}{2003}]{Li03}
Li W., et~al., 2003, PASP, 115, 453
%
\bibitem[\protect\citeauthoryear{Liebert, Bergeron \& Holberg}{Liebert et~al.}{2005}]{Liebert05}
Liebert J., Bergeron P., Holberg J.\,B., 2005, ApJ, 156, 47
%
\bibitem[\protect\citeauthoryear{Lira}{Lira}{1995}]{Lira95}
Lira P., 1995, Master thesis, Univ. Chile
%
\bibitem[\protect\citeauthoryear{Maeda, Mazzali \& Nomoto}{Maeda et~al.}{2006}]{Maeda06}
Maeda K., Mazzali P.\,A., Nomoto K., 2006, ApJ, 645, 1331
%
\bibitem[\protect\citeauthoryear{Maeda et~al.}{Maeda et~al.}{2007}]{Maeda07}
Maeda K., et~al., 2007, ApJ, 666, 1069
%
\bibitem[\protect\citeauthoryear{Maeda et~al.}{Maeda et~al.}{2009}]{Maeda09}
Maeda K., Kawabata K., Li W., Tanaka M., Mazzali P.\,A., Hattori T.,
Nomoto K., Filippenko A.\,V., 2009, ApJ, 690, 1745
%
\bibitem[\protect\citeauthoryear{Marion, Garnavich \& Challis}{Marion et~al.}{2009}]{CBET1776}
Marion H., Garnavich P., Challis P., 2009, CBET, 1776
%
\bibitem[\protect\citeauthoryear{Marion et~al.}{Marion et~al.}{2009a}]{Marion09}
Marion G.\,H., H\"oflich P., Gerardy C.\,L., Vacca W.\,D., Wheeler J.\,C.,
Robinson E.\,L., 2009a, AJ, 138, 727
%
\bibitem[\protect\citeauthoryear{Mazzali, Danziger \& Turatto}{Mazzali et~al.}{1995}]{Mazzali95}
Mazzali P.\,A., Danziger I.\,J., Turatto M., 1995, A\&A, 297, 509
%
\bibitem[\protect\citeauthoryear{Mazzali et~al.}{Mazzali et~al.}{1997}]{Mazzali97}
Mazzali P.\,A., Chugai N., Turatto M., Lucy L.\,B., Danziger I.\,J.,
Cappellaro E., della Valle M., Benetti S., 1997, MNRAS, 284, 151
%
\bibitem[\protect\citeauthoryear{Mazzali et~al.}{Mazzali et~al.}{2001}]{Mazzali01}
Mazzali P.\,A., Nomoto K., Cappellaro E., Nakamura T., Umeda H.,
Iwamoto K., 2001, ApJ, 547, 988
%
\bibitem[\protect\citeauthoryear{Mazzali et~al.}{Mazzali et~al.}{2005}]{Mazzali05}
Mazzali P.\,A., et~al., 2005, ApJ, 623, L37
%
\bibitem[\protect\citeauthoryear{Mazzali, R\"opke, Benetti \& Hillebrandt}{Mazzali et~al.}{2007}]{Mazzali07}
Mazzali P.\,A., R\"opke F.\,K., Benetti S., Hillebrandt W., 2007, Sci, 315, 825
%
\bibitem[\protect\citeauthoryear{McKenzie \& Schaefer}{McKenzie \& Schaefer}{1999}]{McKenzie99}
McKenzie E.\,H., Schaefer B.\,E., 1999, PASP, 111, 964
%
\bibitem[\protect\citeauthoryear{Miyaji et~al.}{Miyaji et~al.}{1980}]{Miyaji80}
Miyaji S., Nomoto K., Yokoi K., Sugimoto D., 1980,
Pub. Astr. Soc. Japan, 32, 303
%
\bibitem[\protect\citeauthoryear{Moriya et~al.}{Moriya}et~al.{2010}]{Moriya10}
Moriya T., Tominaga N., Tanaka M., Maeda K., Nomoto K., 2010, ApJ, 717, L83
%
\bibitem[\protect\citeauthoryear{Morrissey et~al.}{Morrissey et~al.}{2007}]{galex}
Morrissey P., et~al., 2007, ApJS, 173, 682
%
\bibitem[\protect\citeauthoryear{Mould et~al.}{Mould et~al.}{2000}]{Mould00}
Mould J., et~al., 2000, ApJ, 529, 786
%
\bibitem[\protect\citeauthoryear{Nomoto, Filippenko \& Shigeyama}{Nomoto et~al.}{1990}]{Nomoto90}
Nomoto K.., Filippenko A.\,V., Shigeyama T., 1990, A\&A, 240, L1
%
\bibitem[\protect\citeauthoryear{Oke}{Oke}{1990}]{Oke90}
Oke J.\,B., 1990, AJ, 99, 1621
%
\bibitem[\protect\citeauthoryear{Pakmor et~al.}{Pakmor et~al.}{2010}]{Pakmor10}
Pakmor R., Kromer M., R\"opke F.\,K., Sim S.\,A., Ruiter A.\,J.,
Hillebrandt W., 2010, Nature, 463, 61
%
\bibitem[\protect\citeauthoryear{Pastorello et~al.}{Pastorello et~al.}{2007a}]{Pastorello07a}
Pastorello A., et~al., 2007a, MNRAS, 376, 1301
%
\bibitem[\protect\citeauthoryear{Pastorello et~al.}{Pastorello et~al.}{2007b}]{Pastorello07}
Pastorello A., et~al., 2007b, MNRAS, 377, 1531
%
\bibitem[\protect\citeauthoryear{Patat et~al.}{Patat et~al.}{2001}]{Patat01}
Patat F., et~al., 2001, ApJ, 555, 900
%
\bibitem[\protect\citeauthoryear{Pettini \& Pagel}{Pettini \& Pagel}{2004}]{pp04}
Pettini M., Pagel B.\,E.\,J., 2004, MNRAS, 348, L59
%
\bibitem[\protect\citeauthoryear{Pfannes, Niemeyer \& Schmidt}{Pfannes et~al.}{2010}]{Pfannes10}
Pfannes J.\,M.\,M., Niemeyer J.\,C., Schmidt W., 2010, A\&A, 509, A75
%
\bibitem[\protect\citeauthoryear{Phillips, Lira, Suntzeff, Schommer, Hamuy \&
  Maza}{Phillips et~al.}{1999}]{Phillips99}
Phillips M.\,M.,  Lira P.,  Suntzeff N.\,B.,  Schommer R.\,A.,  Hamuy M.,    Maza
  J.,  1999, AJ, 118, 1766
%
\bibitem[\protect\citeauthoryear{Phillips et~al.}{Phillips et~al.}{2007}]{Phillips07}
Phillips M.\,M., et~al., 2007, PASP, 119, 360
%
\bibitem[\protect\citeauthoryear{Pignata et~al.,}{Pignata et~al.}{2004}]{Pignata04b}
Pignata G., et~al., 2004, MNRAS, 355, 178
%
\bibitem[\protect\citeauthoryear{Pinto \& Eastman}{Pinto \& Eastman}{2001}]{Pinto01}
Pinto P.\,A., Eastman R.\,G., 2001, NewA, 6, 307
%
\bibitem[\protect\citeauthoryear{Puckett, Moore \& Newton}{Puckett et~al.}{2009}]{CBET1762}
Puckett T., Moore R., Newton J., 2009, CBET, 1762
%
\bibitem[\protect\citeauthoryear{Quimby et~al.}{Quimby et~al.}{2009}]{sn2009dr}
Quimby R., Kasliwal M.\,M., Nugent P., Howell D.\,A., Rau A., Bhalerao
V., 2009, CBET, 1783
%
\bibitem[\protect\citeauthoryear{Riess et~al.}{Riess et~al.}{1999}]{Riess99}
Riess A.\,G., et~al., 1999, AJ, 118, 2675
%
\bibitem[\protect\citeauthoryear{Riess et~al.}{Riess et~al.}{2007}]{Riess07}
Riess A.\,G., et~al., 2007, ApJ, 659, 98
%
\bibitem[\protect\citeauthoryear{Ruiz-Lapuente et~al.}{Ruiz-Lapuente et~al.}{1992}]{RuizLapuente92}
Ruiz-Lapuente P., Cappellaro E., Turatto M., Gouiffes C., Danziger I.\,J., Della Valle M., Lucy L.\,B., 1992, ApJ, 387, L33
%
\bibitem[\protect\citeauthoryear{Sahu et~al.}{Sahu et~al.}{2008}]{Sahu08}
Sahu D.\,K., et~al., 2008, ApJ, 680, 580
%
\bibitem[\protect\citeauthoryear{Saio \& Nomoto}{Saio \& Nomoto}{1985}]{Saio85}
Saio H., Nomoto K., 1985, A\&A, 150, L21
%
\bibitem[\protect\citeauthoryear{Scalzo et~al.}{Scalzo et~al.}{2010}]{Scalzo10}
Scalzo R.\,A., et~al., 2010, ApJ, 713, 1073
%
\bibitem[\protect\citeauthoryear{Schlegel, Finkbeiner \& Davis}{Schlegel et~al.}{1998}]{Schlegel98}
Schlegel D.\,J., Finkbeiner D.\,P., Davis M., 1998, ApJ, 500, 525
%
\bibitem[\protect\citeauthoryear{Sim et~al.}{Sim et~al.}{2007}]{Sim07}
Sim S.\,A., Sauer D.\,N., R\"opke F.\,K., Hillebrandt W., 2007, MNRAS, 378, 2
%
\bibitem[\protect\citeauthoryear{Silverman et~al.}{Silverman et~al.}{2011}]{Silverman10}
Silverman J.\,M., Ganeshalingam M., Li W., Filippenko A.\,V., Miller A.\,A., Poznanski D., 2011, 
MNRAS, 410, 585
%
\bibitem[\protect\citeauthoryear{Skrutskie et~al.}{Skrutskie et~al.}{2006}]{twomass}
Skrutskie M.\,F., et~al., 2006, AJ, 131, 1163
%
\bibitem[\protect\citeauthoryear{Sollerman et~al.}{Sollerman et~al.}{2000}]{Sollerman00}
Sollerman J., Kozma C., Fransson C., Leibundgut B., Lundqvist P., Ryde
F., Woudt P., 2000, ApJ, 537, L127
%
\bibitem[\protect\citeauthoryear{Spruit}{Spruit}{1998}]{Spruit98}
Spruit H., 1998, A\&A, 333, 603
%
\bibitem[\protect\citeauthoryear{Spyromilio et~al.}{Spyromilio et~al.}{2004}]{Spyromilio04}
Spyromilio J., Gilmozzi R., Sollerman J., Leibundgut B., Fransson C., Cuby J.-G., 2004, A\&A, 426, 547
%
\bibitem[\protect\citeauthoryear{Stanishev}{Stanishev}{2007}]{Stanishev07}
Stanishev V., 2007, AN, 328, 948
%
\bibitem[\protect\citeauthoryear{Stanishev et~al.}{Stanishev et~al.}{2007a}]{Stanishev07a}
Stanishev V., et~al., 2007a, A\&A, 469, 645
%
\bibitem[\protect\citeauthoryear{Stanishev et~al.}{Stanishev et~al.}{2007b}]{Stanishev07b}
Stanishev V., et~al., 2007b, AIP Conference Proceedings, 924, 336
%
\bibitem[\protect\citeauthoryear{Stoughton et~al.}{Stoughton et~al.}{2002}]{stoughton02}
Stoughton C., et~al., 2002, AJ, 123, 485
%
\bibitem[\protect\citeauthoryear{Stritzinger et~al.,}{Stritzinger et~al.}{2002}]{Stritzinger02}
Stritzinger M., et~al., 2002, AJ, 124, 2100
%
\bibitem[\protect\citeauthoryear{Strovink}{Strovink}{2007}]{Strovink07}
Strovink M., 2007, ApJ, 671, 1084
%
\bibitem[\protect\citeauthoryear{Sullivan et~al.}{Sullivan et~al.}{2010}]{Sullivan10}
Sullivan M., et~al., 2010, MNRAS, 406, 782
%
\bibitem[\protect\citeauthoryear{Tanaka et~al.}{Tanaka et~al.}{2008}]{Tanaka08}
Tanaka M., et~al., 2008, ApJ, 677, 448
%
\bibitem[\protect\citeauthoryear{Tanaka et~al.}{Tanaka et~al.}{2010}]{Tanaka09}
Tanaka M., et~al., 2010, ApJ, 714, 1209
%
\bibitem[\protect\citeauthoryear{Taubenberger et~al.}{Taubenberger et~al.}{2008}]{Taubenberger08}
Taubenberger S., et~al., 2008, MNRAS, 385, 75
%
\bibitem[\protect\citeauthoryear{Thomas et~al.}{Thomas et~al.}{2007}]{Thomas07}
Thomas R.\,C., et~al., 2007, ApJ, 654, L53
%
\bibitem[\protect\citeauthoryear{Tominaga et~al.}{Tominaga et~al.}{2005}]{Tominaga05}
Tominaga N., et~al., 2005, ApJ, 633, L97
%
\bibitem[\protect\citeauthoryear{Tremonti et~al.}{Tremonti et~al.}{2004}]{Tremonti04}
Tremonti C.\,A., et~al., 2004, ApJ, 613, 898
%
\bibitem[\protect\citeauthoryear{Turatto et~al.}{Turatto et~al.}{1996}]{Turatto96}
Turatto M., Benetti S., Cappellaro E., Danziger I.\,J., Della Valle M., Gouiffes C., 
Mazzali P.\,A., Patat F., 1996, MNRAS, 283, 1
%
\bibitem[\protect\citeauthoryear{Turatto, Benetti \& Cappellaro}{Turatto et~al.}{2003}]{Turatto03}
Turatto M., Benetti S., Cappellaro E., 2003, in Hillebrandt W., Leibundgut B., eds, Proceedings to the 
ESO/MPA/MPE Workshop ``From Twilight to Highlight: The Physics of Supernovae''. Springer, Berlin, p. 200
%
\bibitem[\protect\citeauthoryear{Umeda \& Nomoto}{Umeda \& Nomoto}{2008}]{Umeda08}
Umeda H., Nomoto K., 2008, ApJ, 673, 1014
%
\bibitem[\protect\citeauthoryear{Valenti et~al.}{Valenti et~al.}{2008}]{Valenti08}
Valenti S., et~al., 2008, ApJ, 673, L155
%
\bibitem[\protect\citeauthoryear{Wang et~al.}{Wang et~al}{2009b}]{Wang09}
Wang X., et~al., 2009b, ApJ, 697, 380
%
\bibitem[\protect\citeauthoryear{Wang et~al.}{Wang et~al.}{2009a}]{Wang09a}
Wang X., et~al., 2009a, ApJ, 699, L139
%
\bibitem[\protect\citeauthoryear{Wegner \& Grogin}{Wegner \& Grogin}{2008}]{wegrog08}
Wegner G., Grogin N., 2008, A\&A, 136, 1
%
\bibitem[\protect\citeauthoryear{Weidemann \& Koester}{Weidemann \& Koester}{1983}]{Weidemann83}
Weidemann V., Koester D., 1983, A\&A, 121, 77
%
\bibitem[\protect\citeauthoryear{Wood-Vasey et~al.}{Wood-Vasey et~al.}{2008}]{WoodVasey08}
Wood-Vasey W.\,M., et~al., 2008, ApJ, 689, 377
%
\bibitem[\protect\citeauthoryear{Yamanaka et~al.}{Yamanaka et~al.}{2009}]{Yamanaka09}
Yamanaka M., et~al., 2009, ApJ, 707, L118
%
\bibitem[\protect\citeauthoryear{Yoon \& Langer}{Yoon \& Langer}{2005}]{Yoon05}
Yoon S.-C., Langer N., 2005, A\&A, 435, 967

\end{thebibliography}

\appendix

\section{Table of S- and K-corrections}

\begin{table*}
\caption{$S$- and $K$-correction added to the zero-point corrected SN
  magnitudes (\textit{instead of} colour-term corrections).}
\label{S-corr}
\begin{center}
\begin{scriptsize}
\begin{tabular}{rrrrrrrrrrrl}
\hline
Epoch$^a \!\!\!\!\!$ & $S_U$\quad\ & $S_B$\quad\ & $S_V$\quad\ & $S_R$\quad\ & $S_I$\quad\ & $K_U$\quad\ & $K_B$\quad\ & $K_V$\quad\ & $K_R$\quad\ & $K_I$\quad\ & Tel.$^b$\\
\hline
 $-8.5$\ \ &  $ 0.001\ \, $ & $ 0.003\ \, $ & $-0.012\ \, $ & $ 0.009\ \, $ & $-0.004\ \, $  &  $-0.021\ \, $ & $ 0.019\ \, $ & $ 0.026\ \, $ & $ 0.048\ \, $ & $ 0.057\ \, $  & NOT  \\
 $-7.5$\ \ &  $ 0.002\ \, $ & $ 0.004\ \, $ & $-0.012\ \, $ & $ 0.011\ \, $ & $-0.005\ \, $  &  $-0.023\ \, $ & $ 0.019\ \, $ & $ 0.026\ \, $ & $ 0.050\ \, $ & $ 0.056\ \, $  & NOT  \\
 $-4.6$\ \ &  $ 0.178\ \, $ & $-0.002\ \, $ & $-0.011\ \, $ & $ 0.021\ \, $ & $-0.006\ \, $  &  $-0.016\ \, $ & $ 0.019\ \, $ & $ 0.027\ \, $ & $ 0.054\ \, $ & $ 0.052\ \, $  & LT   \\
 $-3.7$\ \ &  $-0.009\ \, $ & $ 0.006\ \, $ & $-0.010\ \, $ & $ 0.017\ \, $ & $-0.010\ \, $  &  $-0.014\ \, $ & $ 0.019\ \, $ & $ 0.028\ \, $ & $ 0.056\ \, $ & $ 0.050\ \, $  & NOT  \\
 $-2.6$\ \ &  $ 0.171\ \, $ & $-0.001\ \, $ & $-0.010\ \, $ & $ 0.023\ \, $ & $-0.008\ \, $  &  $-0.020\ \, $ & $ 0.017\ \, $ & $ 0.027\ \, $ & $ 0.055\ \, $ & $ 0.045\ \, $  & LT   \\
  $2.4$\ \ &  $-0.102\ \, $ & $ 0.000\ \, $ & $-0.008\ \, $ & $-0.002\ \, $ & $ 0.017\ \, $  &  $-0.048\ \, $ & $ 0.007\ \, $ & $ 0.025\ \, $ & $ 0.048\ \, $ & $ 0.021\ \, $  & CA   \\
  $2.4$\ \ &  $-0.073\ \, $ & $-0.003\ \, $ & $-0.019\ \, $ & $ 0.019\ \, $ & $-0.015\ \, $  &  $-0.048\ \, $ & $ 0.007\ \, $ & $ 0.025\ \, $ & $ 0.048\ \, $ & $ 0.021\ \, $  & TNG  \\
  $2.5$\ \ &  $ 0.148\ \, $ & $ 0.006\ \, $ & $-0.012\ \, $ & $ 0.020\ \, $ & $-0.021\ \, $  &  $-0.049\ \, $ & $ 0.007\ \, $ & $ 0.025\ \, $ & $ 0.048\ \, $ & $ 0.020\ \, $  & LT   \\
  $3.5$\ \ &  $ 0.147\ \, $ & $ 0.007\ \, $ & $-0.013\ \, $ & $ 0.020\ \, $ & $-0.023\ \, $  &  $-0.055\ \, $ & $ 0.005\ \, $ & $ 0.024\ \, $ & $ 0.047\ \, $ & $ 0.017\ \, $  & LT   \\
  $4.4$\ \ &                & $ 0.006\ \, $ & $-0.010\ \, $ & $-0.003\ \, $ & $ 0.027\ \, $  &                & $ 0.004\ \, $ & $ 0.023\ \, $ & $ 0.046\ \, $ & $ 0.015\ \, $  & CA   \\
  $6.5$\ \ &  $-0.032\ \, $ & $ 0.000\ \, $ & $-0.021\ \, $ & $ 0.019\ \, $ & $ 0.002\ \, $  &  $-0.063\ \, $ & $ 0.004\ \, $ & $ 0.023\ \, $ & $ 0.047\ \, $ & $ 0.010\ \, $  & TNG  \\
  $7.4$\ \ &  $ 0.124\ \, $ & $ 0.011\ \, $ & $-0.016\ \, $ & $ 0.016\ \, $ & $-0.031\ \, $  &  $-0.078\ \, $ & $ 0.003\ \, $ & $ 0.019\ \, $ & $ 0.045\ \, $ & $ 0.007\ \, $  & LT   \\
  $8.4$\ \ &  $-0.026\ \, $ & $ 0.007\ \, $ & $-0.016\ \, $ & $ 0.001\ \, $ & $ 0.047\ \, $  &  $-0.093\ \, $ & $ 0.001\ \, $ & $ 0.015\ \, $ & $ 0.044\ \, $ & $ 0.004\ \, $  & CA   \\
  $9.3$\ \ &  $ 0.094\ \, $ & $ 0.016\ \, $ & $-0.020\ \, $ & $ 0.020\ \, $ & $-0.035\ \, $  &  $-0.108\ \, $ & $-0.007\ \, $ & $ 0.013\ \, $ & $ 0.042\ \, $ & $-0.001\ \, $  & LT   \\
  $9.3$\ \ &  $-0.003\ \, $ & $ 0.015\ \, $ & $-0.020\ \, $ & $ 0.012\ \, $ & $-0.037\ \, $  &  $-0.108\ \, $ & $-0.007\ \, $ & $ 0.013\ \, $ & $ 0.042\ \, $ & $-0.001\ \, $  & NOT  \\
 $11.5$\ \ &  $-0.002\ \, $ & $ 0.015\ \, $ & $-0.023\ \, $ & $ 0.009\ \, $ & $-0.036\ \, $  &  $-0.129\ \, $ & $-0.015\ \, $ & $ 0.007\ \, $ & $ 0.040\ \, $ & $-0.012\ \, $  & NOT  \\
 $12.5$\ \ &  $ 0.067\ \, $ & $ 0.019\ \, $ & $-0.024\ \, $ & $ 0.021\ \, $ & $-0.049\ \, $  &  $-0.131\ \, $ & $-0.019\ \, $ & $ 0.006\ \, $ & $ 0.039\ \, $ & $-0.020\ \, $  & LT   \\
 $12.6$\ \ &  $-0.012\ \, $ & $ 0.015\ \, $ & $-0.023\ \, $ & $ 0.007\ \, $ & $-0.042\ \, $  &  $-0.131\ \, $ & $-0.019\ \, $ & $ 0.006\ \, $ & $ 0.039\ \, $ & $-0.020\ \, $  & NOT  \\
 $16.4$\ \ &                & $ 0.025\ \, $ & $-0.026\ \, $ & $ 0.021\ \, $ & $-0.100\ \, $  &                & $-0.032\ \, $ & $ 0.001\ \, $ & $ 0.035\ \, $ & $-0.038\ \, $  & LT   \\
 $17.3$\ \ &  $ 0.050\ \, $ & $ 0.026\ \, $ & $-0.026\ \, $ & $ 0.021\ \, $ & $-0.115\ \, $  &  $-0.139\ \, $ & $-0.035\ \, $ & $ 0.000\ \, $ & $ 0.035\ \, $ & $-0.042\ \, $  & LT   \\
 $17.6$\ \ &  $ 0.023\ \, $ & $ 0.012\ \, $ & $-0.029\ \, $ & $ 0.004\ \, $ & $ 0.007\ \, $  &  $-0.139\ \, $ & $-0.036\ \, $ & $ 0.000\ \, $ & $ 0.034\ \, $ & $-0.043\ \, $  & TNG  \\
 $19.3$\ \ &  $ 0.050\ \, $ & $ 0.027\ \, $ & $-0.029\ \, $ & $ 0.011\ \, $ & $-0.121\ \, $  &  $-0.142\ \, $ & $-0.040\ \, $ & $-0.006\ \, $ & $ 0.033\ \, $ & $-0.039\ \, $  & LT   \\
 $21.4$\ \ &  $ 0.050\ \, $ & $ 0.028\ \, $ & $-0.032\ \, $ & $ 0.002\ \, $ & $-0.127\ \, $  &  $-0.145\ \, $ & $-0.044\ \, $ & $-0.012\ \, $ & $ 0.031\ \, $ & $-0.034\ \, $  & LT   \\
 $22.6$\ \ &  $ 0.041\ \, $ & $ 0.015\ \, $ & $ 0.023\ \, $ & $ 0.008\ \, $ & $-0.047\ \, $  &  $-0.147\ \, $ & $-0.046\ \, $ & $-0.015\ \, $ & $ 0.030\ \, $ & $-0.032\ \, $  & NTT  \\
 $26.3$\ \ &  $-0.005\ \, $ & $ 0.013\ \, $ & $-0.037\ \, $ & $-0.016\ \, $ & $-0.069\ \, $  &  $-0.143\ \, $ & $-0.055\ \, $ & $-0.028\ \, $ & $ 0.030\ \, $ & $-0.028\ \, $  & NOT  \\
 $29.4$\ \ &  $-0.005\ \, $ & $ 0.013\ \, $ & $-0.042\ \, $ & $-0.018\ \, $ & $-0.067\ \, $  &  $-0.140\ \, $ & $-0.062\ \, $ & $-0.037\ \, $ & $ 0.031\ \, $ & $-0.025\ \, $  & NOT  \\
 $30.3$\ \ &  $ 0.043\ \, $ & $-0.078\ \, $ &               &               &                &  $-0.139\ \, $ & $-0.065\ \, $ &               &               &                & Ekar \\
 $31.5$\ \ &                & $-0.081\ \, $ & $ 0.050\ \, $ & $ 0.010\ \, $ & $-0.008\ \, $  &                & $-0.067\ \, $ & $-0.044\ \, $ & $ 0.031\ \, $ & $-0.023\ \, $  & Ekar \\
 $34.3$\ \ &  $ 0.071\ \, $ & $ 0.041\ \, $ & $-0.054\ \, $ & $-0.033\ \, $ & $-0.104\ \, $  &  $-0.135\ \, $ & $-0.074\ \, $ & $-0.053\ \, $ & $ 0.029\ \, $ & $-0.020\ \, $  & LT   \\
 $35.5$\ \ &  $-0.004\ \, $ & $ 0.056\ \, $ & $-0.068\ \, $ & $ 0.002\ \, $ & $ 0.075\ \, $  &  $-0.134\ \, $ & $-0.077\ \, $ & $-0.056\ \, $ & $ 0.029\ \, $ & $-0.019\ \, $  & CA   \\
 $36.3$\ \ &  $ 0.075\ \, $ & $ 0.042\ \, $ & $-0.057\ \, $ & $-0.037\ \, $ & $-0.115\ \, $  &  $-0.133\ \, $ & $-0.079\ \, $ & $-0.058\ \, $ & $ 0.028\ \, $ & $-0.018\ \, $  & LT   \\
 $40.4$\ \ &  $ 0.074\ \, $ & $ 0.042\ \, $ & $-0.055\ \, $ & $-0.033\ \, $ & $-0.131\ \, $  &  $-0.133\ \, $ & $-0.078\ \, $ & $-0.055\ \, $ & $ 0.030\ \, $ & $-0.019\ \, $  & LT   \\
 $41.4$\ \ &  $-0.006\ \, $ & $ 0.014\ \, $ & $-0.050\ \, $ & $-0.024\ \, $ & $-0.061\ \, $  &  $-0.133\ \, $ & $-0.078\ \, $ & $-0.055\ \, $ & $ 0.030\ \, $ & $-0.019\ \, $  & NOT  \\
 $44.4$\ \ &  $ 0.072\ \, $ & $ 0.041\ \, $ & $-0.054\ \, $ & $-0.029\ \, $ & $-0.146\ \, $  &  $-0.133\ \, $ & $-0.077\ \, $ & $-0.053\ \, $ & $ 0.032\ \, $ & $-0.020\ \, $  & LT   \\
 $48.3$\ \ &  $ 0.071\ \, $ & $ 0.041\ \, $ & $-0.052\ \, $ & $-0.025\ \, $ & $-0.162\ \, $  &  $-0.134\ \, $ & $-0.076\ \, $ & $-0.050\ \, $ & $ 0.033\ \, $ & $-0.021\ \, $  & LT   \\
 $48.5$\ \ &  $-0.006\ \, $ & $ 0.013\ \, $ & $-0.048\ \, $ & $-0.022\ \, $ & $-0.061\ \, $  &  $-0.134\ \, $ & $-0.076\ \, $ & $-0.050\ \, $ & $ 0.034\ \, $ & $-0.021\ \, $  & NOT  \\
 $59.3$\ \ &  $ 0.068\ \, $ & $ 0.038\ \, $ & $-0.047\ \, $ & $-0.013\ \, $ & $-0.204\ \, $  &  $-0.132\ \, $ & $-0.073\ \, $ & $-0.043\ \, $ & $ 0.039\ \, $ & $-0.023\ \, $  & LT   \\
 $60.3$\ \ &  $-0.005\ \, $ & $ 0.012\ \, $ & $-0.044\ \, $ & $-0.019\ \, $ & $-0.063\ \, $  &  $-0.132\ \, $ & $-0.073\ \, $ & $-0.042\ \, $ & $ 0.039\ \, $ & $-0.024\ \, $  & NOT  \\
 $62.4$\ \ &  $ 0.067\ \, $ & $ 0.037\ \, $ & $-0.046\ \, $ & $-0.010\ \, $ & $-0.215\ \, $  &  $-0.131\ \, $ & $-0.073\ \, $ & $-0.040\ \, $ & $ 0.040\ \, $ & $-0.024\ \, $  & LT   \\
 $66.3$\ \ &  $ 0.066\ \, $ & $ 0.038\ \, $ & $-0.044\ \, $ & $-0.006\ \, $ & $-0.214\ \, $  &  $-0.133\ \, $ & $-0.073\ \, $ & $-0.036\ \, $ & $ 0.041\ \, $ & $-0.017\ \, $  & LT   \\
 $69.4$\ \ &  $ 0.010\ \, $ & $ 0.018\ \, $ & $-0.051\ \, $ & $-0.019\ \, $ & $ 0.048\ \, $  &  $-0.135\ \, $ & $-0.074\ \, $ & $-0.033\ \, $ & $ 0.042\ \, $ & $-0.012\ \, $  & TNG  \\
 $74.3$\ \ &  $ 0.064\ \, $ & $ 0.040\ \, $ & $-0.039\ \, $ & $ 0.001\ \, $ & $-0.211\ \, $  &  $-0.137\ \, $ & $-0.074\ \, $ & $-0.027\ \, $ & $ 0.044\ \, $ & $-0.004\ \, $  & LT   \\
 $75.3$\ \ &  $-0.007\ \, $ & $ 0.016\ \, $ & $-0.036\ \, $ & $-0.024\ \, $ & $-0.038\ \, $  &  $-0.138\ \, $ & $-0.074\ \, $ & $-0.026\ \, $ & $ 0.044\ \, $ & $-0.002\ \, $  & NOT  \\
 $78.3$\ \ &  $ 0.063\ \, $ & $ 0.041\ \, $ & $-0.037\ \, $ & $ 0.005\ \, $ & $-0.210\ \, $  &  $-0.139\ \, $ & $-0.075\ \, $ & $-0.023\ \, $ & $ 0.045\ \, $ & $ 0.003\ \, $  & LT   \\
 $81.4$\ \ &  $ 0.051\ \, $ & $ 0.020\ \, $ & $ 0.024\ \, $ & $ 0.007\ \, $ & $-0.005\ \, $  &  $-0.141\ \, $ & $-0.075\ \, $ & $-0.020\ \, $ & $ 0.046\ \, $ & $ 0.008\ \, $  & NTT  \\
 $85.3$\ \ &  $ 0.066\ \, $ & $ 0.043\ \, $ & $-0.031\ \, $ & $ 0.008\ \, $ & $-0.212\ \, $  &  $-0.139\ \, $ & $-0.075\ \, $ & $-0.015\ \, $ & $ 0.046\ \, $ & $ 0.009\ \, $  & LT   \\
 $86.3$\ \ &  $ 0.067\ \, $ & $ 0.043\ \, $ & $-0.031\ \, $ & $ 0.007\ \, $ & $-0.213\ \, $  &  $-0.138\ \, $ & $-0.075\ \, $ & $-0.013\ \, $ & $ 0.047\ \, $ & $ 0.010\ \, $  & LT   \\
 $87.3$\ \ &  $ 0.067\ \, $ & $ 0.044\ \, $ & $-0.030\ \, $ & $ 0.007\ \, $ & $-0.214\ \, $  &  $-0.138\ \, $ & $-0.075\ \, $ & $-0.012\ \, $ & $ 0.047\ \, $ & $ 0.010\ \, $  & LT   \\
 $90.3$\ \ &  $-0.009\ \, $ & $ 0.020\ \, $ & $-0.026\ \, $ & $-0.023\ \, $ & $-0.022\ \, $  &  $-0.136\ \, $ & $-0.076\ \, $ & $-0.008\ \, $ & $ 0.047\ \, $ & $ 0.011\ \, $  & NOT  \\
 $90.3$\ \ &  $ 0.070\ \, $ & $ 0.045\ \, $ & $-0.027\ \, $ & $ 0.007\ \, $ & $-0.216\ \, $  &  $-0.136\ \, $ & $-0.076\ \, $ & $-0.008\ \, $ & $ 0.047\ \, $ & $ 0.011\ \, $  & LT   \\
 $96.3$\ \ &  $-0.005\ \, $ & $ 0.071\ \, $ & $-0.021\ \, $ & $ 0.002\ \, $ & $ 0.020\ \, $  &  $-0.133\ \, $ & $-0.076\ \, $ & $ 0.001\ \, $ & $ 0.048\ \, $ & $ 0.013\ \, $  & CA   \\
$100.4$\ \ &                & $ 0.048\ \, $ & $-0.020\ \, $ & $ 0.007\ \, $ & $-0.222\ \, $  &                & $-0.077\ \, $ & $ 0.006\ \, $ & $ 0.048\ \, $ & $ 0.014\ \, $  & LT   \\
$105.3$\ \ &                & $ 0.050\ \, $ & $-0.019\ \, $ & $ 0.009\ \, $ & $-0.222\ \, $  &                & $-0.079\ \, $ & $ 0.010\ \, $ & $ 0.049\ \, $ & $ 0.013\ \, $  & LT   \\
$109.2$\ \ &                & $ 0.075\ \, $ & $-0.021\ \, $ & $ 0.005\ \, $ & $ 0.008\ \, $  &                & $-0.081\ \, $ & $ 0.013\ \, $ & $ 0.049\ \, $ & $ 0.013\ \, $  & CA   \\
$109.3$\ \ &  $-0.012\ \, $ & $ 0.027\ \, $ & $-0.017\ \, $ & $-0.011\ \, $ & $-0.015\ \, $  &  $-0.126\ \, $ & $-0.081\ \, $ & $ 0.013\ \, $ & $ 0.049\ \, $ & $ 0.013\ \, $  & NOT  \\
$114.2$\ \ &                & $-0.077\ \, $ & $-0.007\ \, $ & $-0.006\ \, $ & $-0.001^c$     &                & $-0.083\ \, $ & $ 0.023\ \, $ & $ 0.047\ \, $ & $ 0.013^c   $  & Ekar \\
$116.2$\ \ &                & $-0.076\ \, $ & $-0.010\ \, $ & $-0.006\ \, $ & $-0.001^c$     &                & $-0.084\ \, $ & $ 0.027\ \, $ & $ 0.046\ \, $ & $ 0.013^c   $  & Ekar \\
$116.3$\ \ &                & $ 0.053\ \, $ & $-0.011\ \, $ & $ 0.013\ \, $ & $-0.222^c$     &                & $-0.084\ \, $ & $ 0.027\ \, $ & $ 0.046\ \, $ & $ 0.013^c   $  & LT   \\
$118.2$\ \ &  $ 0.096\ \, $ &               &               &               &                &  $-0.121\ \, $ &               &               &               &                & LT   \\
$119.3$\ \ &                & $ 0.054\ \, $ & $-0.008\ \, $ & $ 0.015\ \, $ & $-0.222^c$     &                & $-0.086\ \, $ & $ 0.032\ \, $ & $ 0.045\ \, $ & $ 0.013^c$     & LT   \\
$125.2$\ \ &                & $ 0.087\ \, $ & $ 0.000\ \, $ & $ 0.005\ \, $ & $ 0.008^c$     &                & $-0.089\ \, $ & $ 0.044\ \, $ & $ 0.042\ \, $ & $ 0.013^c$     & CA   \\
$126.2$\ \ &                & $ 0.055\ \, $ & $-0.001\ \, $ & $ 0.018\ \, $ & $-0.222^c$     &                & $-0.089\ \, $ & $ 0.046\ \, $ & $ 0.041\ \, $ & $ 0.013^c$     & LT   \\
$130.3$\ \ &                & $ 0.056\ \, $ & $ 0.000\ \, $ & $ 0.014\ \, $ & $-0.222^c$     &                & $-0.091\ \, $ & $ 0.049\ \, $ & $ 0.040\ \, $ & $ 0.013^c$     & LT   \\
$132.3$\ \ &                & $ 0.090\ \, $ & $ 0.004\ \, $ & $ 0.005\ \, $ &                &                & $-0.091\ \, $ & $ 0.050\ \, $ & $ 0.039\ \, $ &                & CA   \\
$137.3$\ \ &                & $ 0.058\ \, $ & $ 0.001\ \, $ & $ 0.004\ \, $ & $-0.222^c$     &                & $-0.093\ \, $ & $ 0.052\ \, $ & $ 0.036\ \, $ & $ 0.013^c$     & LT   \\
$139.2$\ \ &  $ 0.104^c$    &               &               &               &                &  $-0.116^c$    &               &               &               &                & LT   \\
$145.3$\ \ &                &               & $ 0.002\ \, $ & $-0.008\ \, $ &                &                &               & $ 0.055\ \, $ & $ 0.032\ \, $ &                & LT   \\
$151.2$\ \ &                & $ 0.062\ \, $ & $ 0.002\ \, $ & $-0.016\ \, $ & $-0.222^c$     &                & $-0.097\ \, $ & $ 0.057\ \, $ & $ 0.029\ \, $ & $ 0.013^c$     & LT   \\
$166.2$\ \ &                & $ 0.066\ \, $ & $ 0.003\ \, $ & $-0.038\ \, $ & $-0.222^c$     &                & $-0.102\ \, $ & $ 0.062\ \, $ & $ 0.022\ \, $ & $ 0.013^c$     & LT   \\
$169.2$\ \ &                &               & $ 0.004\ \, $ & $-0.042\ \, $ & $-0.222^c$     &                &               & $ 0.063\ \, $ & $ 0.020\ \, $ & $ 0.013^c$     & LT   \\
$262.6$\ \ &                &  $ 0.053\ \, $& $ 0.018\ \, $ & $-0.106\ \, $ &                &                & $-0.130\ \, $ & $ 0.098\ \, $ & $-0.027\ \, $ &                & NOT  \\
$300.7$\ \ &                &  $ 0.040^c$   &  $-0.099^c$   &  $ 0.054^c$   &  $-0.011^c$    &                & $-0.140^c$    &  $ 0.110^c$   & $-0.043^c$    &  $  0.013^c$   & NTT  \\
$313.7$\ \ &                &  $ 0.040^c$   &  $-0.099^c$   &  $ 0.054^c$   &  $-0.011^c$    &                & $-0.140^c$    &  $ 0.110^c$   & $-0.043^c$    &  $  0.013^c$   & NTT  \\
$316.8$\ \ &                &  $ 0.040^c$   &  $-0.099^c$   &  $ 0.054^c$   &                &                & $-0.140^c$    &  $ 0.110^c$   & $-0.043^c$    &                & NTT  \\
$328.6$\ \ &                &               &               &  $-0.126^c$   &  $-0.015^c$    &                &               &               & $-0.043^c$    &  $  0.013^c$   & NOT  \\
$426.3$\ \ &                &               &  $ 0.035^c$   &               &  $  0.059^c$   &                &               &  $ 0.110^c$   &               &  $  0.013^c$   & TNG  \\
\hline
\end{tabular}
\\[1.5ex]
\flushleft
$^a$~Phase in days with respect to $B$-band maximum JD $= 2\,454\,947.1 \pm 0.3$.\quad 
$^b$~See Table~\ref{SN_mags} for details.\quad
$^c$~Constant extrapolation to later epochs.\\
\end{scriptsize}
\end{center}
\end{table*}

\label{lastpage}

\end{document}